\documentclass[11pt,a4paper]{article}

\newcommand{\preprint}[1]{\begin{table}[t]  
             \begin{flushright}               
             {#1}                             
             \end{flushright}                 
             \end{table}}                     
\renewcommand{\title}[1]{\vbox{\center\LARGE{#1}}\vspace{5mm}}
\renewcommand{\author}[1]{\vbox{\center#1}\vspace{5mm}}
\newcommand{\address}[1]{\vbox{\center\em#1}}
\newcommand{\email}[1]{\vbox{\center\tt#1}\vspace{5mm}}

\renewcommand{\date}[1]{\vbox{\center#1}}

\usepackage{amssymb}
\usepackage{amsmath,bm}
\usepackage{amssymb}
\usepackage{graphicx}
\usepackage{amsfonts}         
\usepackage{fancybox}   

\usepackage{enumitem}

\usepackage{slashed}

\usepackage{epstopdf}
\usepackage[usenames,dvipsnames]{xcolor}
\usepackage[pagebackref, bookmarks={false}, pdfauthor={JD}, pdftitle={aitojibonkalida}]{hyperref}
\hypersetup{colorlinks=true, linkcolor=darkblue, citecolor=red, filecolor=OliveGreen, urlcolor=blue, filebordercolor={.8 .8 1}, urlbordercolor={.8 .8 0}}
\usepackage{soul}
\setstcolor{Red}

\usepackage{wrapfig}

\definecolor{jazzberryjam}{rgb}{0.65, 0.04, 0.37}
\definecolor{lust}{rgb}{0.9, 0.13, 0.13}
\definecolor{sandybrown}{rgb}{0.96, 0.64, 0.38}
\definecolor{mountainmeadow}{rgb}{0.19, 0.73, 0.56}
\definecolor{glaucous}{rgb}{0.38, 0.51, 0.71}
\definecolor{chromeyellow}{rgb}{1.0, 0.65, 0.0}
\definecolor{emerald}{rgb}{0.31, 0.78, 0.47}
\definecolor{deepsaffron}{rgb}{1.0, 0.6, 0.2}
\definecolor{darkgreen}{rgb}{0,0.4,0}
\definecolor{darkred}{rgb}{0.4,0,0}
\definecolor{darkblue}{rgb}{0,0,0.4}
\definecolor{lightblue}{rgb}{.6,.6,0.9}

\definecolor{uglybrown}{rgb}{0.8,  0.7,  0.5}

\definecolor{palatinatepurple}{rgb}{0.41, 0.16, 0.38}
\definecolor{celebrationcolor}{rgb}{0.75,  0.0,  0.9}

 \usepackage{mdframed}

\usepackage{framed}
\definecolor{shadecolor}{rgb}{0.90,0.90,0.90}

\usepackage{tkz-euclide}

\usepackage{makeidx}
\usepackage{cite}
\usepackage{bm}
\usepackage{geometry}
\usepackage{braket}
\usepackage[margin=20pt,small]{caption}

\usepackage[toc]{appendix}

\usepackage{tikz}
\usetikzlibrary{calc,decorations.markings,arrows.meta,shapes.misc,decorations.pathmorphing,calc,bending}

\usetikzlibrary{arrows.meta,shapes.misc,decorations.pathmorphing,calc,bending}

\tikzset{
  branch point/.style={cross out,draw=black,fill=none,minimum size=2*(#1-\pgflinewidth),inner sep=0pt,outer sep=0pt}, 
  branch point/.default=5
}
\tikzset{
  branch cut/.style={
    decorate,decoration=snake,
    to path={
      (\tikztostart) -- (\tikztotarget) \tikztonodes
    },
    }
  }

\input epsf

\newlength{\extraspace}
\newlength{\extraspaces}
\def\be{\begin{equation}}
\def\ee{\end{equation}}

\newcommand{\bea}{\begin{eqnarray}}
\newcommand{\eea}{\end{eqnarray}}

\def\p{\partial}

\def\Tr{{{\rm Tr~ }}}
\def\tr{{\rm tr}}

\def\bra#1{{\langle}#1|}
\def\ket#1{|#1\rangle}

\def\II{\relax{I\kern-.10em I}}

\def\IB{\relax{\rm I\kern-.18em B}}

\def\ID{\relax{\rm I\kern-.18em D}}
\def\IE{\relax{\rm I\kern-.18em E}}
\def\IF{\relax{\rm I\kern-.18em F}}
\def\IG{\relax\hbox{$\inbar\kern-.3em{\rm G}$}}
\def\IGa{\relax\hbox{${\rm I}\kern-.18em\Gamma$}}
\def\IH{\relax{\rm I\kern-.18em H}}
\def\II{\relax{\rm I\kern-.18em I}}
\def\IK{\relax{\rm I\kern-.18em K}}

\def\inbar{\,\vrule height1.5ex width.4pt depth0pt}

\def\p{\partial}

\def\IR{\mathbb{R}}

\def\lp10{\ell_p^{10}}
\def\lp11{\ell_p^{11}}
\def\R11{R_{11}}

\def\frac#1#2{{#1 \over #2}}

\newdimen\tableauside\tableauside=1.0ex
\newdimen\tableaurule\tableaurule=0.4pt
\newdimen\tableaustep
\def\phantomhrule#1{\hbox{\vbox to0pt{\hrule height\tableaurule width#1\vss}}}
\def\phantomvrule#1{\vbox{\hbox to0pt{\vrule width\tableaurule height#1\hss}}}
\def\sqr{\vbox{%
  \phantomhrule\tableaustep
  \hbox{\phantomvrule\tableaustep\kern\tableaustep\phantomvrule\tableaustep}%
  \hbox{\vbox{\phantomhrule\tableauside}\kern-\tableaurule}}}
\def\squares#1{\hbox{\count0=#1\noindent\loop\sqr
  \advance\count0 by-1 \ifnum\count0>0\repeat}}
\def\tableau#1{\vcenter{\offinterlineskip
  \tableaustep=\tableauside\advance\tableaustep by-\tableaurule
  \kern\normallineskip\hbox
    {\kern\normallineskip\vbox
      {\gettableau#1 0 }%
     \kern\normallineskip\kern\tableaurule}%
  \kern\normallineskip\kern\tableaurule}}
\def\gettableau#1 {\ifnum#1=0\let\next=\null\else
  \squares{#1}\let\next=\gettableau\fi\next}

 \def\eqnn#1{\xdef #1{(\secsym\the\meqno)}\writedef{#1\leftbracket#1}%
 \global\advance\meqno by1\wrlabeL#1}
 \def\eqna#1{\xdef #1##1{\hbox{$(\secsym\the\meqno##1)$}}
 \writedef{#1\numbersign1\leftbracket#1{\numbersign1}}%
 \global\advance\meqno by1\wrlabeL{#1$\{\}$}}
 \def\eqn#1#2{\xdef #1{(\secsym\the\meqno)}\writedef{#1\leftbracket#1}%
 \global\advance\meqno by1$$#2\eqno#1\eqlabeL#1$$}

\global\newcount\itemno \global\itemno=0

\def\itemaut#1{\global\advance\itemno by1\noindent\item{\the\itemno.}#1}

\def\({\left(}
\def\){\right)}

\def\lsim{\mathrel{\mathstrut\smash{\ooalign{\raise2.5pt\hbox{$<$}\cr\lower2.5pt\hbox{$\sim$}}}}}
\def\gsim{\mathrel{\mathstrut\smash{\ooalign{\raise2.5pt\hbox{$>$}\cr\lower2.5pt\hbox{$\sim$}}}}}

\def\overleftrightarrow#1{\vbox{\ialign{##\crcr
     $\leftrightarrow$\crcr\noalign{\kern-0pt\nointerlineskip}
     $\hfil\displaystyle{#1}\hfil$\crcr}}}
     
     \def\overleftarrow#1{\vbox{\ialign{##\crcr
     $\leftarrow$\crcr\noalign{\kern-0pt\nointerlineskip}
     $\hfil\displaystyle{#1}\hfil$\crcr}}}

\usepackage[yyyymmdd,hhmmss]{datetime}
\newif{\ifeq}           
\eqtrue                 
\newcounter{lecturecounter}

\usepackage[utf8]{inputenc}
\usepackage{comment}
\usepackage{float, xcolor}
\usepackage{physics}
\usepackage{array}
\usepackage{ragged2e}
\usepackage{booktabs}
\usepackage{latexsym}
\usepackage{mathtools}

\usepackage{subcaption}
\usepackage{graphicx}
\usepackage{cleveref}
\usepackage{amsfonts,amsmath,bm}
\usepackage[toc]{appendix}

\numberwithin{equation}{section}
\definecolor{darkgreen}{rgb}{0,0.5,0}
\definecolor{darkblue}{rgb}{0,0,0.6}

\hypersetup{
	citecolor=red,
	colorlinks=true,
	filecolor=red,
	linkcolor=blue,
	linktocpage=true,
	urlcolor=darkgreen
}

\newcount\hour
\newcount\minute
\newtoks\amorpm
\hour=\time\divide\hour by60
\minute=\time{\multiply\hour by60 \global\advance\minute by-\hour}
\edef\standardtime{{\ifnum\hour<12 \global\amorpm={am}%
		\else\global\amorpm={pm}\advance\hour by-12 \fi
		\ifnum\hour=0 \hour=12 \fi
		\number\hour:\ifnum\minute<10 0\fi\number\minute\the\amorpm}}
\edef\militarytime{\number\hour:\ifnum\minute<10 0\fi\number\minute}

\def\draftlabel#1{{\@bsphack\if@filesw {\let\thepage\relax
			\xdef\@gtempa{\write\@auxout{\string
					\newlabel{#1}{{\@currentlabel}{\thepage}}}}}\@gtempa
		\if@nobreak \ifvmode\nobreak\fi\fi\fi\@esphack}
	\gdef\@eqnlabel{#1}}
\def\@eqnlabel{}
\def\@vacuum{}
\def\draftmarginnote#1{\marginpar{\raggedright\scriptsize\tt#1}}

\def\draft{\oddsidemargin -.5truein
		\def\@oddfoot{\sl preliminary draft \hfil
			\rm\thepage\hfil\sl\today\quad\militarytime}
		\let\@evenfoot\@oddfoot \overfullrule 3pt
		\let\label=\draftlabel
		\let\marginnote=\draftmarginnote
		\def\@eqnnum{(\theequation)\rlap{\kern\marginparsep\tt\@eqnlabel}%
			\global\let\@eqnlabel\@vacuum}  }
	
	\def\preprint{\twocolumn\sloppy\flushbottom\parindent 2em
		\leftmargini 2em\leftmarginv .5em\leftmarginvi .5em
		\oddsidemargin -.5in    \evensidemargin -.5in
		\columnsep .4in \footheight 0pt
		\textwidth 10.in        \topmargin  -.4in
		\headheight 12pt \topskip .4in
		\textheight 6.9in \footskip 0pt
		\def\@oddhead{\thepage\hfil\addtocounter{page}{1}\thepage}
		\let\@evenhead\@oddhead \def\@oddfoot{} \def\@evenfoot{} }
	
	\def\numberbysection{\@addtoreset{equation}{section}
		\def\theequation{\thesection.\arabic{equation}}}
	
	\def\underline#1{\relax\ifmmode\@@underline#1\else
		$\@@underline{\hbox{#1}}$\relax\fi}

	\def\titlepage{\@restonecolfalse\if@twocolumn\@restonecoltrue\onecolumn
		\else \newpage \fi \thispagestyle{empty}\c@page\z@
		\def\thefootnote{\fnsymbol{footnote}} }
	
	\def\endtitlepage{\if@restonecol\twocolumn \else \newpage \fi
		\def\thefootnote{\arabic{footnote}}
		\setcounter{footnote}{0}}  

\catcode`@=12
\relax

\def\figcap{\section*{Figure Captions\markboth
		{FIGURECAPTIONS}{FIGURECAPTIONS}}\list
	{Figure \arabic{enumi}:\hfill}{\settowidth\labelwidth{Figure
			999:}
		\leftmargin\labelwidth
		\advance\leftmargin\labelsep\usecounter{enumi}}}
 \relax
\def\tablecap{\section*{Table Captions\markboth
		{TABLECAPTIONS}{TABLECAPTIONS}}\list
	{Table \arabic{enumi}:\hfill}{\settowidth\labelwidth{Table
			999:}
		\leftmargin\labelwidth
		\advance\leftmargin\labelsep\usecounter{enumi}}}
 \relax
\def\reflist{\section*{References\markboth
		{REFLIST}{REFLIST}}\list
	{[\arabic{enumi}]\hfill}{\settowidth\labelwidth{[999]}
		\leftmargin\labelwidth
		\advance\leftmargin\labelsep\usecounter{enumi}}}
 \relax
\makeatletter
\newcounter{pubctr}
\def\publist{\@ifnextchar[{\@publist}{\@@publist}}
\def\@publist[#1]{\list
	{[\arabic{pubctr}]\hfill}{\settowidth\labelwidth{[999]}
		\leftmargin\labelwidth
		\advance\leftmargin\labelsep
		\@nmbrlisttrue\def\@listctr{pubctr}
		\setcounter{pubctr}{#1}\addtocounter{pubctr}{-1}}}
\def\@@publist{\list
	{[\arabic{pubctr}]\hfill}{\settowidth\labelwidth{[999]}
		\leftmargin\labelwidth
		\advance\leftmargin\labelsep
		\@nmbrlisttrue\def\@listctr{pubctr}}}
 \relax
\makeatother
\newskip\humongous \humongous=0pt plus 1000pt minus 1000pt

\newif\ifdtup

\relax

\def\be{\begin{equation}}
	\def\ee{\end{equation}}
\def\ba{\begin{eqnarray}}
	\def\ea{\end{eqnarray}}

\def\p{\pi}

\def\S{\Sigma}

\def\cM{{\cal M}}

 \def\cH{{\cal H}}

\def\cM{{\cal M}}  \def\cO{{\cal O}}

\def\IR{\relax{\rm I\kern-.18em R}}

\def\IR{\relax{\rm I\kern-.18em R}}
\def\IL{\relax{\rm I\kern-.18em L}}

\def\inv{^{\raise.15ex\hbox{${\scriptscriptstyle -}$}\kern-.05em 1}}

\def\cM{{\cal M}}

\def\tr{{\rm tr}}
\def\Tr{{\rm Tr}}

\def\bea{\begin{eqnarray}}
	\def\eea{\end{eqnarray}}

\def\p{\pi} 

  \def\S{\Sigma}

\definecolor{markcolor2}{rgb}{1,0,0}

\definecolor{markcolor3}{rgb}{0,1,0}

\begin{document}
	
	\thispagestyle{empty}

	\title{\Large\bf\boldmath A New Genuine Multipartite Entanglement Measure:\\
 from Qubits to Multiboundary Wormholes }


\author{
Jaydeep Kumar Basak,${}^{1}$
Vinay Malvimat,${}^{2,3}$
Junggi Yoon,${}^{2,3,4}$

}

\address{
{\it ${}^1$
	Department of Physics and Photon Science, Gwangju Institute of Science and Technology, \\
123 Cheomdan-gwagiro, Gwangju 61005, Korea\\}
{\it ${}^2$
	Department of Physics, College of Science, Kyung Hee University, Seoul 02447, Republic of Korea\\}
{\it ${}^3$
	Research Institute for Basic Sciences, Kyung Hee University, Seoul 02447, Republic of Korea\\}
{\it ${}^4$
	International Research Center for Quantum Matter, Kyung Hee University, Seoul 02447, Korea\\}
}

\date{}			

\email{ \href{mailto:jkb.hep@gmail.com}{jkb.hep@gmail.com}, \href{vinaymalvimat@khu.ac.kr}{vinaymalvimat@khu.ac.kr},\\ \href{junggi.yoon@khu.ac.kr}{junggi.yoon@khu.ac.kr}

}

\abstract{We introduce the \textit{Latent Entropy} (L-entropy) as a novel measure to characterize the genuine multipartite entanglement in quantum systems. Our measure leverages the upper bound of reflected entropy and its maximal values attained by 2-uniform states for $n$-party ($n= 4,5$) and GHZ state for 3-party quantum systems. We demonstrate that the measure is a non-negative, local unitary invariant function which vanishes for separable states. We then analyze its interesting characteristics in spin chain models and the Sachdev-Ye-Kitaev (SYK) model. Subsequently, we explore its implications to holography by deriving a Page-like curve for the L-entropy in the CFT dual to a multi-boundary wormhole model. Furthermore, we examine the behavior of L-entropy in Haar random states, deriving analytical expressions and validating them against numerical results.  In particular, we show that for $n =5$, random states approximate 2-uniform states with maximal multipartite entanglement. Furthermore, we propose a potential connection between random states and multi-boundary wormhole geometries. Extending to finite-temperature systems, we introduce the Multipartite Thermal Pure Quantum (MTPQ) state, a multipartite generalization of the thermal pure quantum state, and explore its entanglement properties. By incorporating state-dependent construction of the MTPQ state, we resolve the factorization issue in the random average of the MTPQ state, ensuring consistency with the correlation functions in the holographic dual multiboundary wormhole. Finally, we apply this construction to the multi-copy SYK model and examine its multipartite entanglement structure.

}

\addtocontents{toc}{\protect\setcounter{tocdepth}{2}}
\tableofcontents

\newpage

\section{Introduction }
\label{sec_intro}

The notion of entanglement has been pivotal across diverse domains of physics, encompassing fields from quantum information theory to black hole physics. In the context of a system in a pure quantum state that is partitioned into two distinct segments (bipartite system), the entanglement entropy functions as a unique measure for quantifying entanglement. One of the key breakthroughs in quantum information theory revealed that when there are numerous copies of a given pure state, the entanglement entropy dictates the upper limit on the number of Bell pairs that you can asymptotically extract via local operations \cite{Bennett:1995tk}.

A natural question in this context is whether the idea of entanglement can be broadened to encompass systems involving multiple parties. If so, the immediate inquiry becomes: what are the measures to characterize multipartite entanglement? A pure quantum state involving multiple parties is said to possess genuine multipartite entanglement if it can not be factorized across any bi-partition\footnote{Observe that in this context, bipartition can in general encompass subsystems formed by combining multiple parties as long as we ensure that the two partitions collectively constitute the entire system. However, for most of this article bipartite systems refers to a `two-party system' which is comprised of strictly two separate parties. The context should clarify the specific instance being referenced.}. Quite intriguingly, it was found that even in the simplest example involving three qubits, it was discovered that there are two distinct in-equivalent classes of states with genuine tripartite entanglement \cite{Dur:2000zz}.
This classification arises by enlarging the notion of equivalence based on deterministic local transformations assisted by classical communication (LOCC), where two states are regarded as equivalent if they can be converted into one another with unit probability. Allowing instead stochastic transformations, namely those that succeed with any non-vanishing probability, leads to the broader notion of stochastic local operations and classical communication (SLOCC). The entanglement structures of these two classes of states possessing genuine tripartite entanglement in the 3 qubit example are very different. The GHZ and W states are exemplars of these two categories. The defining characteristic multipartite entanglement in the GHZ state is such that upon reduction of one of the qubits results in vanishing residual bipartite entanglement between the remaining two qubits. On the other hand, the W state retains a substantial degree of bipartite entanglement even after one of the qubit is lost or traced over\cite{Dur:2000zz}.

Within this framework, a pure multipartite state is said to possess genuine multipartite entanglement if it is inseparable across every possible bipartition of the system. Correspondingly, a measure of genuine multipartite entanglement should assign a nonzero value to such states, while vanishing for states that are separable across at least one bipartition. In addition, one generally requires such a measure to satisfy appropriate monotonicity properties under local operations, so that it faithfully captures multipartite entanglement as a resource \cite{Vidal:1998re,Ma:2023ecg,Horodecki:2024bgc,Gadde:2024jfi}. Various measures have been used to characterize tripartite entanglement in a three-qubit scenario including tangle, concurrence fill, and the geometric mean of concurrences etc\cite{PhysRevA.68.042307,PhysRevA.69.062311,5075874,LEE20083157,PhysRevLett.127.040403,PhysRevLett.115.030505,Choi:2022lge,Ge:2022sqp,Khan:2023pny} (see \cite{Ma:2023ecg,Horodecki:2024bgc} and references therein for a more exhaustive list and detailed discussion).

The majority of the previously discussed measures are easily computable only for quantum systems with a limited number of qubits. Nevertheless, for quantum systems characterized by higher-dimensional Hilbert spaces, particularly in the context of Quantum Field Theories (QFTs), where the Hilbert spaces are generally infinite-dimensional, there exists a significant opportunity to conduct comprehensive investigations aimed at developing and proposing novel computable measures to characterize multipartite entanglement. In this context, holography has emerged as a crucial tool and promises to lead to significant progress. Recently, there has been a substantial increase in interest in exploring the multipartite entanglement structure of holographic states. Generic holographic states have been shown to exhibit substantial tripartite entanglement in Ref.~\cite{Akers:2019gcv}. The results in Ref.~\cite{Akers:2019gcv} were based on a particular measure known as reflected entropy. This quantity is defined as the von Neumann entropy of a subsystem $A$ and its copy denoted by $A^*$ in the canonical state $\ket{\sqrt{\rho_{AB}}}$ which serves as a purification for the mixed bipartite state $\rho_{AB}$. Note that the mixed state $\rho_{AB}$ itself may be acquired by tracing out a part of the pure tripartite state $\ket{\psi}_{ABC}$. The authors in Ref.~\cite{Akers:2019gcv}, demonstrated that the difference between reflected entropy and mutual information of $AB$ being large is a sign of the presence of a substantial amount of tripartite entanglement. Subsequently, Ref.~\cite{Hayden:2021gno} provided evidence that this specific difference, referred to as the Markov gap, is linked to the Markov recovery process of a tripartite state $\ket{\psi}_{ABC}$ from its reduced state $\rho_{AB}$ via a quantum channel $R_{B\to BC}$.  This supports the notion that the Markov gap is a measure for characterizing tripartite entanglement in the pure state $\ket{\psi}_{ABC}$. Furthermore, it was shown that in $AdS_3/CFT_2$ this quantity is bounded by the number of boundaries of the wedge cross section of the entanglement (EWCS) in the dual bulk geometry. Quite interestingly, it has also been shown to exhibit certain universal features in condensed matter systems \cite{Zou:2020bly}. More recently, in Ref.~\cite{Gadde:2022cqi}, the authors propose a novel measure termed multi-entropy to characterize multipartite entanglement in a generic quantum system and explore its behavior for holographic states. Additionally, in a series of articles~\cite{Gadde:2023zzj,Gadde:2023zni,Gadde:2024jfi,Gadde:2024taa} the authors examine several generic properties of multipartite entanglement monotones in pure states and establish multi-entropy as a viable measure.  This quantity has been thoroughly investigated in two-dimensional conformal field theories (CFTs) using the replica technique in \cite{Harper:2024ker}.

Note that although the Markov gap mentioned above serves as a good measure to characterize multipartite entanglement in certain classes of states such as the $W$ state, it fails to characterize multipartite entanglement in a whole class of quantum states. For example, even in the simplest example of the three qubit case it uniformly vanishes for a family of states with genuine tripartite entanglement known as the generalized GHZ states. Furthermore, we will demonstrate in this article that there are several such states in 4 qubit case as well. This raises the question whether there exists a different measure related to the reflected entropy that can be utilized to characterize the genuine multipartite entanglement present in a generic $n$-party state. In the present article, we address this issue by introducing a novel measure constructed from the upper bound of the reflected entropy which we refer to as the bipartite latent entropy or the L-entropy. We propose the multipartite generalization of this quantity to be the geometric mean of all bipartite L-entropies. We show that this measure is positive, vanishes for separable states, and is invariant under local unitary transformations. It therefore provides a characterization of genuine multipartite entanglement for $n$-party pure states with $n \leq 5$. For $n>5$, however, genuine multipartite entanglement is characterized by a generalized version of the L-entropy~\cite{newlentropy}.

In the example involving 3 qubits pure states we demonstrate that our measure achieves its maximal possible value for the GHZ state as expected from quantum information theory. Furthermore, this result holds for any three-party pure state. For an $n$-party pure state involving more than three parties, we demonstrate that our measure attains its maximum value for a 2-uniform state, which is defined as a state in which every two-party reduced density matrix is maximally mixed. We also explore the connection between maximal multipartite entanglement and $k$-uniform states.
Once the measure is established, we initially apply it to the Ising model and SYK scenario. We note that both L-entropy and Markov gap display an oscillatory pattern for Ising-like interactions, with this pattern dependent on the initial state and  the interaction type. In contrast, for the SYK, the L-entropy approaches a value close to its peak for sufficiently large number of parties, while the Markov gap becomes negligible. Subsequently, we calculate the multipartite L-entropy for Haar random states involving 3, 4, and 5 parties. Notably, we find that for 3 parties, the L-entropy is an $O(1)$ constant, whereas for 4 parties, it grows with the Hilbert space dimension, yet does not reach its theoretical maximum. Furthermore, we show that the 5 party L-entropy achieves its maximum at the leading order.

 Next, we explore the holographic scenario corresponding to the multipartite entanglement and its manifestation. The components of the bipartite L-entropy possess two different holographic realizations. The holographic entanglement entropy and the reflected entropy are duals to the Ryu-Takayanagi~\cite{Ryu:2006bv} surface and the entanglement wedge cross sections~\cite{Dutta:2019gen} respectively. We use these two holographic quantities to construct the bipartite L-entropy and subsequently the multipartite L-entropy. However, within the framework of multipartite entanglement in holography, the multi-boundary wormhole models have become the most crucial~\cite{Krasnov:2000zq,Skenderis:2009ju,Balasubramanian:2014hda,Balasubramanian:2024ysu}. This model provides the holographic manifestation of a natural multiparty generalization of the Thermo Field Dynamics/Double (TFD) state $|\Sigma\rangle_n$ in the Hilbert space $\mathcal{H}^{\otimes n}$ where each Hilbert space $\mathcal{H}$ represents a boundary of the spacetime geometry. 
Recently in Ref.~\cite{Akers:2019nfi}, this model has also been used to understand the black hole information loss paradox by considering one of the boundaries as an evaporating black hole. Here we utilize this particular scenario with a three boundary wormhole and explore the evolution of the tripartite L-entropy corresponding to the black hole evaporation procedure. Interestingly, we observe a novel characteristic of the multipartite entanglement in this process where the L-entropy attains the maximum when all three boundaries or subsystems are of equal sizes. The maximum L-entropy reflects that all degrees of freedom in each subsystem contribute fully to the construction of tripartite entanglement within the system. However, the situation is more complicated with multi-boundary wormholes with four or more boundaries as there are more parameters involved in the characterization of L-entropy. We adapt an analytical procedure to understand the nature of multipartite entanglement in different parameter regions of a four boundary wormhole scenario.

We then describe how the notion of temperature can be ascribed to a multipartite variation of the thermal pure quantum state (TPQ) as a method to study the multipartite entanglement at finite temperature. In order to do this we consider an extension of the canonical TPQ state which we term as the multipartite thermal pure quantum state (MTPQ). Furthermore, we propose a generalization of the notion of a $k$-uniform state for finite temperatures which we refer to as a thermal $k$-uniform state. We then emphasize on a state-dependent construction of the MTPQ state, to resolve the factorization issue in its random average, ensuring consistency with the correlation functions of the holographic dual to a multi-boundary wormhole. Lastly, we apply this framework to the multi-copy SYK model and analyze its multipartite entanglement structure. By analyzing the entanglement entropy, relative entropy, and energy eigenvalues of the Hamiltonian, we demonstrate that $3$-party, $4$-party, and $5$-party MTPQ states exhibit thermal behavior at the level of each individual party. In the $5$-party case, we further show that the L-entropy aligns with the behavior of a thermal 2-uniform state.

The structure of the paper is as follows. In \cref{rev_all}, we review the reflected entropy and Markov gap, describe the essential properties of a genuine multipartite entanglement (GME) measure, and discuss why the Markov gap fails to qualify as one. In \cref{set_up}, we introduce the bipartite and tripartite L-entropy, compute this measure for various three-party entangled states, and demonstrate its invariance under local unitaries and that its vanishing behaviour for any separable state. This section concludes with an exploration of tripartite L-entropy in spin chain systems. In \cref{N_party_L}, we extend L-entropy to mixed states using the convex roof extension and propose reflected negativity as an entanglement monotone for mixed states. After a brief review of $k$-uniform state we show that bipartite L-entropy is maximized for 2-uniform states and discuss its application to spin chain systems. Section \ref{holography} examines multipartite L-entropy for random states and its holographic realization in a multi-boundary scenario, where we also derive the Page curve for L-entropy in black hole evaporation. 
In \cref{sec_temp_multi}, we introduce temperature to multipartite states via the multipartite thermal pure quantum (MTPQ) state, analyze the dynamics of L-entropy in the multi-copy SYK model, and investigate the notion of thermal $k$-uniform states. Finally, \cref{Summary and Discussion} summarizes our findings and presents the conclusions.

\section{Canonical purification and genuine multipartite entanglement}\label{rev_all}
\subsection{Canonical purification and  Markov gap}\label{sec_rev}
In this section, we provide a brief overview of the notion of canonical purification and its connection to reflected entropy as described in \cite{Dutta:2019gen}. We then summarize the findings of \cite{Hayden:2021gno}, where it was proposed that the Markov gap, a quantity derived from the lower bound of the reflected entropy, serves as a measure of tripartite entanglement. Following this, in the next section we describe the limitations of the Markov gap in capturing the tripartite entanglement of the GHZ state, suggesting that it is not a measure of genuine tripartite entanglement.

Consider a  mixed state $\rho_{A}$ defined on a subsystem $A$. The process of constructing a pure state $\ket{\psi}_{AA^*}$ in a higher-dimensional Hilbert space associated with a system $AA^*$, such that tracing over the auxiliary subsystem $A^*$ yields the original density matrix $\rho_A$, is known as \textit{purification}. However, this procedure is not unique, implying that multiple pure states can correspond to the same mixed state upon tracing over the auxiliary subsystem $A^*$. The most familiar form of purification is the \textit{canonical purification}, where $A^*$ is an identical copy of $A$. This method is notably used in the purification of the thermal state $\rho_L^{\text{th}}$ into the thermofield dynamics state $\ket{\text{TFD}}_{LR}$. In such a purification, the entanglement entropy of the subsystem $L$ is equivalent to the thermal entropy due to the specific nature of the purification.

The authors of \cite{Dutta:2019gen} observed that this procedure could be generalized by considering a mixed state $\rho_{AB}$ on a bipartite system. Expressing $\rho_{AB}$ in an orthonormal basis $\ket{\phi_{i}}$, we have:
\begin{align}
    \rho_{\textrm{\tiny AB}}=\sum_i p_i \ket{\phi_{i}}\bra{\phi_i}
\end{align}
where $p_i$ are probabilities such that $\sum_i p_i=1$
Such a mixed state could be purified canonically by considering the following pure state in the doubled the Hilbert space as follows
\begin{align}
    \ket{\sqrt{\rho_{{\textrm{\tiny AB}}}}}_{\textrm{\tiny $ABA^*B^*$}}=\sum_i \sqrt{p_i} \ket{\phi_{i}}\ket{\phi_{i}}
\end{align}
In this canonical purified state the reflected entropy is defined as the entanglement entropy of $AA^*$
\begin{align}\label{def_SR}
S_{R}(A:B)=S_{AA^*}=-tr(\rho_{A}\log\rho_A)
\end{align}
The reflected entropy exhibits several interesting properties, as noted in\cite{Dutta:2019gen}

\begin{itemize}
    \item It vanishes for a factorized state
    \begin{align}
        \rho_{AB}=\rho_{A}\otimes \rho_{B}\implies S_{R}(A:B)=0
    \end{align}
    \item For a pure state it is twice the entanglement entropy of the individual subsystems
     \begin{align}
        \rho_{AB}=\ket{\psi}\bra{\psi} \implies S_{R}(A:B)=2 S_A
    \end{align}

    \item The most important property is that it is bounded from above and below
    \begin{align}\label{lub}
	 	2	\min \{ S(A),  S(B)\} \geq S_R(A: B) \geq I(A: B).
    \end{align}
    The lower bound is just re-stating of the strong subadditivity for subsystems $A,A^*,B$ whereas the upper bound arises by considering the subadditivity associated with the subsystems $A$ and $A^*$, and, $B$ and $B^*$.
\end{itemize}

In \cite{Hayden:2021gno}, the authors proposed a new quantity known as the Markov gap which is based on the lower bound for reflected entropy in eq.\eqref{lub} and is defined as follows
\begin{equation}
	h_{AB}=S_R(A:B)-I(A:B).
\end{equation}
Note that because the lower bound is derived from the strong subadditivity this quantity is nothing about the conditional mutual information
\begin{align}
  h_{AB}=  I(A:B^*|B).
\end{align}
Furthermore, the authors utilized certain theorem from quantum information which relates the conditional mutual information to the Markov recovery process or reconstruction of a tripartite state $\rho_{ABB^*}$  through a quantum channel $=\mathcal{R}_{B \rightarrow B B^{*}}$ acting on $\rho_{AB}$
\begin{align}
			h(A: B) \geq  -\log F_{max}\left(\rho_{A B B^{*}}, \tilde{\rho}_{A B B^{*}}=\mathcal{R}_{B \rightarrow B B^{*}}\left(\rho_{A B}\right)\right) \notag
\end{align}
where $F$-Fidelity, ${\cal R}$ is quantum map that tries to reconstruct $\rho_{ABB^*}$. This led the authors in \cite{Hayden:2021gno} to propose that this measure characterizes tripartite entanglement. Following which it has been explored in several interesting quantum systems \cite{Zou:2020bly}.

It should also be noted that in \cite{Hayden:2023yij} it was shown that for a generic state $\rho_{ABC}$ the reflected entropy can violate monotonicity under partial trace.  This raises the question of how a measure based on reflected entropy can serve as a measure of multipartite entanglement. However, we wish to clarify a subtlety here. The density matrix $\rho_{ABC}$ utilized as a counterexample in \cite{Hayden:2023yij} is a mixed state. In fact, it is quite straightforward to demonstrate monotonicity under partial trace for a pure tripartite state $\ket{\psi}_{ABC}$. To illustrate this, consider the subadditivity relations for $A,A^*$ and $B,B^*$, the very bounds upon which L-entropy was formulated.
\begin{align}
    I(A:A^*)\geq0 \implies 2 S(A)\geq S_R(A:B)\\
    I(B:B^*) \geq 0 \implies 2 S(B) \geq S_R(A:B)
\end{align}
Since $\ket{\psi}_{ABC}$ is a pure state 
\begin{align}
    S_{R}(A:BC)=2 S(A)\\
    S_R(B:AC)=2 S(B)
\end{align}
Utilizing the above expressions one can immediately conclude that
\begin{align}
    S_R(A:BC)\geq S_R(A:B)\\
    S_R(B:AC) \geq S_R(A:B)
\end{align}
which are the required conditions for $S_R(A:B)$ to be a correlation measure.
Therefore, it is evident that when as long as $ABC$ is described by a pure state, the reflected entropy adheres to monotonicity under partial trace. Given that we are currently focusing on pure states in this article, the above finding shows that there is no inconsistency.

\subsection{A genuine multipartite entanglement measure}\label{mult_ent}
In this subsection, we discuss the concept of a genuine tripartite entanglement measure in quantum information theory and explain why Markov gap fails to qualify as such a measure due to its inadequacy to capture the entanglement structure which is characteristic of GHZ-type states.

Before introducing the notion of genuine multipartite entanglement, we first recall that a bipartite pure state $\ket{\psi}_{AB}$ is defined as separable if it can be expressed in the form
\begin{align}
   \ket{\psi}_{AB}=\ket{\phi_A}\otimes \ket{\phi_B}
\end{align}
$\ket{\phi_A}$ and $\ket{\phi_B}$ are quantum states in the Hilbert spaces ${\cal H}_A$ and ${\cal H}_B$ respectively. Any state which can not be expressed in the above form is said to posess bipartite entanglement. However, the concept of separability becomes more nuanced in the context of multipartite states. To illustrate this, consider an $N$-partite state $\ket{\psi}$. Such a state is considered fully separable if it can be written as a completely factorized state:
\begin{align}
\ket{\psi}_{A_1,A_2...A_N}=\ket{\phi_1}_{A_1}\otimes \ket{\phi_2}_{A_2}\otimes \ket{\phi_3}_{A_3} \cdots \cdots \ket{\phi_N}_{A_N}\nonumber
\end{align}
However, this is not the only type of separability as it could be $k$-seprable when it can be expressed as follows
\begin{align}
	\ket{\psi}_{A_1,A_2...A_N}=\ket{\phi_1}_{B_1}\otimes \ket{\phi_2}_{B_2}\otimes \ket{\phi_3}_{B_3} \cdots \cdots \ket{\phi_N}_{B_k}\nonumber
\end{align}
where $\cup_j B_j=\cup_i A_i$ such that each $A_i$ occurs only once. An $N$-partite state is said to exhibit genuine multipartite entanglement if and only if it cannot be decomposed into a product state across any bipartition of the $N$ parties
\begin{align}
\ket{\psi}_{A_1,A_2...A_N}=\ket{\phi_1}_{B_1}\otimes \ket{\phi_2}_{B_2}\nonumber
\end{align}where $ B_1\cup B_2=\cup_i A_i$. Note that this definition encompasses all kinds of $k$-separability considered earlier. For example, a 3-party pure state $\ket{\psi}_{ABC}$ is said to posses genuine tripartite entanglement when it is not expressible  in any of the following four forms
\begin{align}
	\ket{\psi}_{ABC}&\neq\ket{\phi_{A}}\otimes \ket{\phi_{B}}\otimes \ket{\phi_{C}}\nonumber\\
	\ket{\psi}_{ABC}&\neq\ket{\phi_{AB}}\otimes  \ket{\phi_{C}}\nonumber\\
	\ket{\psi}_{ABC}&\neq\ket{\phi_{A}}\otimes \ket{\phi_{BC}}\nonumber\\
\ket{\psi}_{ABC}&\neq\ket{\phi_{AC}}\otimes \ket{\phi_{B}}\nonumber
\end{align}
In the first case, the state is termed fully separable, while in the other cases, it is referred to as bi-separable. For example, in the second state above which is bi-separable, $A$	and $B$ have bipartite entanglement between them where as they do not have any entanglement with subsystem $C$.  Although various measures in quantum information theory, such as the geometric mean of concurrence and fidelity of teleportation, work well for three-qubit systems, they become either ill-defined or computationally challenging as the dimensions of the Hilbert spaces increase, particularly in quantum field theories where the Hilbert spaces are often infinite-dimensional. In this article, we focus on measures based on reflected entropy, which are computationally feasible for qubit systems, certain quantum field theories such as conformal field theories (CFTs), and in the context of holography. For a comprehensive review of various multipartite entanglement measures, see \cite{Ma:2023ecg}.	As described in detail in \cite{Ma:2023ecg,PhysRevLett.127.040403}, a genuine multipartite entanglement measure, $\mathcal{E}$, is defined by the following properties:

\begin{enumerate}
	\item $\mathcal{E}$ should be zero for any fully separable state. This condition ensures that the measure correctly identifies states with no entanglement.
	\item $\mathcal{E}$ should be zero for any biseparable state, indicating the absence of genuine multipartite entanglement. This ensures that the measure distinguishes between genuinely multipartite entangled states and those that are only entangled in a biseparable manner.
	\item $\mathcal{E}$ should be strictly positive for all non-biseparable states. This is crucial as a good measure must be sensitive to any state exhibiting genuine multipartite entanglement.
	\item $\mathcal{E}$ should be non-increasing on average under local operations and classical communication (LOCC). A measure that obeys this property is also referred to as an LOCC monotone or entanglement monotone, reflecting that local operations and classical communication cannot increase the measure.
	\item In 3 qubit case $\mathcal{E}$ should rank the GHZ state higher than the W state. This criterion is supported by the fact that the GHZ state is more capable of teleporting any arbitrary single qubit state compared to the W state.
\end{enumerate}

Conditions (1), (2), (3), and (4) are necessary for a measure to accurately characterize genuine multipartite entanglement, whereas condition (5) is a weaker condition that provides additional insight into the measure's behavior with respect to specific entangled states. A measure satisfying all five conditions is known as a proper genuine multipartite entanglement measure \cite{PhysRevLett.127.040403}. 

Nevertheless, it is not essential that every useful entanglement quantifier satisfy all of the above axioms. For example, quantities such as the tangle and the Markov gap may fail to identify certain classes of genuinely multipartite entangled states. As a result, they are not necessarily strictly positive on the entire GME set, but may instead be positive only for those entangled states that they successfully capture. Similarly, failure to satisfy full LOCC monotonicity does not preclude a quantity from being of operational relevance. In many situations, the minimal expectations are that it vanish on biseparable states, be non-negative, and remain invariant under local unitary operations. These basic properties ensure that the quantity is physically well defined, basis independent, and capable of serving as a meaningful indicator of genuine multipartite entanglement. In the present paper, however, we shall not make a sharp distinction between quantities satisfying all of ($1$)--($4$) and those meeting only these minimal conditions, and we will refer to both collectively as GME measures whenever no ambiguity arises.
\subsubsection*{Entanglement entropy is not enough}
It is natural to wonder whether some suitable combination of  entanglement entropies corresponding to various subsystems, or more generally of quantities depending only on the spectra of reduced density matrices, could still capture genuine multipartite entanglement. The following example of iso-spectral states shows that this is not the case in general. Consider the two three qubit states
\begin{align}
    |W\rangle &= \frac{1}{\sqrt{3}}\left(|001\rangle + |010\rangle + |100\rangle\right) \, , \\
    |\psi\rangle &= \frac{1}{\sqrt{3}}|000\rangle + \sqrt{\frac{2}{3}}|111\rangle \, .
\end{align}
For these two states, both the one party and the two party reduced density matrices have identical spectra. Consequently, the states are indistinguishable by any quantity that depends only on the spectra of the reduced density matrices, including measures constructed solely from bipartite entanglement entropies. Nevertheless, their multipartite entanglement structures are fundamentally different: $|W\rangle$ is of W type, whereas $|\psi\rangle$ belongs to the GHZ class.

By contrast, the reflected entropy of the corresponding bipartite reduced density matrices distinguishes the two states, since it is sensitive to the full operator structure of the reduced density matrix and not merely to its eigenvalue spectrum
\begin{align}
    S_R^{\ket{W}}(A:B)&= \frac{6\log [6]-\left(2+\sqrt{3}\right) \log \left(2+\sqrt{3}\right)-\left(2-\sqrt{3}\right) \log \left(2-\sqrt{3}\right)}{6\log [2]}\approx 1.488 \\S_R^{\ket{\psi}}(A:B)&=\frac{\log [3]}{\log [2]}-\frac{2}{3}\approx 0.9183
\end{align}

\subsection{Limitations of the Markov Gap in characterizing GME }
Having reviewed the exact properties any genuine multipartite entanglement measure should be satisfying we will now describe why Markov gap does not satisfy all the properties required.
\begin{figure}
	\centering
	\includegraphics[width=.55\linewidth]{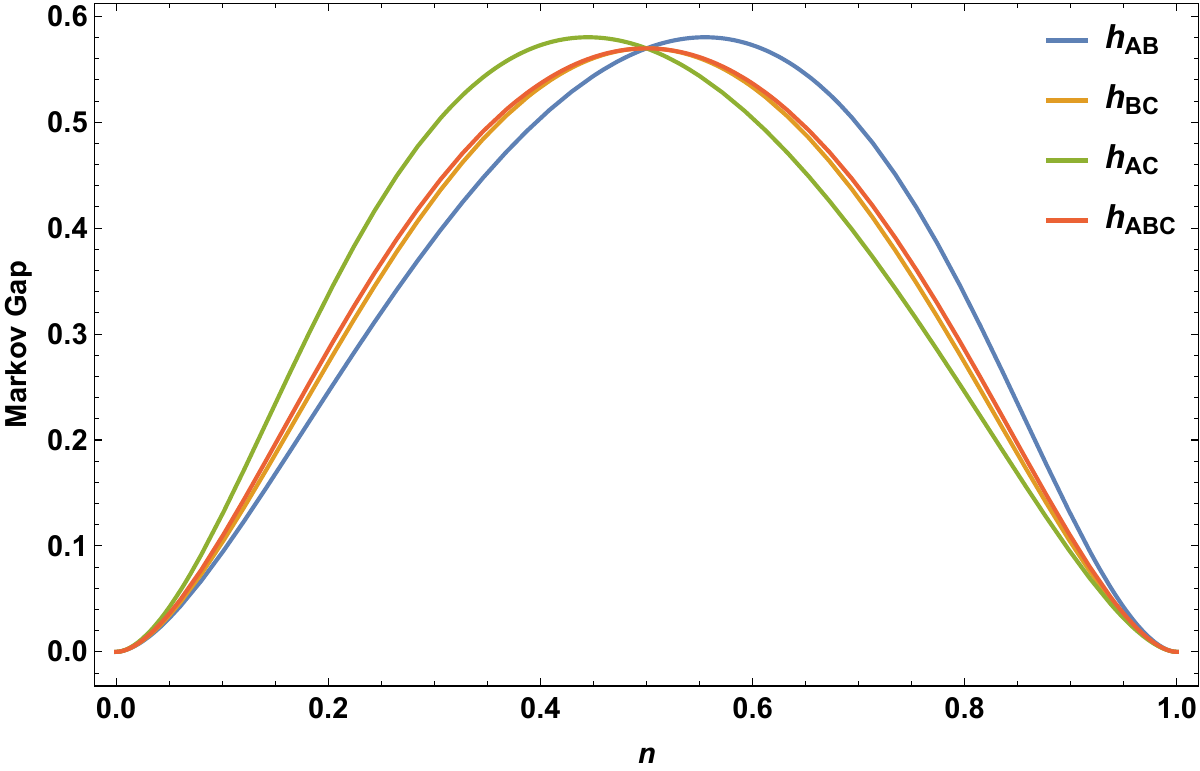}
	\caption{Here $h_{AB}$ (blue), $h_{BC}$ (orange), $h_{AC}$ (green) and $h_{ABC}$ (red) is plotted wrt $n$  for the state given by \cref{mg_prob_1}.}  
	\label{markovprob}
\end{figure}
For simplicity we will use the 3-qubit example. As described in the introduction, for the 3 qubit case a complete classification of the Hilbert space based on SLOCC revealed two nonequivalent types of genuine entangled states and the representative of these two classes the GHZ and the W states given by
\begin{align}
    |W\rangle&=\frac{1}{\sqrt{3}}\left(|001\rangle+|010\rangle+|100\rangle\right)\label{3qbitW}\\
    |GHZ\rangle&=\frac{1}{\sqrt{2}}\left(|000\rangle+|111\rangle\right)\label{3qbitGHZ}
\end{align}
Note that the $GHZ$ and the $W$ states are the maximally entangled states in their respective classes. Considering the following state,
\begin{equation}\label{mg_prob_1}
	|\psi\rangle=\sqrt{\frac{2}{3}}\sin\frac{n \pi}{2}|001\rangle+ \sqrt{\frac{2}{3}}\cos\frac{n \pi}{2}|010\rangle+\frac{1}{\sqrt{3}}|100\rangle
\end{equation}
which reduces to the $W$ state described earlier for $n=\frac{1}{2}$ . We compute the Markov gaps for \cref{mg_prob_1} . $h_{AB}$ and $h_{AC}$ show maximum value $.58$ at $n=.55$ and $n=.44$ respectively (in $\log[2]$ units).  However, the state in \cref{mg_prob_1} becomes maximally entangled at $n=.5$. However, $h_{BC}$ still maximizes at $n=.5$ but with a different magnitude $.57$. This indicates that the Markov gap corresponding to each bipartition gives a different answer and leads to the question of which one characterizes the multipartite entanglement. It is easy to rectify this issue by proposing a three party measure \textit{generalized Markov gap}  by taking all the bipartitions into account as follows,
\begin{equation}
	h_{ABC}=\big[h_{AB}h_{BC}h_{AC}\big]^{\frac{1}{3}}.
\end{equation}
In Fig.~\ref{markovprob} we plot $h_{AB},$ $h_{BC}$, $h_{AC}$ and $h_{ABC}$ where the maximal entanglement in the the state $|\psi\rangle$ is detected correctly at $n=.5$ with the magnitude .57 (in $\log[2]$ units).

Having this modification is not enough to justify the Markov gap to be considered as a measure of multipartite entanglement. To demonstrate this let us take the following state from the GHZ class,
\begin{equation}\label{gen_ghz}
	|\psi\rangle=\sin\frac{n \pi}{2}|000\rangle+ \cos\frac{n \pi}{2}|111\rangle,
\end{equation}
which reduces to the $GHZ$ state for $n=\frac{1}{2}$. For the state in \cref{gen_ghz}, the Markov gap as well as the generalized Markov gap becomes $0$. However, as described earlier $GHZ$ considered as a state having maximum amount of genuine tripartite entanglement. Therefore although the Markov gap or the generalized Markov gap captures the multipartite entanglement of the $W$ type it fails characterize the entanglement in the $GHZ$-type states described about and therefore it does not satisfy condition $(3)$ for a genuine entanglement measure listed earlier. To overcome this limitation, in the next section, we introduce a new measure designed to detect both $W$-type and $GHZ$-type states, thus extending the scope of entanglement detection to both classes within a single framework.

\section{Tripartite Latent Entropy (L-entropy) as a new GME}\label{set_up}
In this section we introduce a novel measure for genuine tripartite entanglement for 3-party pure states and demonstrate that it satisfies all the required conditions described earlier. In a subsequent section we will generalize our measure to characterize genuine $N$-partite entanglement in a n-party pure state. Let us recall that the measure of Markov gap was constructed from the lower bound for reflected entropy. Here we instead consider the upper bound for the reflected entropy and propose a new measure \textit{``L-entropy"} by considering all the bipartitions as follows
\begin{equation}\label{genl}
	\ell_{ABC}=\big[\ell_{AB}\ell_{BC}\ell_{AC}\big]^{\frac{1}{3}},
\end{equation} 
where, $\ell_{AB}$ is expressed as,
\begin{equation}\label{ldef}
	\ell_{AB}=\textrm{Min}\{2S(A),2S(B)\}-S_R(A:B)=\textrm{Min}\{I(A:A^*),I(B:B^*)\}
\end{equation} 
and similarly for $\ell_{BC}$ and $\ell_{AC}$. Note that the Markov gap was related to the conditional mutual information whereas this quantity is related to the minimum of the mutual informations of the individual subsystems and their respective copies.

\subsection{L-entropy as a genuine tripartite entanglement measure}
\subsubsection{Vanishes for a separable state}
One of the necessary property of a multipartite entangled state is that it has to vanish for any bi-separable state. We will now demonstrate that the L-entropy is zero for such a state. Without loss of generality let us consider the biseparable tripartite pure state of the form
\begin{align}
	|\psi\rangle_{ABC}&=|\phi\rangle _{AB}\otimes |\tilde{\phi}\rangle_C\nonumber\\
	\rho_{AB}&=\ket{\phi}_{AB \, AB}\bra{\phi}\nonumber\\
	\rho_{BC}&=\rho_{B}\otimes \rho_C\nonumber\\
	\rho_{AC}&=\rho_{A}\otimes \rho_C
\end{align}	
Since $\rho_{AB}$ is pure its reflected entropy is given by twice the entanglement entropy leading to a vanishing L-entropy
\begin{align}
	S_R(A:B)=2S(A)=2S(B) 	\implies \ell_{AB}=0
\end{align}
But for $\rho_{BC}$ and $\rho_{AC}$, $S_R(B:C)=S_R(A:C)=0$ as the density matrices are factorized. Furthermore, min$\{ S(B),S(C)\}=$ min$\{ S(A),S(C)\}=S(C)=0$ because $\rho_C$ corresponds to a pure state. Therefore, $\ell_{BC}=\ell_{AC}=0$. Note that the fully separable state is special case of the above when $\ket{\phi}_{AB}=\ket{\psi}_{A}\otimes \ket{\tilde{\psi}}_{B}$ and therefore even entanglement entropies vanish i.e $S(A)=S(B)=S(C)=0$.

\subsubsection{For three party case GHZ has maximum L-entropy}

\begin{figure}[h]
	\centering
	\includegraphics[width=.55\linewidth]{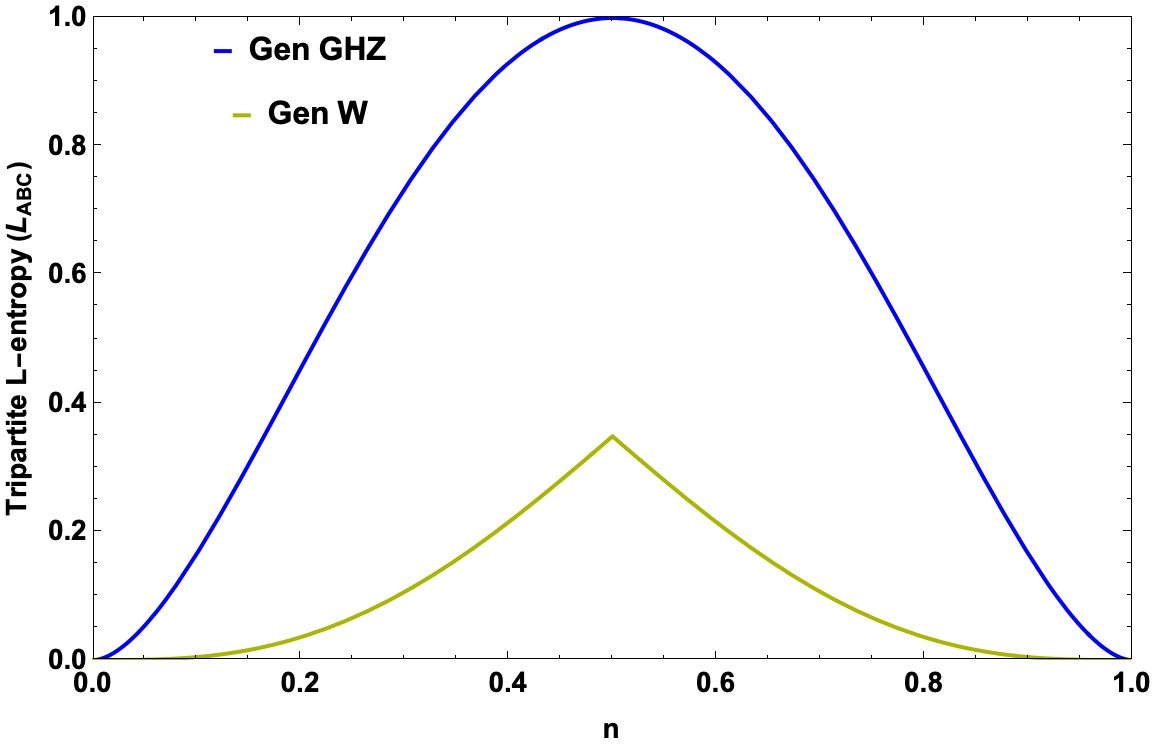}
	\caption{Here $\ell_{ABC}$ (in $\log[2]$ units) is plotted wrt $n$ for generalized $GHZ$ (blue) and generalized $W$ (green) states described in \cref{gen_ghz} and \cref{mg_prob_1}. }  
	\label{gen_GHZ_W}
\end{figure}

We now demonstrate that the maximum value of $\ell_{AB}$ is $log[d]$ for three party pure states of $ABC$. Furthermore we will show that GHZ obeys this bound indicating that it has maximum genuine tripartite entanglement as characterized by L-entropy. To this end, note that the following inequality holds on the account that reflected entropy is bonded from below by mutual information
\begin{align}
	\ell_{AB}=2 \, \textrm{Min}\{S(A),S(B)\}-S_R(A:B)\leq 2 \, \textrm{Min}\{S(A),S(B)\}-I(A:B)
\end{align}
Utilizing the definition of $I(A:B)$ we have
\begin{align}\label{labi1}
	\ell_{AB}\leq 2\, \textrm{Min}\{S(A),S(B)\}-S(A)-S(B)+S(AB)
\end{align}
 It is also straightforward to show that
\begin{align}
	2 \textrm{Min}\{S(A),S(B)\}-S(A)-S(B)\leq 0
\end{align}
Note that since we are only concerned pure states of $ABC$, $S(AB)=S(C)$. Utilizing this result and the above in \cref{labi1}
and denoting 2 $\textrm{Min}\{S(A),S(B)\}-S(A)-S(B)=-\epsilon$ (where $\epsilon\geq 0$) we obtain
\begin{align}
	\ell_{AB}\leq S(C)-\epsilon
\end{align}
Therefore, the upper bound for L-entropy across all states in the Hilbert space will be determined by the maxima of $S(C)$ when $\epsilon=0$ provided at least a single state attains this value.
\begin{align}\label{dcmax}
\ell_{AB}\leq S^{max}(C)=\log[d_C]
\end{align}
Furthermore, the mutual informations themselves are bounded by twice the entanglement entropies whose maximum values are as follows
\begin{align}
	I(A:A^*)&\leq 2 \log[d_A]\\
	I(B:B^*)&\leq 2 \log[d_B]
\end{align}
Hence the L-entropy obeys the following bound as well
\begin{align}
    \ell_{AB}= \textrm{Min}\{I(A:A^*),	I(B:B^*)\}&\leq 2 \textrm{Min}\{\log[d_A],\log[d_B]\}\label{dabmax}
\end{align}
Taking the bounds in \cref{dcmax} and \cref{dabmax} into account we have
\begin{align}\label{labbound3p}
	\ell_{AB}\leq \textrm{Min}\{2\log[d_A],2\log[d_B],\log[d_C]\}
\end{align}
For 3 qubits we have $d_A=d_B=d_C=2$ and hence
\begin{align}
	\ell_{AB}\leq \log[2]
\end{align}
The L-entropy for GHZ saturates this condition. More generally, if  $d_A=d_B=d_C=d$ an
\begin{align}
	\ell_{AB}\leq \log[d]
\end{align}
Hence, the full L-entropy is given by
\begin{align}
	\ell_{ABC}\leq \log[d]
\end{align}
We now demonstrate that the GHZ satisfies this condition. The three party generalized GHZ state when $A,B,C$ have dimensions $d_{A}=d_{B}=d_{C}=d$ may be expressed as
\begin{align}\label{gghz_mult_qubit}
	|\psi\rangle_{GGHZ}=\sum_{j=1}^{d}\lambda_j |j_Aj_Bj_C\rangle.
\end{align}
where $j_A,j_B,j_C$ is the basis of the Hilbert space of the subsystems $A,B$ and $C$ respectively.
Hence the reflected entropies for the different bi-partitions are given by
\begin{align}
    S_R(A:B)=S_R(B:C)=S_R(A:C)=-\sum_{j=1}^{d} |\lambda_j|^2 \log|\lambda_j|^2
\end{align}
The corresponding L-entropies are therefore
\begin{align}
\ell_{AB}&=\ell_{BC}=\ell_{AC}=-\sum_{j=1}^{d} |\lambda_j|^2 \log|\lambda_j|^2\nonumber\\
\ell_{ABC}&=-\sum_{j=1}^{d} |\lambda_j|^2 \log|\lambda_j|^2\nonumber\\
\end{align}
For the GHZ state $\lambda_j=1/\sqrt{d}$,
\begin{align}\label{ghz_mult_qubit}
	|\psi\rangle_{GGHZ}=\frac{1}{\sqrt{d}}\sum_{j=1}^{d}|j_Aj_Bj_C\rangle \implies \ell_{ABC}=\log d
\end{align}
In Fig.~\ref{gen_GHZ_W}, the $ L $-entropy is plotted for the 3 qubit generalized GHZ and W states in \cref{mg_prob_1}  and  \cref{gen_ghz}, as a function of $ n $. It is observed that both states exhibit the maximum amount of multipartite entanglement in their respective family of generalized states at $ n = \frac{1}{2} $. This is expected, as at $ n = \frac{1}{2} $, the states in  \cref{mg_prob_1,gen_ghz}  reduce to the standard GHZ and W states, respectively. Notably, our results also indicate that the GHZ state exhibits significantly more tripartite entanglement, with $ \ell_{ABC}(|GHZ\rangle) = 1 $ (in Log[2] units), compared to the W state, where $ \ell_{ABC}(|W\rangle) = 0.35 $. Interestingly, other genuine multipartite entanglement (GME) measures in the literature, such as the concurrence fill $ F_{123} $ \cite{PhysRevLett.127.040403} and the GME-concurrence $C_{GME}$ \cite{PhysRevA.83.062325}, also describe the W state with lower values, i.e., $ F_{123}(|W\rangle) = \frac{8}{9} = C_{GME}(|W\rangle) $, compared to the GHZ state, where $ F_{123}(|GHZ\rangle) = 1 = C_{GME}(|GHZ\rangle) $. In this context, our measure further corroborates that the W state possesses less genuine entanglement. However, $ L $-entropy suggests an even lower amount of genuine entanglement in the W state compared to the results  in\cite{PhysRevLett.127.040403,PhysRevA.83.062325}.

A generalization of the three party W state for subsystems with generic Hilbert space dimensions $d_A=d_B=d_c=d$ can be written as,
\begin{align}
\left|W\right\rangle=\frac{1}{\sqrt{3(d-1) }} \sum_{j\neq a}(|j a  a\rangle+|a j a  \rangle+|a a j\rangle)
\end{align}
where $a$ can be any element from the basis and the sum is over all the basis states excluding $a$. For $d=2$ it reduces to the three qubit W state in \cref{3qbitW}. The L-entropy for this case is given by
\begin{align}
\ell_{ABC}=\frac{1}{\sqrt{3}}\log \left(2+\sqrt{3}\right)+\log (3)-\frac{7}{3}  \log (2)\approx0.2416\nonumber
\end{align}
As evident from the results above, the L-entropy for the three-party W state is independent of the dimension dd, in contrast to the GHZ state. Consequently, for any three-party state, the L-entropy of the W state is consistently lower than that of the GHZ state, where it reaches its maximum. This outcome is expected, as the GHZ state lacks bipartite entanglement—its bipartite reduced density matrices are maximally mixed, which can be easily verified using vanishing of measures such as negativity in this scenario. In contrast, for the W state, the bipartite reduced density matrix is given by
\begin{align}
	\rho_{AB}=\frac{2}{3}\sum_j p_j |\psi_j^+\rangle \langle \psi_j^+|+\frac{1}{3}|aa\rangle \langle aa|
\end{align}	
where $p_j=\frac{1}{\sqrt{d-1}}$ and $|\psi^+\rangle$ denotes the Bell state
\begin{align}
	|\psi^+\rangle=\frac{1}{\sqrt{2}}(|a j\rangle+|j a\rangle)
\end{align}
Therefore, the W state exhibits a degree of tripartite entanglement along with a significant amount of residual bipartite entanglement, as described in \cite{Dur:2000zz}.
\subsubsection{Invariance under Local Unitaries and attempt at LOCC monotonicity }
In this section, we illustrate two key properties of L-entropy: its invariance under local unitaries and its non-increasing behavior on average under local operations and classical communication (LOCC monotonicity). 

\subsection*{LU invariance}
To prove the former it is worth noting that local unitaries applied to the mixed state $\rho_{AB}$ result in corresponding local unitaries on the purified state, as shown below
\begin{align}
	\rho_{AB}&\to \rho_{AB}'= U\rho_{AB} U^{\dagger} \qquad U=U_{A} \otimes U_B \nonumber
\end{align}
It can be easily checked that this in terms implies the following for the canonically purified state
\begin{align}
	\ket{\sqrt{\rho_{AB}}}\to\ket{\sqrt{\rho_{AB}'}}_{ABA^*B^*}=U_{A} \otimes U_B\otimes U_{A^*} \otimes U_{B^*} \ket{\sqrt{\rho_{AB}}}
\end{align}
where $U_{A^*}$ and $U_{B^*}$ are copies of the unitaries $U_{A}$ and $U_{B}$ respectively. Firstly, it is clear from the above that the L entropy is invariant under local unitaries because both $I(A:A^*)$ and $I(B:B^*)$ are made up of reflected entropies and entanglement entropies both of which are local unitary invariants.

\subsection*{Attempt at LOCC Monotonicity}
We now make an attempt to determine whether L-entropy obeys LOCC monotonicity. As described in \cite{Gadde:2024jfi} in order for a local unitary invariant function $f(\ket{\psi}_{\textrm{\tiny{ABC}}})$ to be a pure state entanglement monotone it has to be concave under local operations
\begin{align}\label{mono1}
	f(|\psi\rangle_{\textrm{\tiny{ABC}}}) \geq \sum_i p_i f\left(\left|\psi_i\right\rangle_{\textrm{\tiny{ABC}}}\right)
\end{align}
where $|\psi_i\rangle_{\textrm{\tiny{ABC}}}$ are the states obtained  after a local operation on one of the parties denoted by the map $\Lambda$
\begin{align}\label{Lam}
	\Lambda(\rho)=\sum_i p_i|\psi_i \rangle\langle\psi_i|, \quad p_i:=| E_i^{(A) }\ket{\psi}|^ 2, \quad \ket{\psi_i}:E_i^{(A)}|\psi\rangle / \sqrt{p_i}
\end{align}
where $E_i^{A}$ is a linear local operation on one of the parties that preserves trace (let us choose the party to be $A$)
\begin{align}
 \sum_i E_i^{\dagger (A)} E_i^{(A)}=\mathbb{I}
\end{align}
Let us denote $\ket{\phi}$ as the purification of the reduced density matrix (denoted as $\rho_{\textrm{\tiny{AB}}}$), obtained by tracing out one of the parties (denoted as $C$) from the state $\ket{\psi}$. Similarly, $\ket{\phi_i}$ denotes the purification of the reduced density matrix (denoted as $\rho_{i,\textrm{\tiny{AB}}} = \mathrm{Tr}_C(\ket{\psi_i})$). Note that $\ket{\phi}$ resides in a Hilbert space ${H}_{AB} \otimes H_{\tilde{A}}$, where $H_{\tilde{A}}$ has a dimension at least as large as the rank of the reduced density matrix $\rho_{\textrm{\tiny{AB}}}$. Since $f$ is invariant under local unitaries, and different purifications of the bipartite system including the original state $\ket{\psi}$ are related by local unitary transformations, it follows that $f$ remains the same for all such purifications.
\begin{align}
    f(\ket{\phi})=f(\ket{\psi}),\quad  f(\ket{\phi_i})=f(\ket{\psi_i})
\end{align}
Hence the statement in \cref{mono1} may be re-expressed in terms of the canonically purified states as
\begin{align}
	f(\ket{\phi}) \geq \sum_i p_i f(\ket{\phi_i})
\end{align}
We now can apply this result to our measure based on the canonical purification where $\ket{\phi}$ lives in a Hilbert space ${H}_{AB}\otimes H_{A^*B^*}$ where $H_{A^*B^*}$ has the dimension exactly same as that of the rank of the reduced density matrix $AB$. In terms of canonical purification, we need to demonstrate that 
\begin{align}
	f(\ket{\sqrt{\rho_{\textrm{\tiny{AB}}}}}) \geq \sum_i p_i  f(\ket{\sqrt{\rho_{\textrm{i,\tiny{AB}}}}})
\end{align}
where $\ket{\sqrt{\rho_{\textrm{\tiny{AB}}}}}$ and $\ket{\sqrt{\rho_{\textrm{i,\tiny{AB}}}}}$ denotes the canonical purification of the reduced density matrices   $\rho_{\textrm{\tiny{AB}}}$and $\rho_{i,\textrm{\tiny{AB}}}$ respectively.
Note that the effect of map $\Lambda$ acting on $\rho$ in \cref{Lam} can in turn be thought of as the $\tilde{\Lambda}$ map on the canonical purified state  $ \ket{\sqrt{\rho_{\textrm{\tiny{AB}}}}}$ which results in states  $\ket{\sqrt{\rho_{\textrm{i,\tiny{AB}}}}}$ with probabilities $p_i$. This is expressed as follows
\begin{align}
	\tilde{\Lambda}( \ket{\sqrt{\rho_{\textrm{\tiny{AB}}}}})=\sum_i p_i \ket{\sqrt{\rho_{\textrm{i,\tiny{AB}}}}}\bra{\sqrt{\rho_{\textrm{i,\tiny{AB}}}}}.
\end{align}
 We will now demonstrate that $\tilde{\Lambda}$ is nothing more than a local operation $\Lambda$ on A and its reflected copy $A^*$. To this end, consider the reduced density matrix $\rho_{i,AB}$ after a local operation $E_i$ on A alone
\begin{align}\label{rhoi}
	\rho_{i,AB}&=Tr_{C}(\rho_{\psi_i})\notag\\
	&=\frac{E_i^{(A)}Tr_C(\rho_{\psi})E_i^{\dagger(A)}}{p_i}\\
	&=\frac{E_i^{(A)}\rho_{AB}E_i^{\dagger(A)}}{p_i}
\end{align}
Note that since $E_i^{(A)}$ acts locally only on A,  we could push the trace over $C$ inside the operation in the second line.
Now let us say that the reduced density matrix $\rho_{AB}$ can be decomposed in terms of pure states as follows
\begin{align}\label{redexp}
	\rho_{AB}=\sum_i q_i \ket{\lambda_i}\bra{\lambda_i}
\end{align}
Its canonical purification is simply
\begin{align}
	\ket{\sqrt{\rho_{\textrm{\tiny{AB}}}}}=\sum_i \sqrt{q_i} \ket{\lambda_i}\ket{\lambda_i}
\end{align}
Furthermore from \cref{rhoi} and \cref{redexp} we have
\begin{align}
	\rho_{i,AB}=\sum_j q_j \frac{E_i^{(A)}}{\sqrt{p_i}}\ket{\lambda_j}\bra{\lambda_j}\frac{E_i^{\dagger(A)}}{\sqrt{p_i}}
\end{align}   
Motivated by the logic underlying local-unitary invariance, one may be tempted to expect that the canonical purification after a Kraus operation on a local subsystem should be obtainable from the original canonical purification through a corresponding local operation. This expectation is valid for local unitary operations, and more generally for maps that take orthogonal states to another set of orthogonal states. In general, however, such a relation does not hold. The reason is that the canonical purification depends nonlinearly on $\sqrt{\rho}$, so that a local operation acting on $\rho$ does not, in general, lift to a corresponding local operation acting on $|\sqrt{\rho}\rangle$\footnote{We thank  Abhijit Gadde for bringing this important issue to our attention.}. 

Furthermore,  consider the following state
\begin{align}
|\phi_i\rangle := \frac{(K_i \otimes I)|\sqrt{\rho}\rangle}{\sqrt{p_i}} \, .
\end{align}
It is straightforward to verify that $|\phi_i\rangle$ is a legitimate purification of $\rho_i$. However, in general, the issue is that it does not coincide with the canonical purification $|\sqrt{\rho_i}\rangle$. In view of the above discussed obstruction encountered in attempting to prove LOCC monotonicity analytically, it is useful to examine the issue numerically. For this purpose, we studied the behaviour of $L$-entropy under local Kraus operations by applying complete sets of random Kraus operators to Haar-random states and comparing the averaged post-measurement $L$-entropy with its initial value.

\begin{figure}[h]
    \centering
    \begin{subfigure}[t]{0.32\textwidth}
        \centering
        \includegraphics[width=\textwidth]{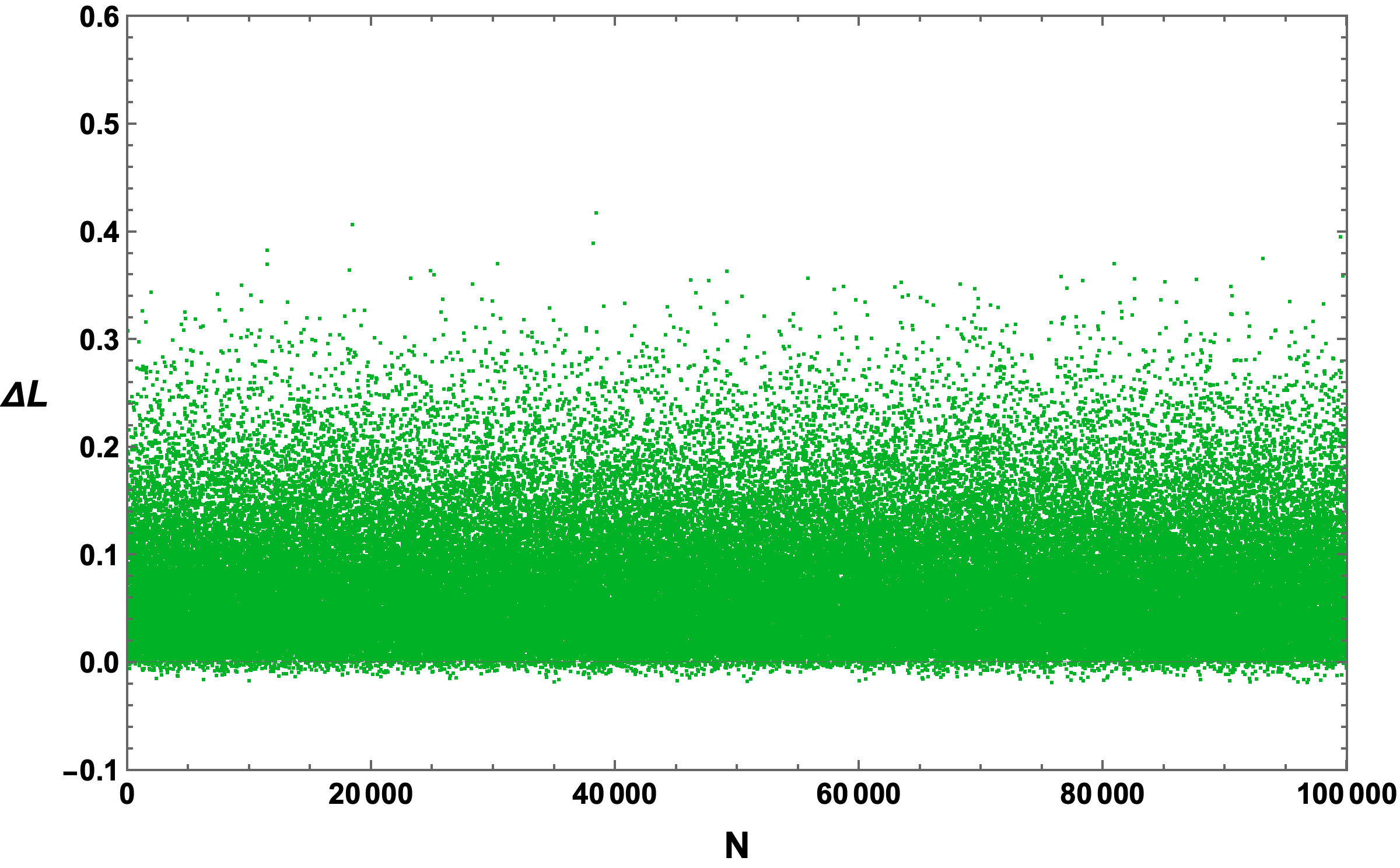}
        \caption{Three-qubit random states acted upon by a complete set of random Kraus operators.}
        \label{fig:3p_locc}
    \end{subfigure}
    \hfill
    \begin{subfigure}[t]{0.32\textwidth}
        \centering
        \includegraphics[width=\textwidth]{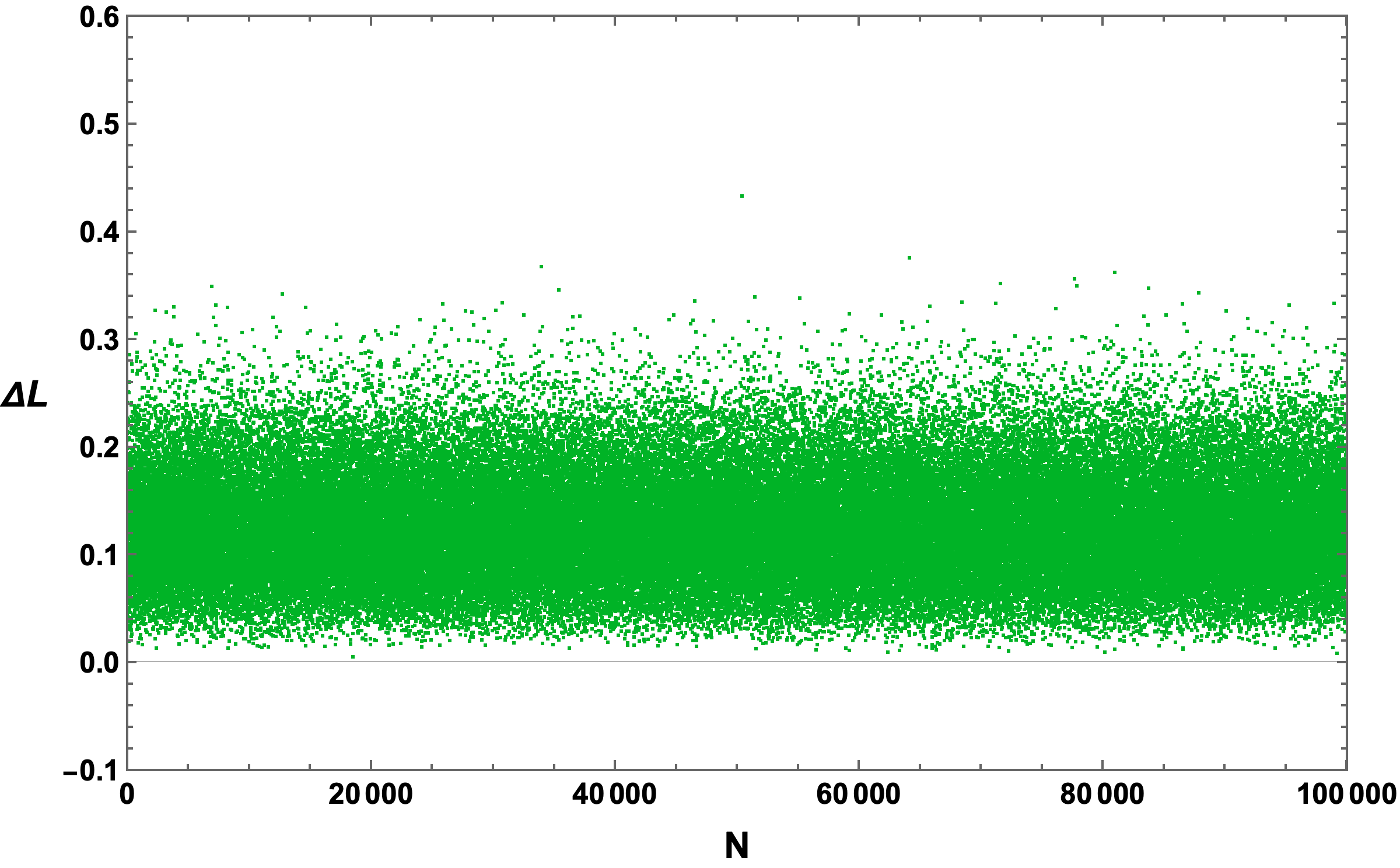}
        \caption{Four-qubit random states acted upon by a random set of two Kraus operators.}
        \label{fig:4p_locc}
    \end{subfigure}
    \hfill
    \begin{subfigure}[t]{0.32\textwidth}
        \centering
        \includegraphics[width=\textwidth]{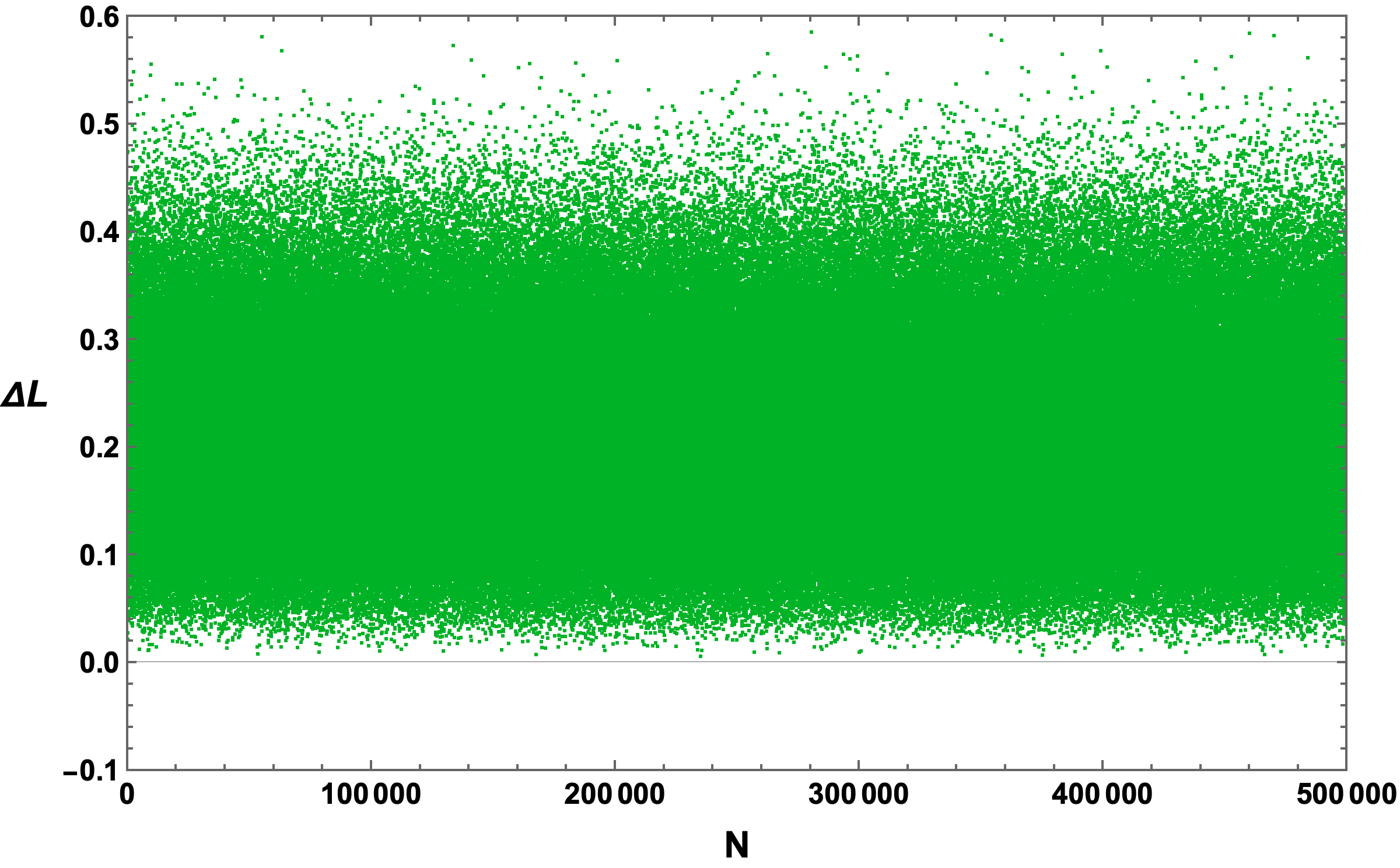}
        \caption{Five-qubit random states acted upon by a random set of two Kraus operators.}
        \label{fig:5p_locc}
    \end{subfigure}
    \caption{Numerical tests of the monotonicity of $L$-entropy under local Kraus operations for random three-, four-, and five-qubit states. For the three-qubit case, a small fraction of states violate monotonicity, with $\Delta L<0$, whereas no such violations were observed for the four- and five-qubit samples considered here.}
    \label{fig:locc_all}
\end{figure}

\begin{figure}[h]
    \centering
    \begin{subfigure}[t]{0.3\textwidth}
        \centering
        \includegraphics[width=\textwidth]{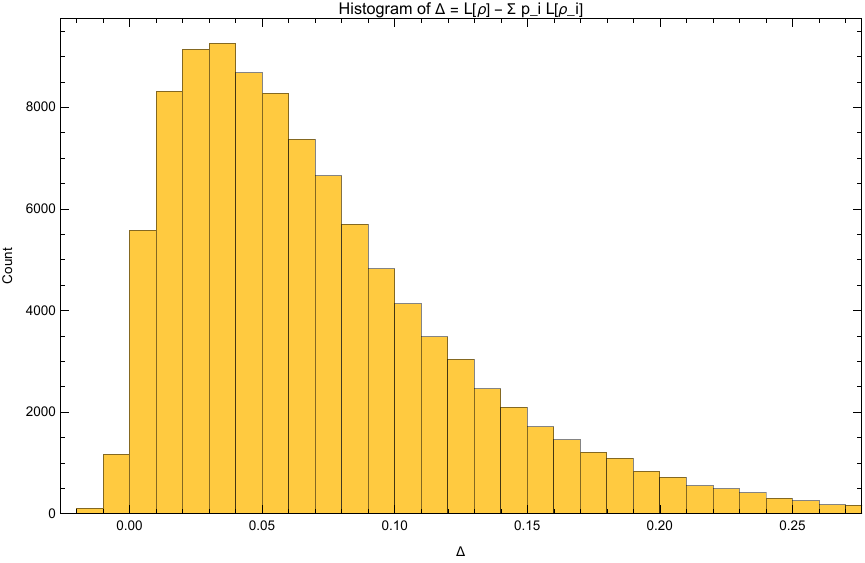}
        \caption{Histogram of $\Delta L$ for three-qubit states. Approximately $1\%$ of the sampled states have $\Delta L<0$, with only small negative values.}
        \label{fig:3p_hist}
    \end{subfigure}
    \hfill
    \begin{subfigure}[t]{0.3\textwidth}
        \centering
        \includegraphics[width=\textwidth]{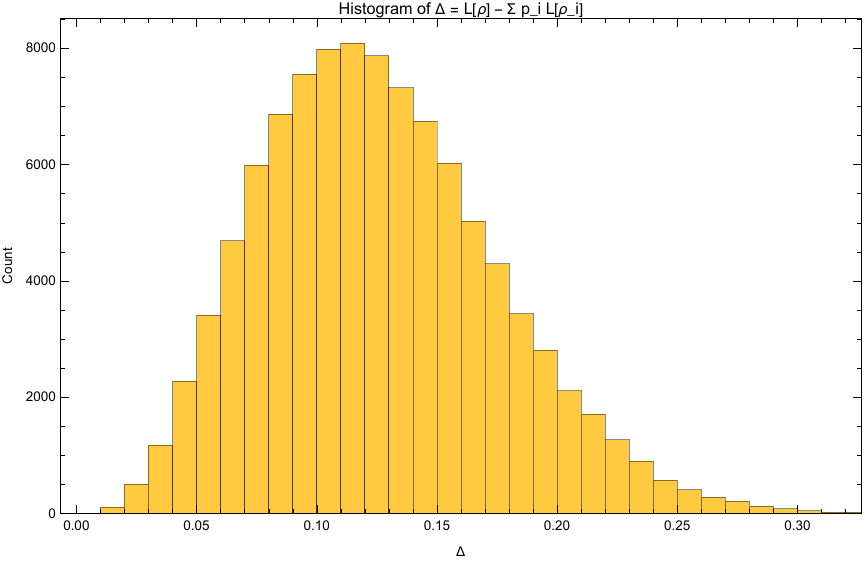}
        \caption{Histogram of $\Delta L$ for four-qubit states. No states with $\Delta L<0$ were found in the sample.}
        \label{fig:4p_hist}
    \end{subfigure}
    \hfill
    \begin{subfigure}[t]{0.3\textwidth}
        \centering
        \includegraphics[width=\textwidth]{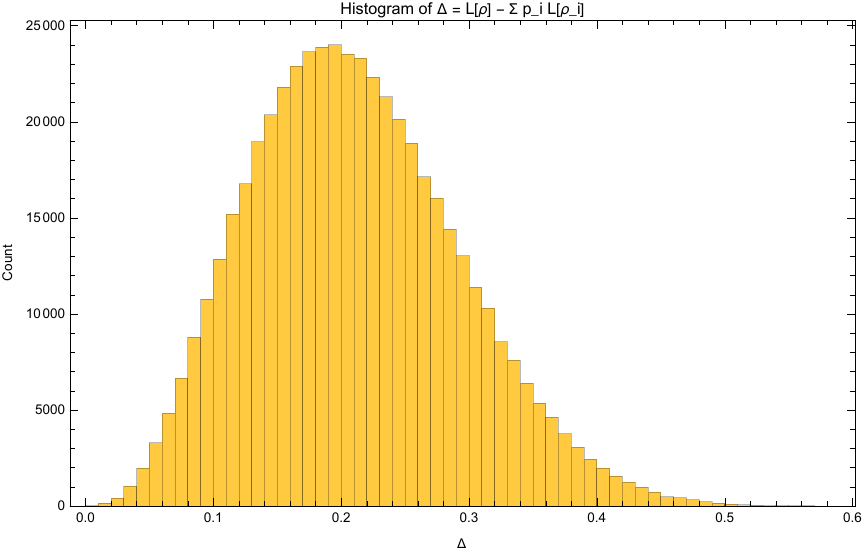}
        \caption{Histogram of $\Delta L$ for five-qubit states. Among $500000$ sampled states, no instances with $\Delta L<0$ were found.}
        \label{fig:5p_hist}
    \end{subfigure}
    \caption{Histograms of $\Delta L$ for random three-, four-, and five-qubit states under local Kraus operations. The three-qubit case exhibits a small tail at negative $\Delta L$, while the four- and five-qubit cases show no observed violations of monotonicity in the samples considered.}
    \label{fig:hist_all}
\end{figure}

For the three-qubit case, we found a small fraction of states, approximately $1\%$, that appear to exhibit a violation of monotonicity. However, the corresponding values of $\Delta L$ are exceedingly small, as illustrated in figures~\cref{fig:locc_all,fig:hist_all}, and their presence appears to be somewhat sensitive to numerical precision. By contrast, for the four- and five-qubit cases, no such violations were observed, even after repeating the analysis with higher numerical precision; see figures~\cref{fig:locc_all,fig:hist_all}.

\subsection{Tripartite L-entropy in spin chain and SYK model}

 Here we examine the dynamics of the three-party L-entropy for Hamiltonians corresponding to various spin-chain models, such as the Ising and also the SYK model. 

\begin{figure}[H]
	\centering
	\begin{subfigure}[h]{0.45\textwidth}
		\centering
		\includegraphics[width=1\textwidth]{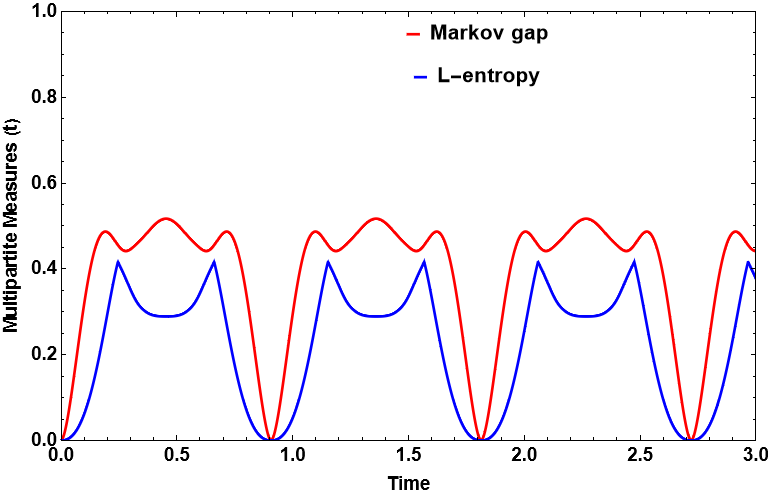}
		\caption{Initial state $\ket{Bell}\otimes\ket{0} .$}
		\label{NNising_1_bell0}
	\end{subfigure}
	\hfill
	\begin{subfigure}[h]{0.45\textwidth}
		\centering
		\includegraphics[width=1\textwidth]{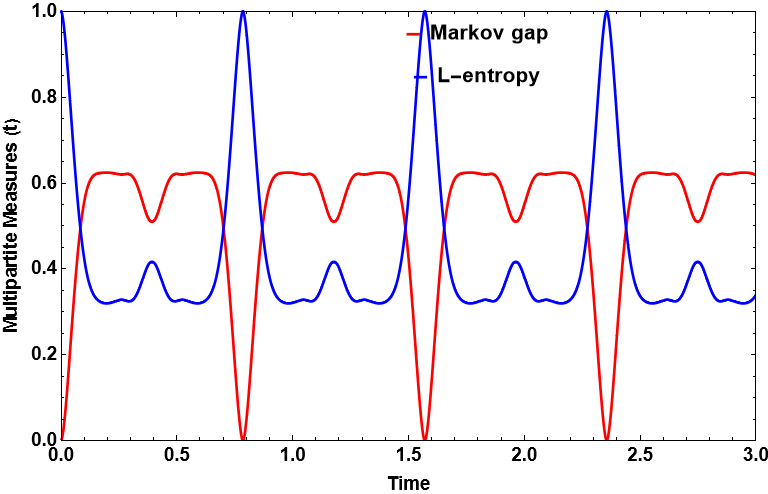}
		\caption{Initial state  $ \ket{GHZ}$}
		\label{NNising_1_ghz}
	\end{subfigure}
	\caption{ Plots of L-entropy and Markov gap for the Hamiltonian $H=\sum_i \sigma_x^{i}\sigma_x^{i+1}+\sigma_y^{i}\sigma_y^{i+1}$  .}
	\label{NNising_1}
\end{figure}

\begin{figure}[H]
	\centering
	\begin{subfigure}[h]{0.32\textwidth}
		\centering
		\includegraphics[width=1\textwidth]{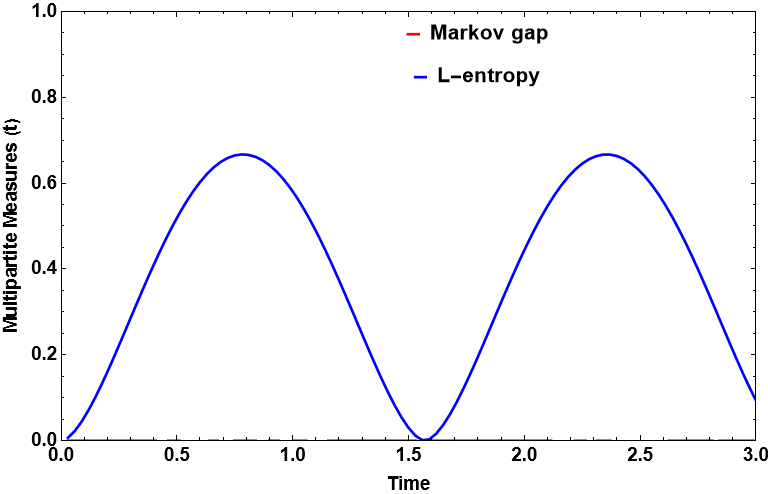}
		\caption{Initial state fully separable}
		\label{NNising_2_000}
	\end{subfigure}
	\hfill
	\begin{subfigure}[h]{0.32\textwidth}
		\centering
		\includegraphics[width=1\textwidth]{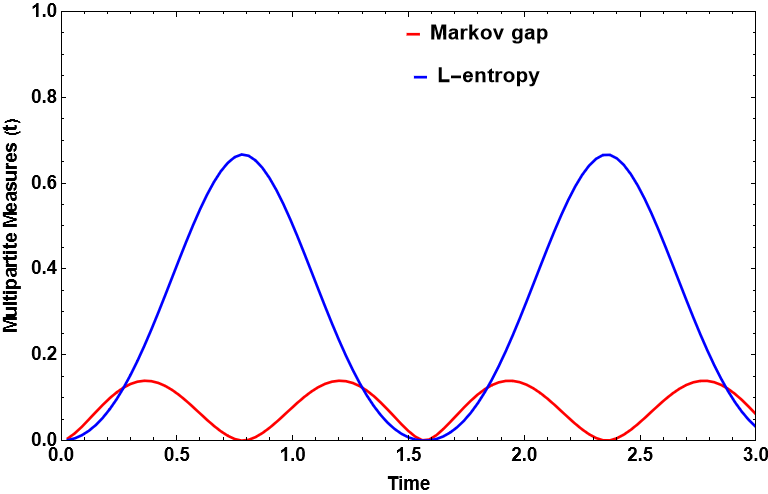}
		\caption{Initial state $\ket{Bell}\otimes\ket{0}$}
		\label{NNising_2_bell0}
	\end{subfigure}
	\hfill
	\begin{subfigure}[h]{0.32\textwidth}
		\centering
		\includegraphics[width=1\textwidth]{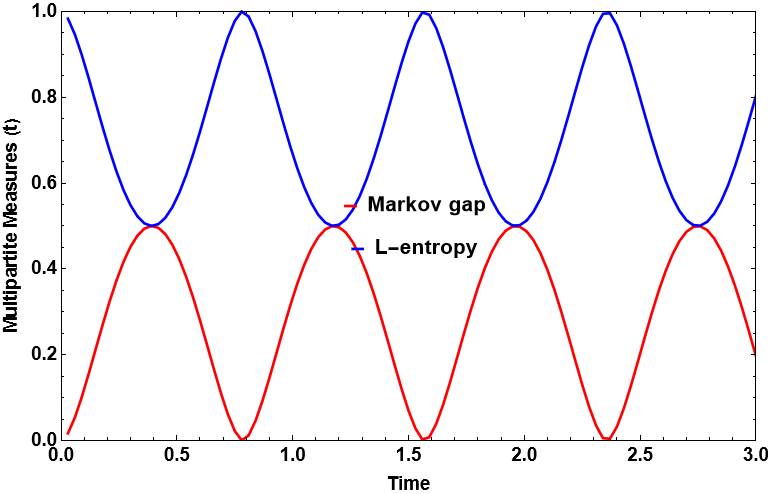}
		\caption{Initial state  is GHZ}
		\label{NNising_2_ghz}
	\end{subfigure}
	\caption{  Plots of tripartite L-entropy and tripartite Markov gap for the Hamiltonian $H=\sum_i \sigma_x^{i}\sigma_x^{i+1}$ for n=9 qubits partitioned into 3 parties of 3 qubits each.}
	\label{NNising_2}
\end{figure}

\begin{figure}[H]
	\centering
	\begin{subfigure}[h]{0.32\textwidth}
		\centering
		\includegraphics[width=1\textwidth]{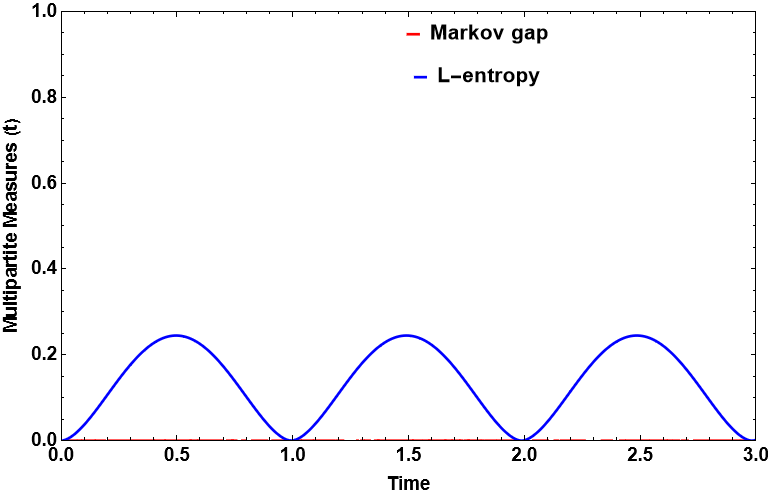}
		\caption{Initial state fully separable}
		\label{NNRH_1}
	\end{subfigure}
	\hfill
	\begin{subfigure}[h]{0.32\textwidth}
		\centering
		\includegraphics[width=1\textwidth]{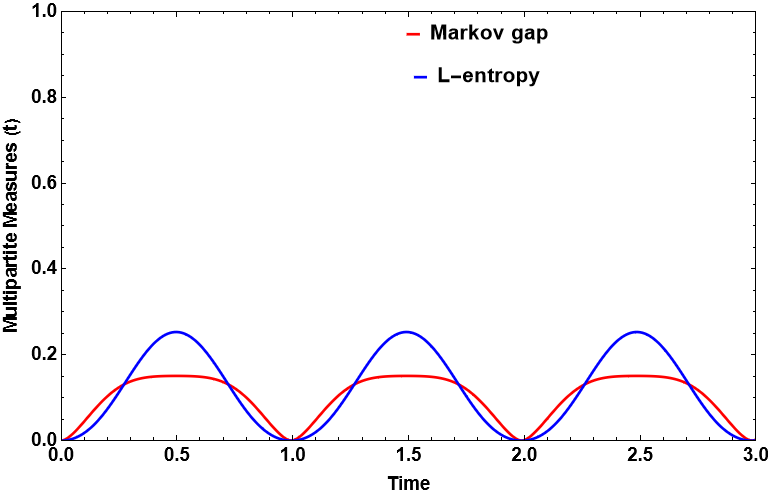}
		\caption{Initial state  \\$\ket{Bell}\otimes\ket{0}$}
		\label{ising_N32}
	\end{subfigure}
	\hfill
	\begin{subfigure}[h]{0.32\textwidth}
		\centering
		\includegraphics[width=1\textwidth]{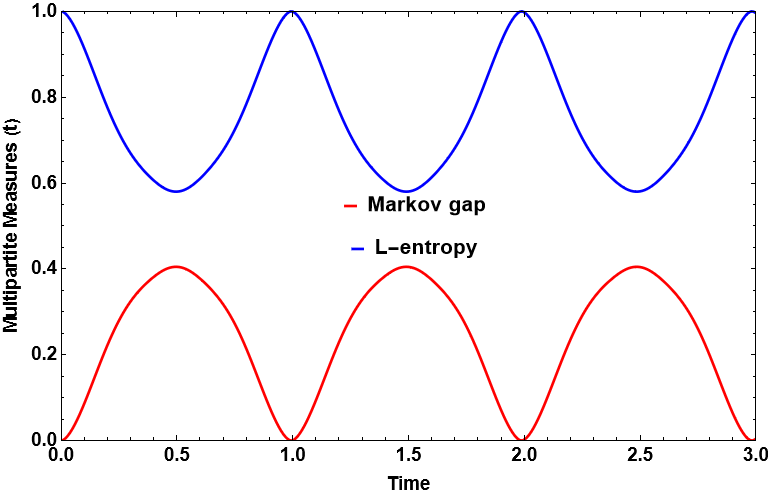}
		\caption{Initial state GHZ}
		\label{NNRH_2}
	\end{subfigure}
	\caption{ N=9 Plots of tripartite L-entropy and tripartite Markov gap  for the random Hamiltonian $H=\sum_i h^{i} h^{i+1}+h^{i}h ^{i+1}$.}
	\label{NNRH}
\end{figure}

\begin{figure}[H]
	\centering
	\begin{subfigure}[h]{0.32\textwidth}
		\centering
		\includegraphics[width=1\textwidth]{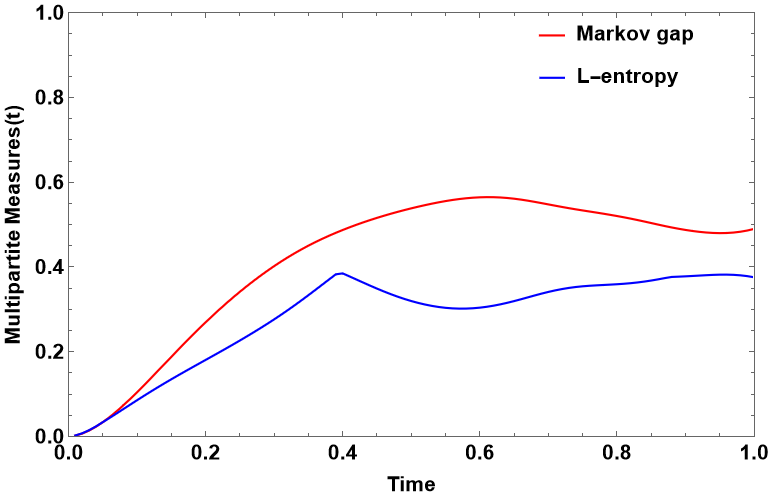}
		\caption{$N=6$}
		\label{syk_N6}
	\end{subfigure}
	\hfill
	\begin{subfigure}[h]{0.32\textwidth}
		\centering
		\includegraphics[width=1\textwidth]{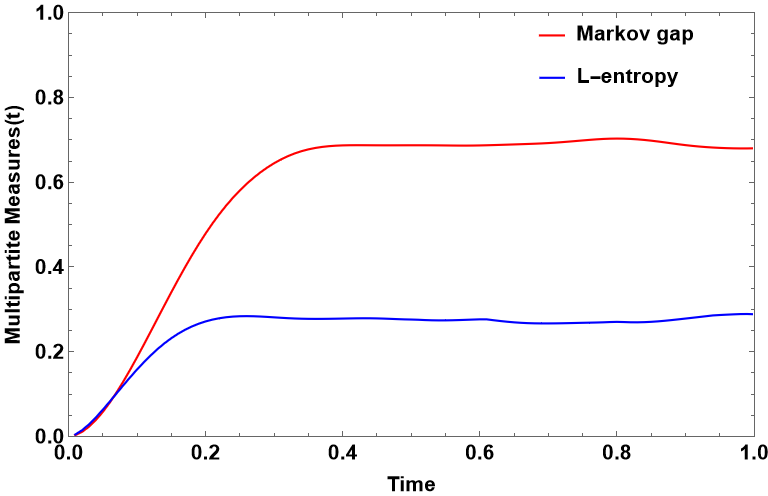}
		\caption{$N=9$}
		\label{syk_N9}
	\end{subfigure}
	\hfill
	\begin{subfigure}[h]{0.32\textwidth}
		\centering
		\includegraphics[width=1\textwidth]{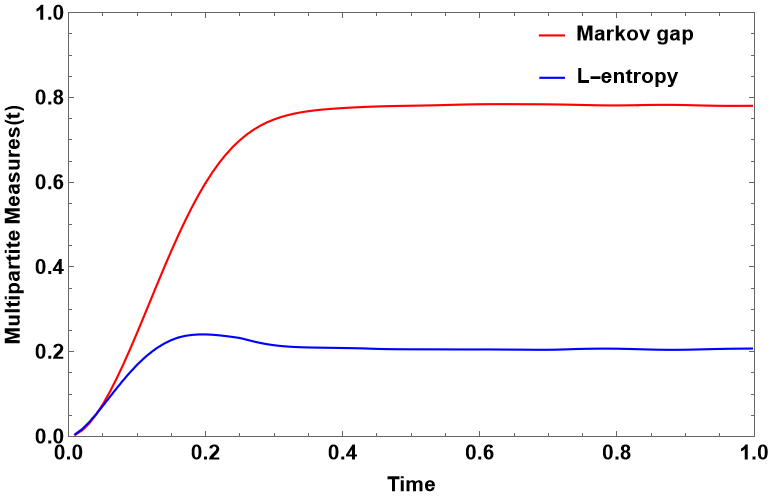}
		\caption{$N=12$}
		\label{syk_N12}
	\end{subfigure}
	\caption{
	}
	\label{syk_plots}
\end{figure}

\begin{figure}[H]
	\centering
	\begin{subfigure}[h]{0.32\textwidth}
		\centering
		\includegraphics[width=1\textwidth]{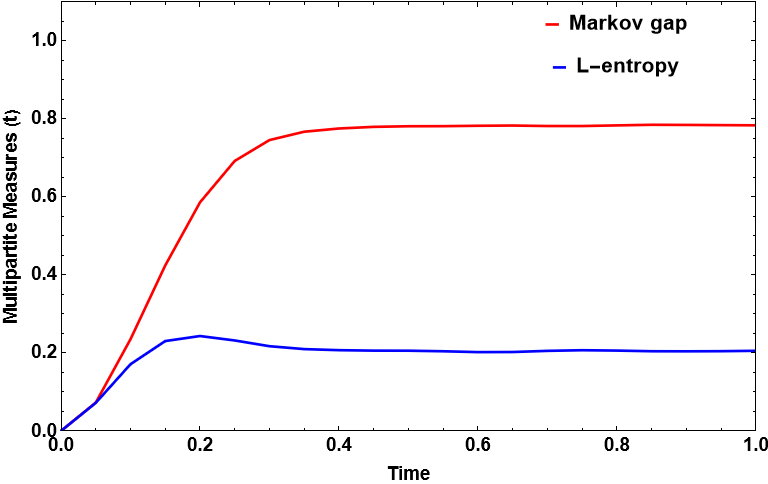}
		\caption{$J_0=1 K_0=0.01$}
		\label{syk_JK1}
	\end{subfigure}
	\hfill
	\begin{subfigure}[h]{0.32\textwidth}
		\centering
		\includegraphics[width=1\textwidth]{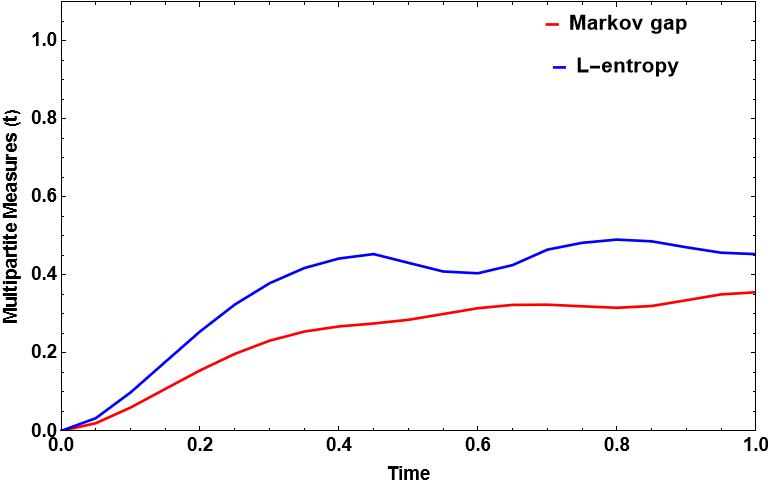}
		\caption{$J_0=0.1, K_0=1$}
		\label{syk_JK3}
	\end{subfigure}
	\hfill
	\begin{subfigure}[h]{0.32\textwidth}
		\centering
		\includegraphics[width=1\textwidth]{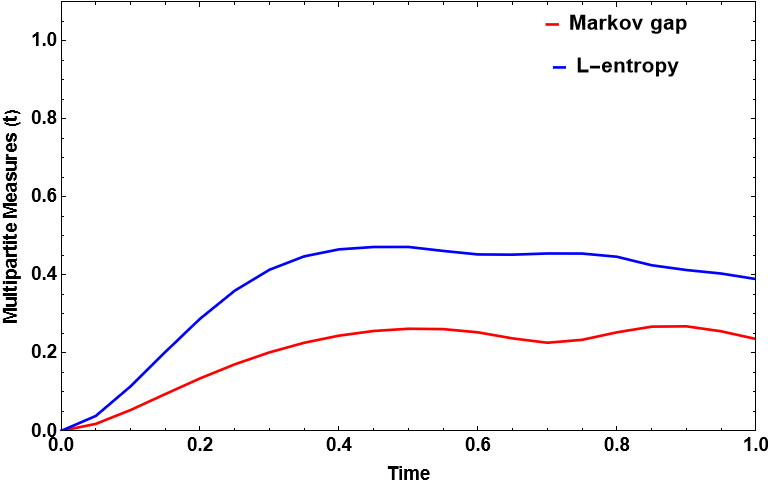}
		\caption{$J_0=0.001, K_0=1$}
		\label{syk_JK5}
	\end{subfigure}
	\caption{
	}
	\label{msyk_plots}
\end{figure}

 The plots for the evolution of L-entropy through unitaries corresponding to the spin chain Hamiltonian involving nearest-neighbor interactions are depicted in Fig.~\ref{NNising_1} and Fig.~\ref{NNising_2}. Following that, we have also examined the behavior of L-entropy when the state is being evolved with a unitary generated by a random Hermitian matrix in Fig.~\ref{NNRH}. Subsequently, we obtain the same when the state is being evolved with a unitary corresponding to the SYK Hamiltonian in Fig.~\ref{syk_plots} and the mass deformed SYK in Fig.~\ref{msyk_plots}. Quite interestingly, in all the models with nearest neighbour interactions we have examined in the present article the behaviour of L-entropy is oscillator whereas in the SYK-model it saturates to a constant value after an initial growth. However, unlike the entanglement entropy the saturation value is not very close to the maximum L-entropy indicating that the tripartite entanglement in a n-party pure state is close to its peak value at large-$N$ ($N=2n$). We will see in the following section that the generalized L-entropy characterizing $n$-partite entanglement of a n-party pure state is close to its maxima at  large -$N$.
\subsubsection{Nearest neighbor Ising}

 \subsubsection{Nearest Neighbour Random Hamiltonian}

\subsubsection{SYK model}\label{syk}

\section{Generalization to higher number of parties}\label{N_party_L}
 In the preceding section,  we introduced and defined the measure of L-entropy, which serves to characterize the genuine three-party entanglement and examined its behavior within certain simple quantum systems, including the Ising model and the SYK model. Moving forward, we shall expand the notion of L-entropy to encompass the characterization of  genuine $n$-party entanglement in a general $n$-party pure state $|\psi\rangle_{A_1A_2\cdot\cdot\cdot A_n}$ for $n=4,5$.

\subsection{A n-partite L-entropy}
 Our construction involves determining all possible two party L-entropies  ({}$^nC_2$ number of bipartitions $A_i,A_j$ without combining multiple parties ),
\begin{equation}
	\ell_{A_i,A_j}=\textrm{Min}\{ 2S[A_i],2S[A_j] \}-S_{R}(A_i:A_j),
\end{equation}
where, $i=1,2,\cdot\cdot\cdot,n$ and $j=i+1,i+2,\cdot\cdot\cdot,n$. Finally, we define the n-party generalized L-entropy\footnote{Here, we restrict our analysis up to five parties for two reasons. First, for $n > 5$, it is straightforward to construct separable states for which the multipartite $L$-entropy, as defined in \cref{multil}, does not vanish. Second, and more interestingly, the bipartite $L$-entropies attain their peak for $k$-uniform states with $k \geq 2$, but they fail to distinguish between a $k = 2$-uniform state and any higher $k>2$ uniform state. 
 Hence, extending the analysis to higher-party systems would require a generalization of the basic $L$-entropy which maximizes for higher uniform states. In particular, one must examine the purifications of all $\lfloor n/2 \rfloor$ density matrices for a $n$-party pure state and utilize multipartite reflected entropy, which becomes significantly more intricate\cite{newlentropy}.} as,
\begin{equation}\label{multil}
	\ell_{A_1A_2\cdot\cdot\cdot A_n}=\left(\prod_{i<j}\ell_{A_iA_j}\right)^{\frac{2}{n(n-1)}}.
\end{equation}
It is easy to verify that when $|\psi\rangle_{A_1A_2\cdots A_n}$ is biseparable, at least one of the $\ell_{A_i,A_j}$ equals zero, causing the product in the above expression to vanish. This ensures compliance with conditions (1) and (2) in \cref{mult_ent} for n-party states. Condition (4) is satisfied because each $\ell_{A_i,A_j}$ is an entanglement monotone, and the geometric mean of entanglement monotones is also a monotone, as explained in \cite{Li:2021hhj}.  The proof turns out to be quite simple as described below
\begin{align}
\ell_{A_1A_2\cdot\cdot\cdot A_n}(\ket{\psi}_{A_1A_2\cdot\cdot\cdot A_n})&=\left(\prod_{i=1}^{n}\prod_{j=i+1}^{n}\ell_{A_iA_j}(\ket{\psi}_{A_iA_j\overline{A_iA_j}})\right)^{\frac{2}{n(n-1)}}\nonumber\\&\geq \left(\prod_{i=1}^{n}\prod_{j=i+1}^{n}\sum_k p_k \ell_{A_{i} A_{j}}\left(\ket{\psi_k}_{A_iA_j\overline{A_iA_j}}\right)\right)^{\frac{2}{n(n-1)}}. \nonumber\\
& \geq \sum_k p_k\left(\prod_{i=1}^{n}\prod_{j=i+1}^{n} \ell_{A_{i} A_{j}}\left(\ket{\psi_k}_{A_iA_j\overline{A_iA_j}}\right)\right)^{\frac{2}{n(n-1)}}.
\end{align}
Considering the n-party GHZ and W states of n-qubits, where it can be verified that $\ell_{A_1A_2\cdots A_n}$ ranks the GHZ state higher than the W state, conforming to condition (5) in \cref{mult_ent}. For example, in the four party case we have
\begin{align}
	\ell_{\textrm{\tiny{ABCD}}}=(\ell_{\textrm{\tiny{AB}}}\ell_{\textrm{\tiny{AC}}}\ell_{\textrm{\tiny{AD}}}\ell_{\textrm{\tiny{BC}}}\ell_{\textrm{\tiny{BD}}}\ell_{\textrm{\tiny{CD}}})^{\frac{1}{6}}
\end{align}
Note that the maxima for the n-party case changes slightly from the three party case because the bound becomes
\begin{align}\label{labboundn0p}
\ell_{A_iA_j}\leq \textrm{Min}\{2\log[d_{A_i}],2\log[d_{A_j}],\log[d_{\overline{A_{i}A_{j}}}]\}
\end{align}
where $\overline{A_iA_j}$ refers to the rest of the system with respect to the bipartite system $A_iA_j$. Once again if the Hilbert space dimensions of all parties are equal i.e $d_{A_1}=d_{A_2}=d_{A_3}=\cdots=d$ then the bound becomes
\begin{align}\label{labboundnp}
\ell_{A_iA_j}\leq 2\log[d]
\end{align}
Hence the full multipartite L-entropy defined in \cref{multil} also obeys the same bound
\begin{equation}
	\ell_{A_1A_2\cdot\cdot\cdot A_k}\leq 2\log[d].
\end{equation}

It is widely recognized in quantum information theory that the Hilbert space of pure four-qubit states can be categorized into numerous distinct classes of entangled states, unlike the simpler two-class system in the three-qubit scenario \cite{PhysRevA.65.052112}. A survey of such multipartite entangled states was carried out in \cite{Enríquez_2016}. In \cref{tab:sample}, we present the numerical values of tripartite information, the Markov gap, and L-entropy for various such states. Our findings reveal that, among these notable states, the cluster state exhibits the highest numerical value for four-party L-entropy.
\begin{table}[h]
    \centering
    \begin{tabular}{|c|c|c|c|}
        \hline
        \textbf{ 4 qubit states} & \textbf{Tripartite Information}& \textbf{Markov Gap}  & \textbf{L-entropy} \\
        \hline
      $\ket{GHZ}$ & +0.5& 0 & 0.5 \\
        \hline
        $\ket{W}$& 0.1226& 0.2896 & 0.2104 \\ 
        \hline
     $\ket{C1},\ket{C2},\ket{C3}$& -0.5& 0 & 0.7937 \\
        \hline
         $\ket{D4}$& -0.6887& 0.1266 & 0.7696 \\
         \hline
         $\ket{M}$& -0.6887& 0.1266 & 0.7696  \\
        \hline
          $\ket{BSSB4}$& -6009& 0 & 0.6394 \\
        \hline
          $\ket{HD}$& -0.3774& 0.3962 & 0.3962 \\
        \hline
           $\ket{YC}$&  -0.5& 0 & 0.7937 \\
        \hline
           $\ket{L}$& -0.3774& 0.3962 & 0.3962 \\
        \hline
     $\ket{\psi}$& -0.5& 0 & 0.7937 \\
        \hline
    \end{tabular}
    \caption{Different states involving four qubits displaying genuine four-party entanglement alongside their respective values of the tripartite information, four-party L-entropy and Markov gap. We have normalized all the quantities by $2\log[d]=2\log[2]$ such that maximal allowed value of L-entropy is 1.}
    \label{tab:sample}
\end{table}
The states in the above table are given by
\begin{align}
\ket{GHZ}&= \frac{1}{\sqrt{2}}(\ket{0000}+\ket{1111})\nonumber\\ 
\ket{W}&= \frac{1}{2}(\ket{0001}+\ket{0010}+\ket{0100}+\ket{0001})\nonumber\\
\left|\mathrm{C1}\right\rangle&=\frac{1}{2}(|0000\rangle+|0011\rangle+|1100\rangle-|1111\rangle)\nonumber\\
\left|B S S B_4\right\rangle&=\frac{1}{2 \sqrt{2}}(|0110\rangle+|1011\rangle+i(|0010\rangle+|1111\rangle)+(1+i)(|0101\rangle+|1000\rangle))\nonumber\\
|H D\rangle&=\frac{1}{\sqrt{6}}(|1000\rangle+|0100\rangle+|0010\rangle+|0001\rangle+\sqrt{2}|1111\rangle),\nonumber\\
|D_4\rangle&=\frac{1}{\sqrt{6}}\left[|0011\rangle+|1100\rangle+w(|0101\rangle+|1010\rangle)+w^2(|0110\rangle+|1001\rangle)\right], \quad w=\exp (2 i \pi / 3)\nonumber\\
|Y C\rangle&=\frac{1}{2 \sqrt{2}}(|0000\rangle-|0011\rangle-|0101\rangle+|0110\rangle+|1001\rangle+|1010\rangle+|1100\rangle+|1111\rangle) \nonumber\\
|\psi\rangle&= \frac{z_0+z_1}{2}|0000\rangle+\frac{z_0-z_1}{2}|0011\rangle+\frac{z_2+z_3}{2}|0101\rangle+\frac{z_2-z_3}{2}|0110\rangle+ \nonumber\\
& \frac{z_2-z_3}{2}|1001\rangle+\frac{z_2+z_3}{2}|1010\rangle+\frac{z_0-z_1}{2}|1100\rangle+\frac{z_0+z_1}{2}|1111\rangle .\nonumber\\
|L\rangle&=\frac{1}{2 \sqrt{3}}((1+\omega)(|0000\rangle+|1111\rangle)+(1-\omega)(|0011\rangle+|1100\rangle)+\nonumber \\
& \quad \quad \left.\omega^2(|0101\rangle+|0110\rangle+|1001\rangle+|1010\rangle)\right), \quad \omega=\exp (2 i \pi / 3)
\end{align}

\subsection{k-uniform states  and L-entropy}
\label{k_unif}
This section provides a concise review of  the notion of a $k$-uniform state within a multiparty quantum system. Subsequently, we will illustrate that, in such systems, the bipartite L-entropies achieve their peak values, leading to the maximum possible value for the entire multiparty L-entropy. Following that, we will describe how a cluster state becomes $2$-uniform when more than four qubits ($n > 4$) are incorporated.

A quantum state involving $n$ parties is called a $k$-uniform state when every reduced density matrix, involving any $k$-party subset of these $n$ parties, is maximally mixed \cite{GISIN19981,HIGUCHI2000213,Arnaud_2013,Enríquez_2016}. It is important to note that the spectra of a $k$-party reduced density matrix coincide with the spectra of an $n-k$ party density matrix. Consequently, it is possible to have no more than $n/2$ reduced states that are maximally mixed. In other words, the value of $k$ for a $k$ uniform state is subject to the following bound
\begin{align}
    k\leq \lfloor \frac{n}{2}\rfloor\label{eq: k uniform state condition}
\end{align}
where $\lfloor X\rfloor$ denotes the greatest integer less than X. If all $\lfloor \frac{n}{2}\rfloor$ are maximally mixed then such a state is called absolutely maximally entangled (AME) state. Note that the partial trace of a maximally mixed state always results in a maximally mixed state, and hence any $k$-uniform state is also a $k-1$ uniform state, but the other way around need not be true. For the purpose of the present article we will restrict ourselves to examining whether or not states are 2-uniform. Note that a 2-uniform state essentially means all the bipartite states are maximally mixed, and it is easy to check that the bipartite L-entropy is maximal for a maximally mixed bipartite state
\begin{align}
S_R(A_i:A_j)=0&,\qquad S(A)=S(B)=\log[d]\\
\ell_{A_iA_j}&=2\log[d]
\end{align}
This in turn implies that the full multipartite L-entropy  is maximal for a $2$-uniform state. 
\begin{align}
    \ell_{A_1A_2\cdot\cdot\cdot A_n}=2\log[d]
\end{align}
Note that one of the best examples for $2$-uniform states for $n> 4$ qubits turns out to be the cluster state defined as
\begin{align}
    |Cl_n\rangle=e^{-\frac{i \pi}{4}\sum_{i}\sigma_z^i\sigma_z^{i+1}}\underbrace{\ket{++\cdots+}}_{n-qubits}
\end{align}
where
\begin{align}
    \ket{+}=\frac{1}{\sqrt{2}}(\ket{0}+\ket{1}).
\end{align}


%

%
\subsubsection{2-uniform states by L-entropy optimization}\label{optimization}
\begin{figure}[H]
	\centering
	\begin{subfigure}[h]{0.45\textwidth}
		\centering
		\includegraphics[width=1\textwidth]{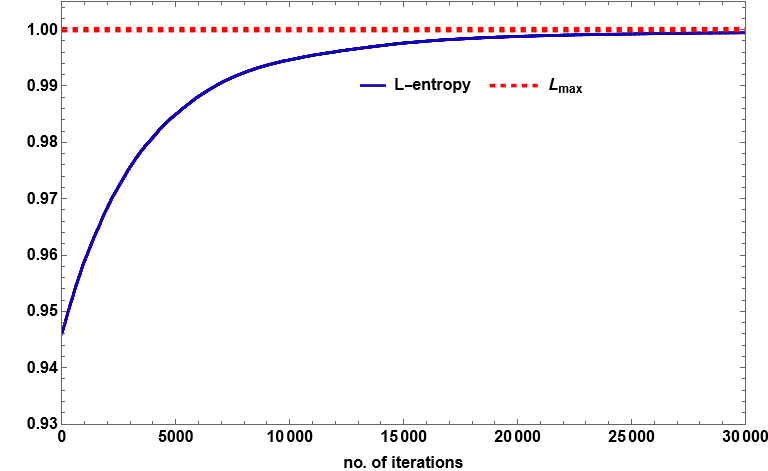}
		\caption{8 party d=2}
		\label{fig:opti1}
	\end{subfigure}
    \begin{subfigure}[h]{0.45\textwidth}
		\centering
		\includegraphics[width=1\textwidth]{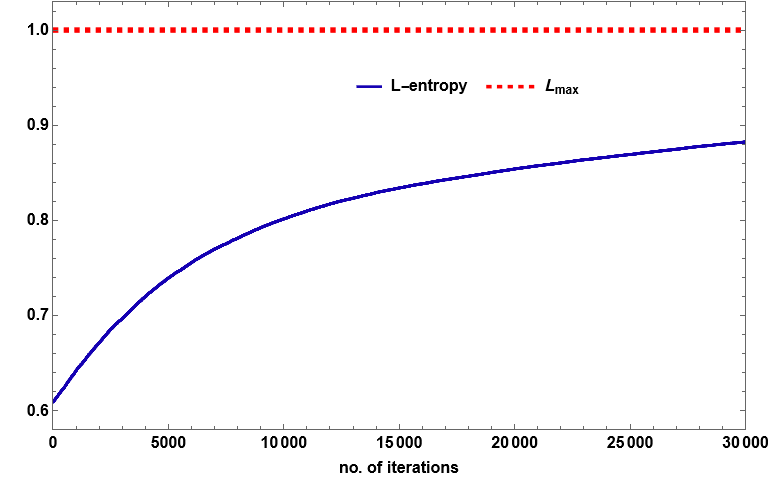}
		\caption{4 party d=6}
		\label{fig:opti2}
	\end{subfigure}	
    \caption{Optimization to reach maximal L-entropy state}
\end{figure}

The L-entropy serves as an effective measure of multipartite entanglement and is maximized by 2-uniform states. Consequently, optimizing the L-entropy can provide a systematic approach to discovering 2-uniform states. Starting from an initial random state $|\psi_0\rangle$, we introduce a small random perturbation $|\epsilon\rangle$ to generate candidate states. The next state $|\psi_1\rangle$ is chosen as the one with the highest L-entropy among three possibilities: the original state $|\psi_0\rangle$, the positively perturbed state $|\psi_0 + |\epsilon\rangle|$, and the negatively perturbed state $|\psi_0 - |\epsilon\rangle|$\footnote{Considering both $\pm|\epsilon\rangle$ perturbations improves the efficiency of the optimization process.}.

\begin{align}
    c_{0,-}\big(|\psi_0\rangle -  | \epsilon\rangle\big)\;\;,\quad |\psi_0\rangle \;\;,\quad c_{0,+}\big(|\psi_0\rangle +  | \epsilon\rangle\big)
\end{align}
where $c_{0,\pm}$ is the normalization constant.
This procedure can be repeated iteratively to generate a sequence of states $|\psi_j\rangle$ ($j=0,1,2,\dots$). The method closely resembles a zero-temperature Monte Carlo simulation. As an example, we optimize the L-entropy starting from an 8-partite random state where the dimension of each party is $d=2$ (see Fig.~\ref{fig:opti1}). The L-entropy of the initial random state is $0.946034$, which exceeds the value $0.920593$ estimated using the resolvent technique~\eqref{Lenrandom} (in units of $2\log d$). This discrepancy arises due to the small dimension $d=2$. After 30,000 iterations of the optimization procedure, we obtain a state with an L-entropy of $0.999447$, which serves as a good approximation to a $2$-uniform state. Although the L-entropy is very close to its maximum value, the resulting state is not exactly $2$-uniform. However, in some cases, the optimization process may fortuitously yield exact $2$-uniform states. A variety of exact $2$-uniform states obtained through this optimization are presented in \cref{2unioptim}.

On the other hand, the optimization process does not perform well for a 4-partite random state with $d=8$. In Fig.~\ref{fig:opti2}, the L-entropy of the initial random state is $0.608663$, which is close to the estimated value of $0.610653$ obtained using the resolvent technique. After 30,000 iterations of the optimization procedure, we obtain a state with an L-entropy of $0.882374$, which is still significantly below the maximum value of $1$. Although a 4-partite 2-uniform state $(n=4,\, k=2)$ satisfies the condition in Eq.~\eqref{eq: k uniform state condition}, in some dimensions such as $d=6$ the existence or non-existence of such a state remains unproven \cite{goyeneche2014genuinely,Pang:2019wod}. This limitation in Fig.~\ref{fig:opti2} might be attributed to the optimization process becoming trapped in local extrema, preventing further progress. While the optimization method cannot definitively confirm the non-existence of a 4-partite 2-uniform state, a more exhaustive search—such as increasing the number of iterations and exploring diverse initial configurations—would be necessary to achieve a conclusive result.

\subsection{Generalization to mixed states}\label{mixed}

Note that a pure state entanglement monotone can be extended to mixed states via a common approach called the convex roof extension \cite{Ma:2023ecg,Ge:2022sqp}. Consider the n-party density matrix $\rho$ which is expressed as a mixture of $n$-partite pure states as described below
\begin{align}
	\rho=\sum_i p_i\left|\psi_i\right\rangle\left\langle\psi_i\right|
\end{align}
where $p_i$ are classical probabilities such that $\sum_i p_i=1$.
Observe that such a decomposition is not unique. Therefore, the convex roof extension of a the pure state multiparty entanglement measure such as L-entropy is defined as follows
\begin{align}
\operatorname{CR}_\ell(\rho):=\min _{M_\rho} \sum_i p_i \ell\left(\left|\psi_i\right\rangle\right), \quad 
\end{align}
where $\min _{M_\rho}$ denotes the minimization over all possible  pure state  decompositions of the density matrix.
Note that here we resorted to convex roof extension because we were able to prove that the $\ell$ entropy is a pure state entanglement monotone. However, we can avoid this by defining quantities akin to multipartite L-entropy, using a mixed state bipartite entanglement measure such as log-negativity, a known mixed state entanglement monotone.
\begin{align}
L^{\cal E}_{AB}=\textrm{Min}\{{\cal E}(A:A^*),{\cal E}(B:B^*)\}
\end{align}
where ${\cal E}$ denotes mixed state entanglement measures such as log-negativity.

The $n$-partite generalized measure can then be defined as
\begin{equation}
	L^{\cal E}_{A_1A_2\cdot\cdot\cdot A_k}=\left(\prod_{i=1}^{k}\prod_{j=i+1}^{k}L^{\cal E}_{A_iA_j}\right)^{\frac{2}{k(k-1)}}.
\end{equation}

\subsection{Multipartite L-entropy in spin chain and SYK model}

Upon defining the multipartite form of the L-entropy, we have investigated its behavior in the context of the evolution of the nearest neighbors Ising model. As before, we observe an oscillatory pattern for the multipartite L-entropy. For comparison, we have also included a plot of the multipartite form of the Markov gap, which is derived using the geometric mean of all the bipartite Markov gaps.
\subsubsection{Ising}
\begin{figure}[H]
	\centering
	\begin{subfigure}[h]{0.32\textwidth}
		\centering
		\includegraphics[width=1\textwidth]{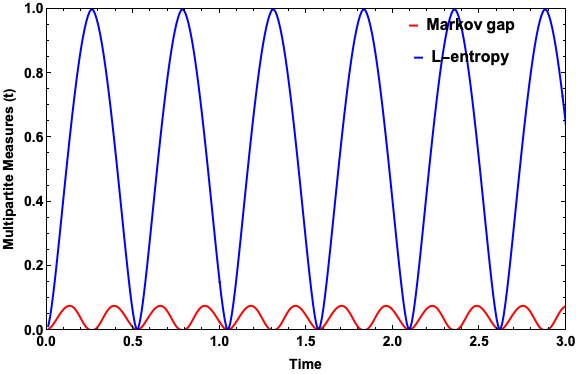}
		\caption{Initial state fully separable}
		\label{ising_N310}
	\end{subfigure}
	\hfill
	\begin{subfigure}[h]{0.32\textwidth}
		\centering
		\includegraphics[width=1\textwidth]{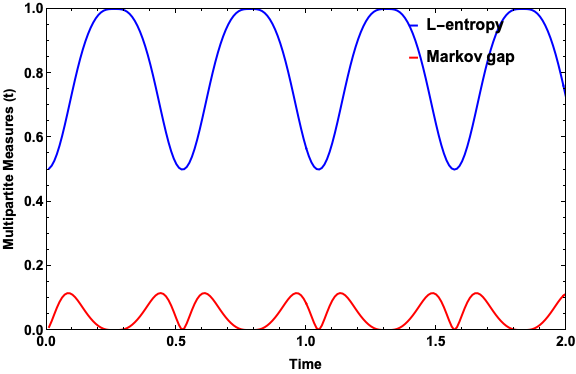}
		\caption{Initial state is  $\ket{GHZ}_5$}
		\label{ising_N320}
	\end{subfigure}
	\hfill
	\begin{subfigure}[h]{0.32\textwidth}
		\centering
		\includegraphics[width=1\textwidth]{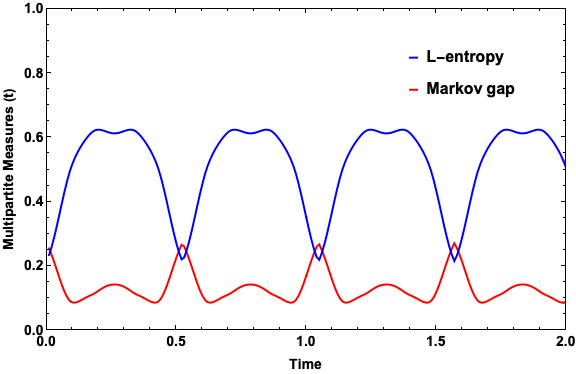}
		\caption{Initial state is $\ket{W}_5$}
		\label{ising_N330}
	\end{subfigure}
	\caption{Graphs illustrating the 8-partite L-entropy and Markov gap for a state evolving under a unitary generated by a Hamiltonian involving nearest neighbor interactions, defined by $H=J_{y}\sum_i \sigma_y^{i}\sigma_y^{i+1}$.}
	\label{ising_plots60}
\end{figure}
\begin{figure}[H]
	\centering
	\begin{subfigure}[h]{0.32\textwidth}
		\centering
		\includegraphics[width=.9\textwidth,height=4cm]{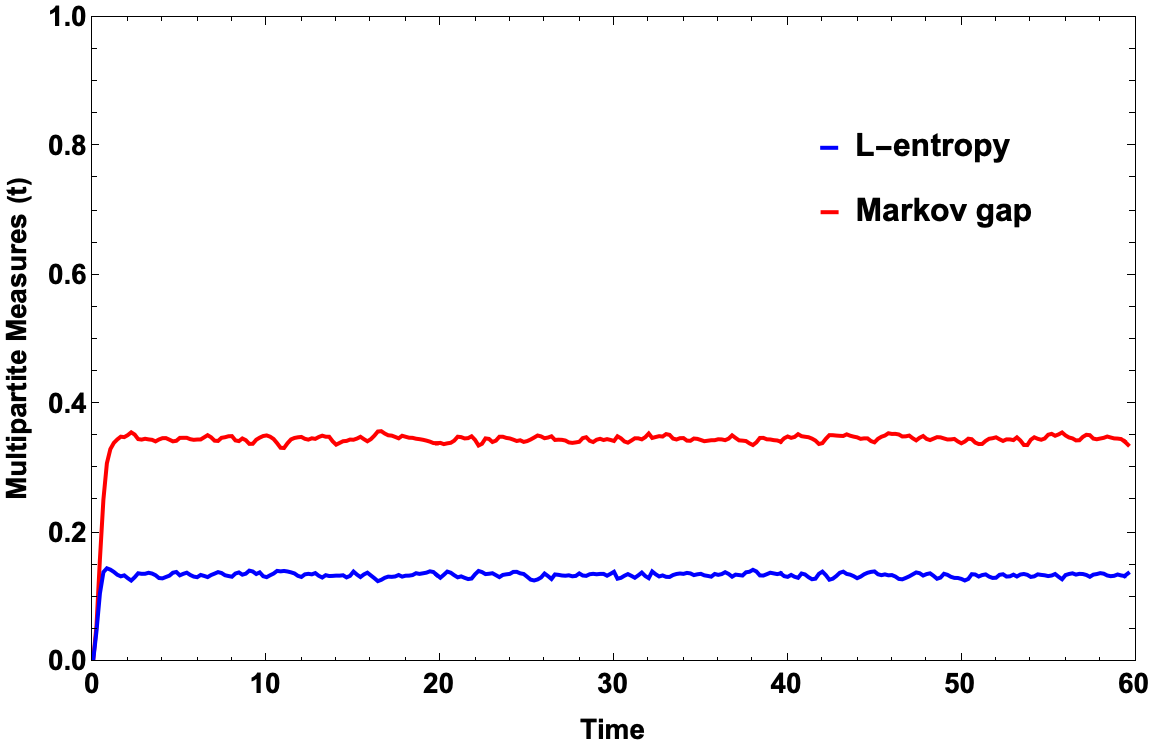}
		\caption{3-Party}
		\label{syk_N60}
	\end{subfigure}
	\hfill
	\begin{subfigure}[h]{0.32\textwidth}
		\centering
		\includegraphics[width=.9\textwidth,height=4cm]{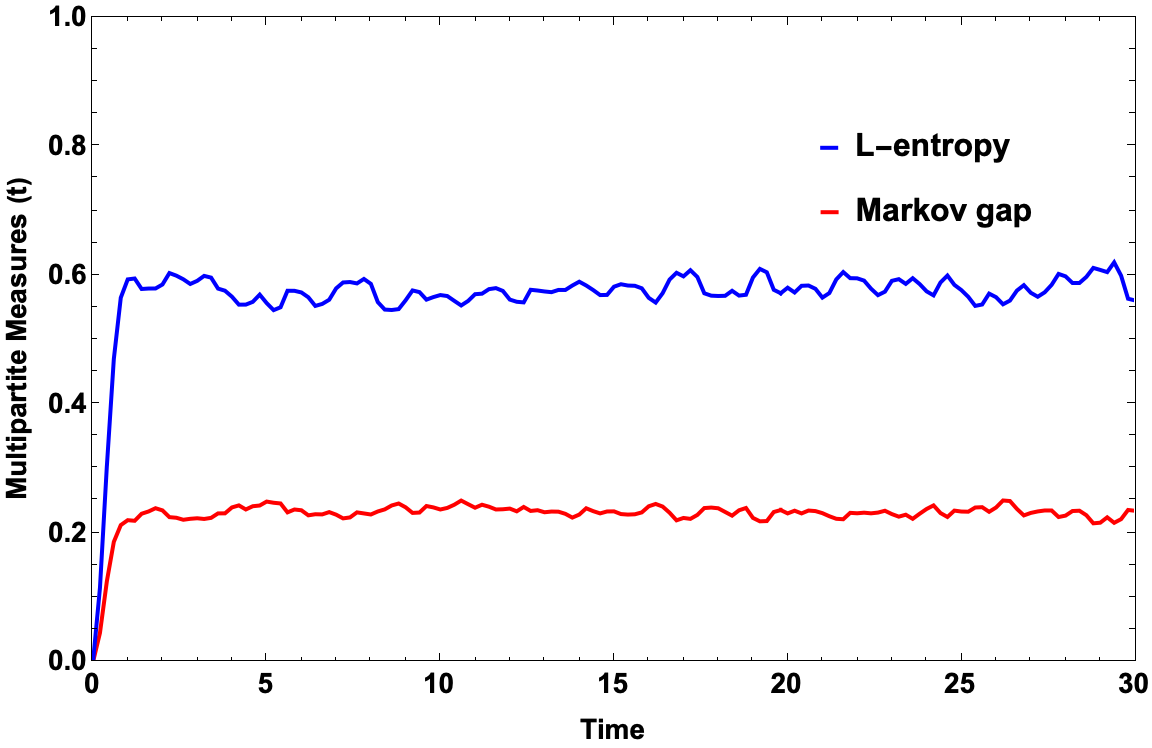}
		\caption{4-Party}
		\label{syk_N80}
	\end{subfigure}
	\hfill
	\begin{subfigure}[h]{0.32\textwidth}
		\centering
		\includegraphics[width=.9\textwidth,height=4cm]{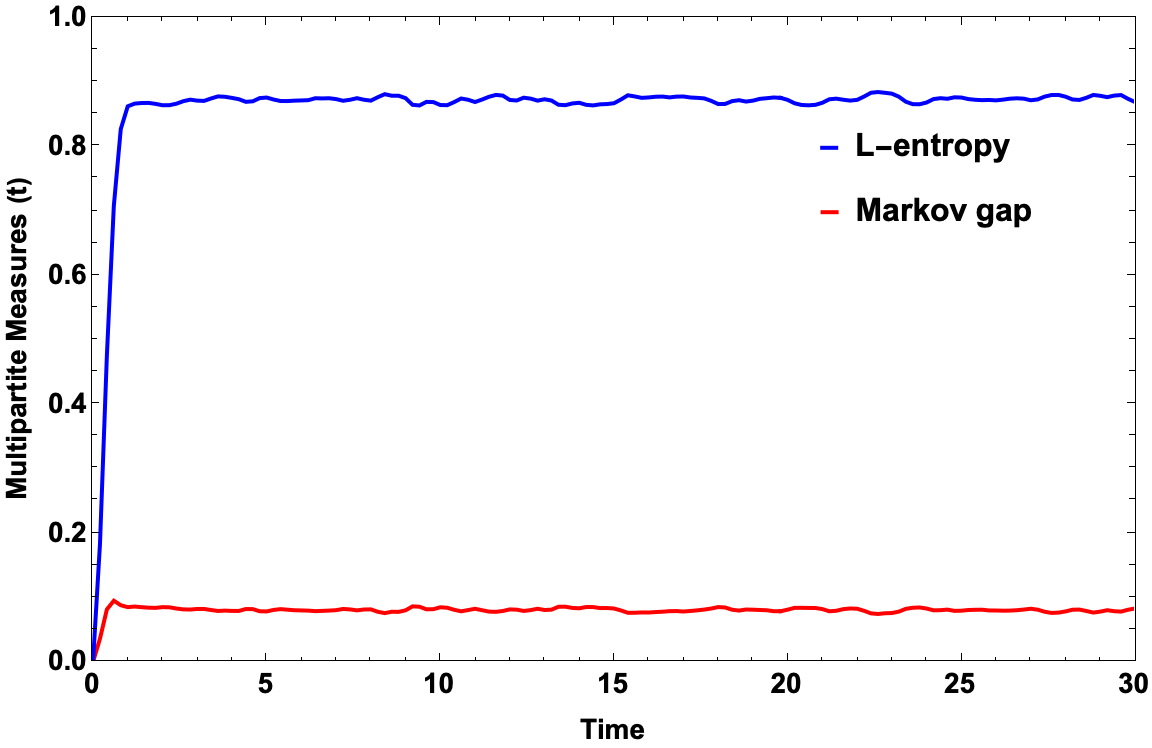}
		\caption{5-Party}
		\label{syk_N100}
	\end{subfigure}
	\caption{Plots of $n$-partite L-entropy and Markov gap  state evolving under the unitary generated SYK Hamiltonian
	}
	\label{syk_plotsM}
\end{figure}

Furthermore, we also plot the same for the SYK model in Fig.~\ref{syk_plotsM} and quite interestingly, in this case, we see that for large enough-$N$ the L-entropy grows to almost its maxima and saturates whereas the Markov gap simply vanishes. In \cref{randomsec} we will demonstrate such a behavior for random state analytically.
\subsubsection{SYK}
\subsubsection{L-entropy of energy eigenstate of SYK model}
\label{sec: l entropy of energy eigenstate}

\begin{figure}[H] 
	\centering
	\begin{subfigure}[h]{0.32\textwidth}
		\centering
		\includegraphics[width=1\textwidth]{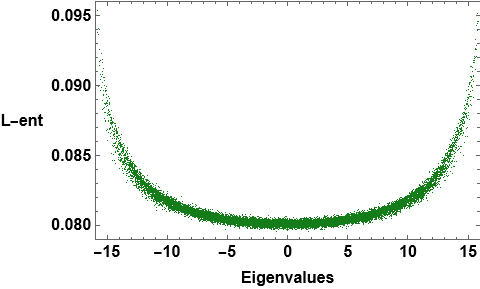}
		\caption{3-Party}
		\label{LEEeigen3}
	\end{subfigure}
	\begin{subfigure}[h]{0.32\textwidth}
		\centering
		\includegraphics[width=1\textwidth]{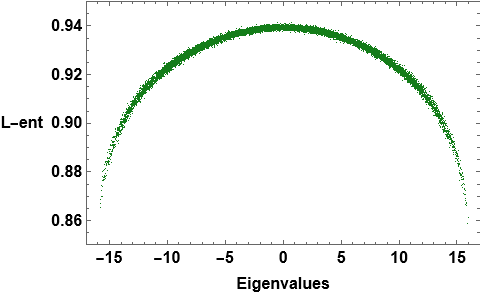}
		\caption{5-party}
		\label{LEEeigen5}
	\end{subfigure}
    \caption{L-entropy (normalized by $2\log[d]$ such that its maximum is set to 1) plotted against the energy eigenvalues for a total of $15$ qubits states in SYK model divided into $3$ and $5$-party states.}
    \label{LESYKeig}
\end{figure}

We numerically evaluate $3$ and $5$-partite L-entropy of the energy eigenstate of SYK model. The plots are depicted in Fig.~\ref{LEEeigen3}, Fig.~\ref{LEEeigen5} for 3 and 5 party quantum systems respectively. We see that for the 5 party case most of the eigen states lie in the region where the L-entropy is close to its maximal value. Quite interestingly, the $3$ party case seems to be special as the L-entropy plot seems to be inverted relative to the $5$ and $15$ party case with a very small value of L-entropy.

\section{Random states and holography}\label{holography}

\subsection{Multipartite L-entropy in random states}\label{randomsec}

Here we briefly review the results for the reflected entropy and entanglement entropy of a random state \cite{Akers:2021pvd}.  Utilizing them we derive the expression for the L-entropy for a random pure state. Following that, we will demonstrate that our L-entropy calculation aligns precisely with the numerical data.

In \cite{Akers:2021pvd}, the authors computed the reflected entropy for a random state and obtained the corresponding Page-curves for the same utilizing the resolvent technique. They demonstrated that the reflected entropy is given by the following expression
\begin{align}\label{refran}
    S_R(A:B)\approx-p_0(q) \ln p_0(q)-p_1(q) \ln p_1(q)+p_1(q)\left(\ln d_A^2-\frac{d_A^2}{2 d_B^2}\right)
\end{align}
where $q=\frac{d_A d_B}{d_{\overline{AB}}}$, $\overline{AB}$-denotes the complement of the subsystem-$AB$ such that $A,B,\overline{AB}$ together form the full system in the random pure state. Notice that the initial two terms, when combined, exhibit a form analogous to the Shannon entropy for a single bit where the probabilities $p_0$ and $p_1$ are functions dependent on the variable $q$. These probabilities are expressed in terms of the q-Catalan number, providing a direct connection to the combinatorial structure.
\begin{align}\label{piq}
   p_0(q)&\equiv \frac{C_{m / 2}\left(q^{-1}\right)^2}{C_m\left(q^{-1}\right)} \\p_1(q)&\equiv\frac{C_m\left(q^{-1}\right)-C_{m / 2}\left(q^{-1}\right)^2}{C_m\left(q^{-1}\right)} 
\end{align}
Observe that the last term in \cref{refran} resembles the entanglement entropy of the subsystem $AA^*$ in a Haar random state on $AA^*BB^*$. The $C_{n }\left(x\right)$ appearing in the above are generalization of Catalan numbers known as the $q$-Catalan numbers and they admit an analytical continuation in terms of the Hypergeometric functions as follows
\begin{align}
    C_n(q^{-1}) \equiv \begin{cases}\frac{1}{q} \hspace{0.1cm} {}_2 F_1(1-n,-n ; 2 ;\frac{1}{q}), & q \geq 1 \\ \frac{1}{q^n}\hspace{0.1cm} { }_2 F_1(1-n,-n ; 2 ; q), & q<1\end{cases}
\end{align}
For the case we are interested in $d_A=d_B=d$, it is to be noted that the entanglement entropy on the other hand is given by
\begin{align}\label{EErand}
S(A)=S(B)=\log[d]-\frac{1}{2 d_{\overline{AB}}}
\end{align}
Therefore the L-entropy for the random state ($d_A=d_B=d$)  may be computed by  utilizing \cref{EErand,refran} in \cref{ldef}
 \begin{align}\label{lenran}
    \ell_{AB}=\log[d]-\frac{1}{ d_{\overline{AB}}}+p_0(q) \ln p_0(q)+p_1(q) \ln p_1(q)-p_1(q)\left(2\ln d-\frac{1}{2 }\right)
\end{align}

\begin{figure}[H]
    \centering
    \includegraphics[width=0.5\textwidth]{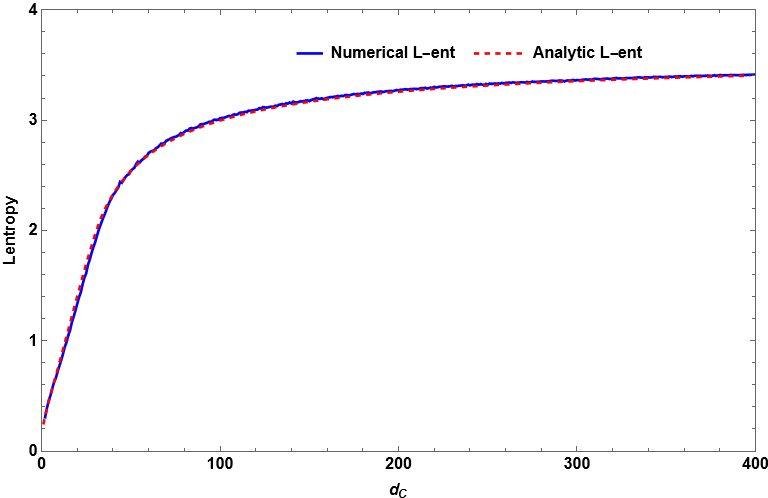}
	\caption{The dependence of L-entropy on $d_C=d_{\overline{AB}}$ is examined for the case $d_A=d_B=d=6$. The blue curve in the graph represents the numerically calculated L-entropy values whereas the red dashed line illustrates the L-entropy values derived from the analytical expression presented in \cref{lenran}.} 
 \label{random}
\end{figure}

\subsubsection{3-party}
\begin{figure}[H]
	\centering
	\begin{subfigure}[h]{0.3\textwidth}
		\centering
		\includegraphics[width=53mm]{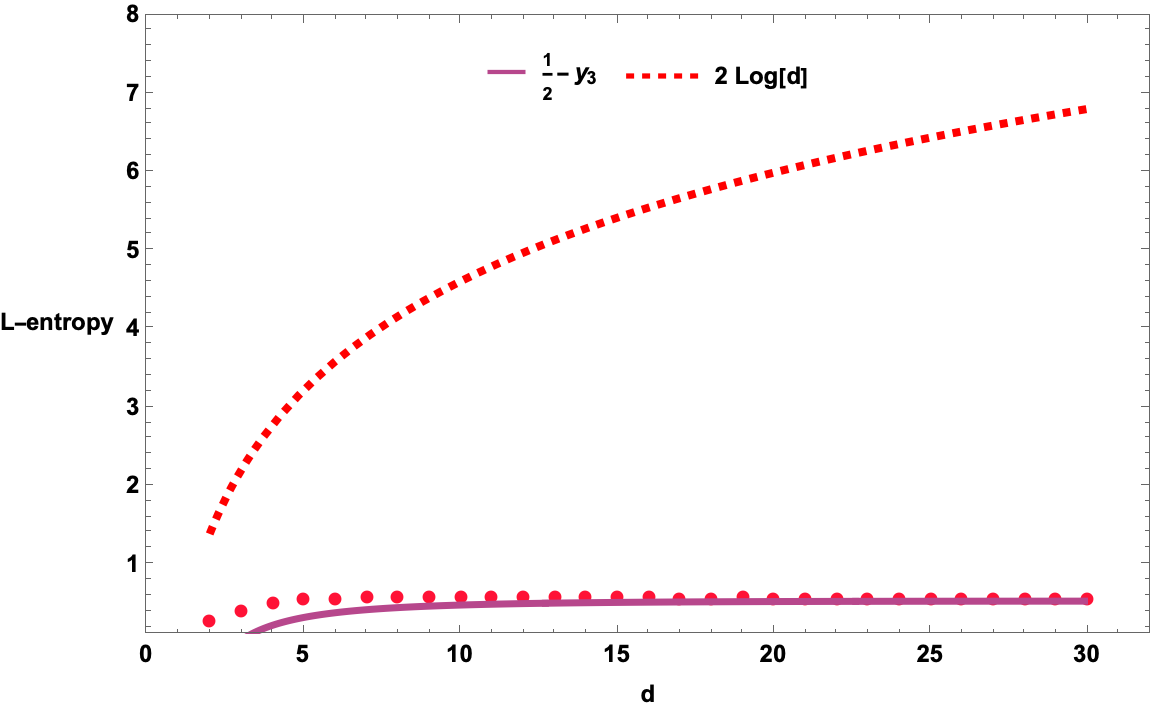}
		\caption{3-Party}
		\label{Ran_3}
	\end{subfigure}
	\hspace{0.05cm}
	\begin{subfigure}[h]{0.3\textwidth}
		\centering
		\includegraphics[width=53mm]{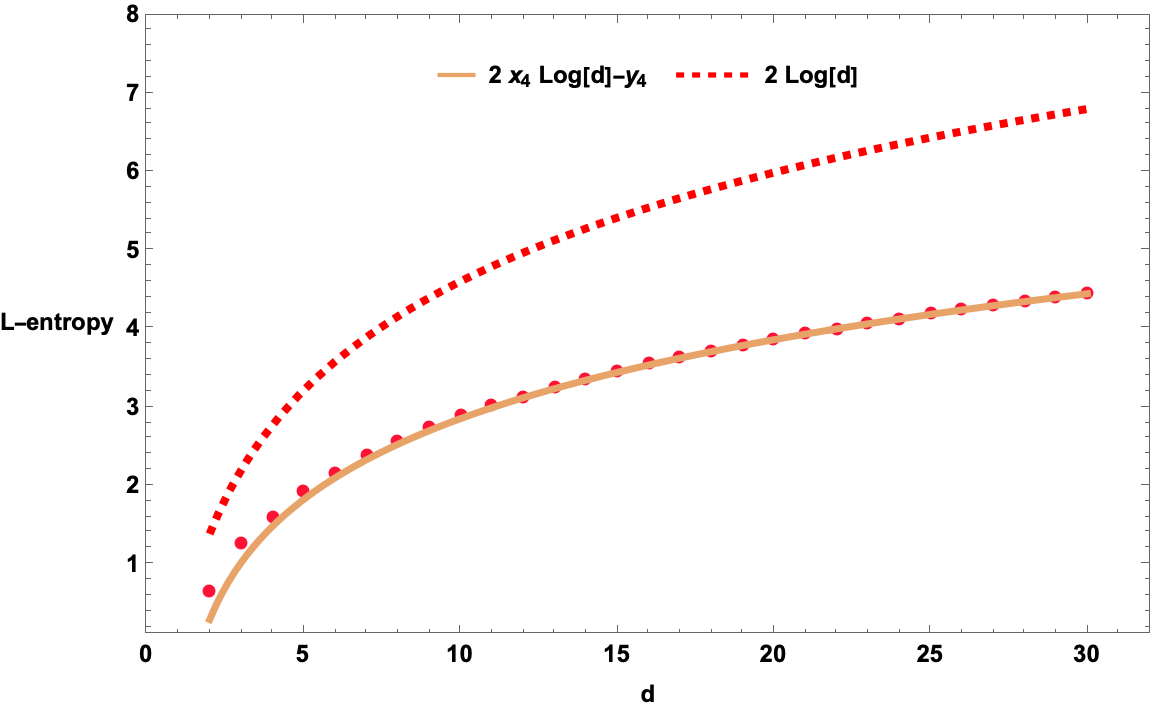}
		\caption{4-Party}
		\label{Ran_4}
	\end{subfigure}
 \hspace{0.05cm}
	\begin{subfigure}[h]{0.3\textwidth}
		\centering
		\includegraphics[width=53mm]{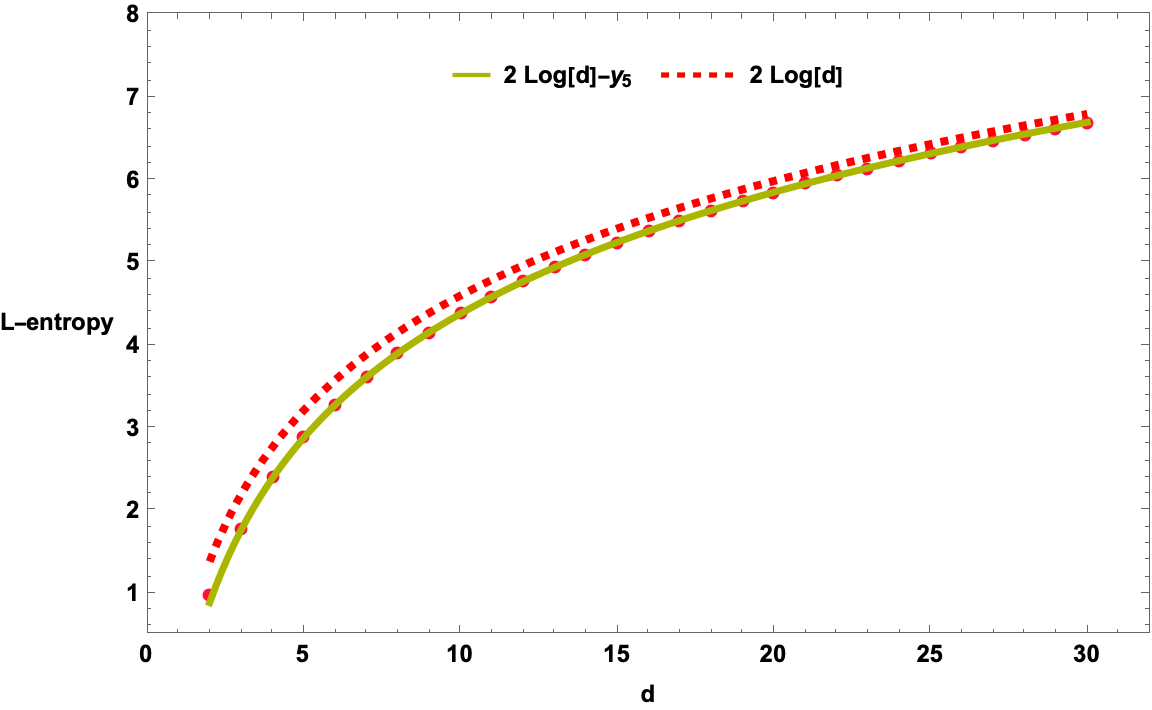}
		\caption{5-Party}
		\label{Ran_5}
  	\end{subfigure}

	\caption{ L-entropy as a function of $d_A=d_B=d$ for 3, 4 and 5-party. In all three cases the analytic results match quite well with the  numerically calculated L-entropies. Y}
	\label{345random}
\end{figure}

Consider the case involving $d_{\overline{AB}} \leq d_{A}d_{B}$, the functions $p_i(q)$ described in \cref{piq} are in their initial phase. We will first explore the situation where $d_{\overline{AB}} \leq d_{AB}$ and separately address the scenario where $d_{\overline{AB}}=d_A d_B$. Assuming all parties possess identical dimensions, the unique situation for $d_{\overline{AB}} < d_{AB} = d^2$ occurs with three parties. In this instance, w expand the reflected entropy as a series in terms of $1/d$ as follows
\begin{align}
  S_R(A:B)=  2 \log[d]-\frac{1}{2} + \frac{3-2 d \log[d]}{d}+O(\frac{1}{d^2})
\end{align}
The entanglement entropy is given by
\begin{align}\label{EErand_3}
S(A)=S(B)=\log[d]-\frac{1}{2 d}
\end{align}
Hence, we obtain the L-entropy to be
\begin{align}
 \ell_{AB}=  \frac{1}{2} +\frac{2\log [d]-5}{2 d}+O(\frac{1}{d^2})
\end{align}

\subsubsection{4-party}

The case of $d_{\overline{AB}}=d_Ad_B$ is special because in this case $q=1$ and hence $p_i(q)$ in \cref{piq} become just numbers independent of $d_{A},d_{B}$ and if we choose all the parties to have same Hilbert space dimension this essentially reduces to 4-parties. In this case we find the reflected entropy to be
\begin{align}
  S_R(A:B)= S_0(1)+p_1(1)(2\log[d]-\frac{1}{2})
\end{align}
where $S_0$ corresponds to the Shannon entropy term for $q=1$ in 
The entanglement entropy is given by
\begin{align}\label{EErand_4}
S(A)=S(B)=\log[d]-\frac{1}{2 d^2}
\end{align}
Hence, we obtain the L-entropy to be
\begin{align}
 \ell_{AB}=  (2 x_0 \log[d])+y_0+O(\frac{1}{d^2})\label{eq: l entropy 4 partite random state}
\end{align}
where $x_0,y_0$ are constants given by
\begin{align}
    x_0&=p_0(1)\approx 0.720\nonumber\\
    y_0&=-S_0(1)+\frac{p_1(1)}{2}\approx-0.453
\end{align}
Observe that, given $x_0<1$, the bipartite L-entropies in this phase do not attain their peak values. Consequently, the multipartite L-entropy is also less than its maximum possible value.

\subsubsection{5-party}

In the context of the present article, we will now focus on the more interesting phase where $q\leq 1$ ($d_{\overline{AB}}>d_A d_B$).  More specifically, we will set $d_A = d_B = d$ and investigate  the behaviour of L-entropy upon increasing $d_{\overline{AB}}$  to values significantly greater than 1 i.e $d_{\overline{AB}}>>1$. In this asymptotic regime, the expression for the q-Catalan numbers mentioned above may be utilized to expand \cref{refran} in terms of $\frac{1}{d_{\overline{AB}}}$. 
\begin{align}\label{SRrand_5}
 S_{R}(A:B) = \frac{d^2+4 d^2 \log (d)-2 d^2 \log \left(\frac{d^2}{4 d_{\overline{AB}}}\right)}{8 d_{\overline{AB}}}+O\left(\frac{1}{d_{\overline{AB}}^2}\right)
\end{align}
Note that in this phase the reflected entropy vanishes at the leading order and the above expression is the first order correction. Hence, utilizing \cref{SRrand_5,refran} in the definition for L-entropy in \cref{ldef} we obtain
\begin{align}\label{Lenrandom}
    \ell_{AB}=2\log[d]-\frac{8+d^2+4 d^2 \log (d)-2 d^2 \log(\frac{d^2}{4 d_{\overline{AB}}})}{8 d_{\overline{AB}}}+O\left(\frac{1}{d_{\overline{AB}}^2}\right)
\end{align}
If we keep only terms upto $O(1/d)$ then the  L-entropy is given by
\begin{align}\label{Lenrandom5}
    \ell_{AB}=2\log[d]-\frac{4 \log (d)+2 \log (d)+1+4 \log (2)}{8 d}+O\left(\frac{1}{d^2}\right)
 \end{align}   
The L-entropy's behavior for three, four, and five parties, along with its numerically computed values for a random quantum state, are illustrated in Fig.~\ref{345random}. In the scenario involving three parties, L-entropy remains largely independent of $d$ when considered to the leading order in $1/d$, resulting in a relatively low value. Conversely, in the four-party scenario, L-entropy scales with $\log(d)$ at the leading order. It, however, does not achieve its maximum theoretical value of $2\log(d)$. In the case of a 5-party system, L-entropy does reach the maximal value of $2\log(d)$ at leading order in $1/d$.

\begin{figure}[h]
    \centering
    \includegraphics[width=0.5\textwidth]{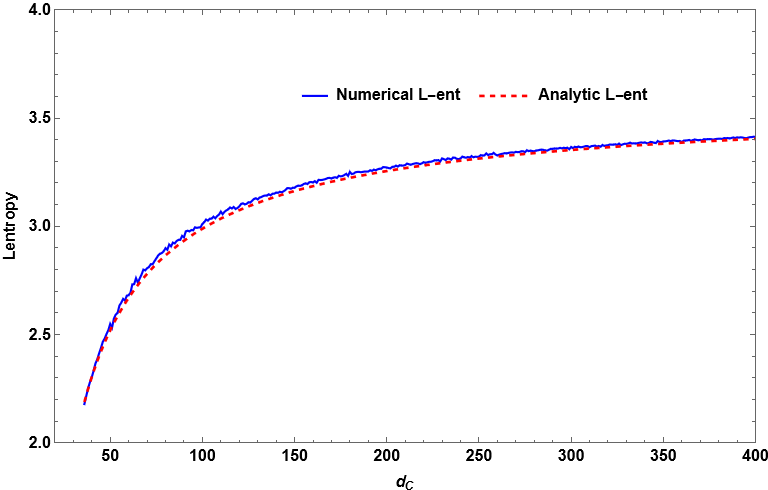}
	\caption{L-entropy as a function of $d_C$ in the second phase with $d_C>d_{AB}$ and $d_A=d_B=d=6$. The blue curve represents numerically calculated L-entropy, while the red dashed line denotes L-entropy derived from the expression given in \cref{Lenrandom}.} 
	\label{random_cgcb}
\end{figure}

The behavior of the aforementioned expression for L-entropy in the required phase ($d_{\overline{AB}}>d_A d_B$) is illustrated in Fig.~\ref{random_cgcb}, along with the numerically evaluated value of the same for a random state. Notably, the leading term in \cref{Lenrandom} corresponds to the maximum L-entropy value, which is obtained for the maximally mixed state $\rho_{AB}=\frac{1}{d^2}{\cal I}$. Additionally, considering the 5-party pure state, it is observed that all bipartite L-entropies attain their maximum values, and the bipartite density matrices are all maximally mixed and hence proportional to the Identity matrix. This particular state is referred to as a 2-uniform state in quantum information theory.

\subsection{Holography and multiboundary wormholes}

In this section, we explore the holographic understanding of the multipartite entanglement utilizing L-entropy. In \cref{set_up}, L-entropy has been defined as the difference between the entanglement entropy and the reflected entropy. In the $AdS/CFT$ correspondence, the entanglement entropy of a subsystem in the $CFT$ is given by the area of the homologous codimension-two bulk minimal surface. These surfaces are also known as the Ryu-Takayanagi (RT) surfaces \cite{Ryu:2006bv,Hubeny:2007xt}. However, the holographic dual of the reflected entropy is proposed to be the entanglement wedge cross section (EWCS) which has a more involved geometry. Note that, EWCS has also been proposed as a bulk dual of different measures as the entanglement of purification (EoP) \cite{Takayanagi:2017knl}, odd entropy \cite{Tamaoka:2018ned}, balanced partial entanglement (BPE) \cite{Wen:2021qgx} and the entanglement of negativity \cite{Kudler-Flam:2018qjo,KumarBasak:2020eia}. 

\begin{figure}[h]
	\centering
	\includegraphics[width=.4\linewidth]{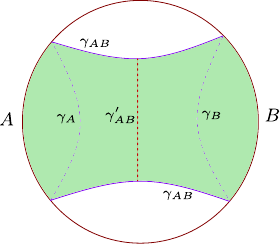}
	\caption{Entanglement wedge cross section.}  
	\label{wedge}
\end{figure}

In this discussion, we consider the configurations of two subsystems $A$ and $B$ situated on the $CFT$ at the boundary.  The dual of the density matrix $\rho_{AB}$ corresponding to these intervals is defined as the entanglement wedge which is a bulk region with boundaries $A$, $B$ and the RT surface for $A\cup B$, $\gamma_{AB}$. The EWCS $\gamma^\prime_{AB}$ can be defined as the minimal cross sectional area of the entanglement wedge as,
\begin{align}
    E_W(A:B)=\frac{\text{Area}(\gamma^\prime_{AB})}{4G_N},
\end{align}
where $G_N$ is the Newton constant. In Fig.~\ref{wedge}, the RT surfaces corresponding to $A,B$ and the EWCS corresponding to the bipartite system $AB$ are depicted. 
Note that, for a disconnected wedge where $\gamma_{AB}=\gamma_A\cup \gamma_B$, the EWCS is zero. This phase dominates when the subsystems are far away from each other. The authors in \cite{Dutta:2019gen}, demonstrated the holographic description of the canonical purification of a mixed state $\rho_{AB}$ by gluing the entanglement wedges of $A\cup B$ and $A^\star\cup B^\star$ along the RT surfaces $\gamma_{AB}$ and $\gamma_{A^\star B^\star}$ where $A^\star$ and $B^\star$ are the canonical conjugates of $A$ and $B$ respectively. Following the definition in \cref{def_SR}, the RT surface of the subsystem $A\cup A^\star$ in the glued geometry is the holographic dual of the reflected entropy. Interestingly, the RT surface can be expressed as the union of two identical EWCS $\gamma^\prime_{AB}$ and $\gamma^\prime_{A^\star B^\star}$. Finally, the holographic dual of the reflected entropy is shown to be twice the area of the EWCS,
\begin{align}
    S_R(A:B)=2E_W(A:B).
\end{align}
Utilizing the above relation, the holographic L-entropy can be written as,
\begin{align}
    \ell_{AB}=\frac{\text{Min}\left[\text{Area}(\gamma_A),\text{Area}(\gamma_B)\right]-\text{Area}(\gamma^\prime_{AB})}{2G_N}.
\end{align}
In the pure state limit of $\rho_{AB}$, $\ell_{AB}$ is zero indicating the absence of tripartite entanglement. However, for subsystems possessing a disconnected wedge, the EWCS $\text{Area}(\gamma^\prime_{AB})=0$ which yields the highest possible value of $\ell_{AB}$. Interestingly, the L-entropy $\ell_{ABC}$ depends on the L-entropies of all possible pair of parties $\ell_{AB}$, $\ell_{BC}$ and $\ell_{AC}$. It indicates the fact that all L-entropies corresponding to the pair of subsystems have to achieve the maximum in order to show the maximum tripartite information in state $\rho_{ABC}$. In the following subsections, we will discuss some of the holographic scenarios where we obtain the L-entropy.

\subsection*{Multiboundary wormhole}

In this section, we will analyze the L-entropy for a multiboundary wormhole \cite{Krasnov:2000zq,Skenderis:2009ju,Balasubramanian:2014hda}. First, considering a three-boundary wormhole, we calculate the L-entropy and obtain the Page curve corresponding to a black hole evaporation process. Furthermore, we consider a four-boundary wormhole and analyze the characteristics of L-entropy.

\subsubsection{Three-boundary wormhole}

\begin{figure}[h]
	\centering
	\includegraphics[width=.4\linewidth]{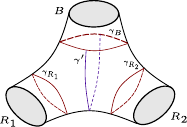}
	\caption{Three-boundary wormhole.}  
	\label{mb_wh}
\end{figure}

Here we consider the three-boundary wormhole following the model in \cite{Akers:2019nfi} where one of the boundaries can be thought of as a black hole and the other two as the radiation regions. Utilizing the L-entropy, we will evaluate the genuine tripartite entanglement between the black hole and the two radiation regions along the evaporation of the black hole. In Fig.~\ref{mb_wh}, $\gamma_{R_1},~\gamma_{R_2}$ and $\gamma_{B}$ are the HRT surfaces corresponding to the two radiation region and the black hole respectively. The entanglement wedge for the total radiation region ($R_1\cup R_2$) is the interior bulk region of the wormhole bounded by $\gamma_{R_1},~\gamma_{R_2}$ and $\gamma_{B}$. The plausible entanglement wedge cross sectional surfaces corresponding to this wedge geometry are $\gamma_{R_1},~\gamma_{R_2}$ and $\gamma^\prime$. 
Following the constructions in \cite{Hayden:2021gno,Akers:2022max,Akers:2022zxr}, the area of the surface $\gamma^\prime$ can be computed in terms of the area of the HRT surfaces as,
\begin{align}\label{ewcs_mbwh}
    \mathcal{A}_{\gamma^\prime}=2 \sinh ^{-1}\left[\textrm{csch}\left(\frac{\mathcal{A}_B}{2}\right) \sqrt{2 \cosh \left(\frac{\mathcal{A}_B}{2}\right) \cosh \left(\frac{\mathcal{A}_{R_1}}{2}\right) \cosh \left(\frac{\mathcal{A}_{R_2}}{2}\right)+\frac{\cosh \left(\mathcal{A}_{R_1}\right)}{2}+\frac{\cosh \left(\mathcal{A}_{R_2}\right)}{2}+1}\right]
\end{align}
where $\mathcal{A}_j$ are the area of the HRT $\gamma_j$. Here we further simplify the calculation by considering the total energy of the spacetime to be fixed. Consequently, utilizing the energy-entropy relation of the holography $S=2\pi\sqrt{cE/3}$, the areas of the HRT surfaces follow the relation,
\begin{align}
    \mathcal{A}_{B_0}^2=\mathcal{A}_{R_1}^2+\mathcal{A}_{R_2}^2+\mathcal{A}_{B}^2,
\end{align}
where $\mathcal{A}_{B_0}$ is the initial horizon area of the black hole where no radiation region existed. Now the bipartite L-entropies can be computed for three possible pairs of boundaries. Finally, the L-entropy is obtained by following the \cref{genl}.
\begin{figure}[h]
	\centering
	\includegraphics[width=.5\linewidth]{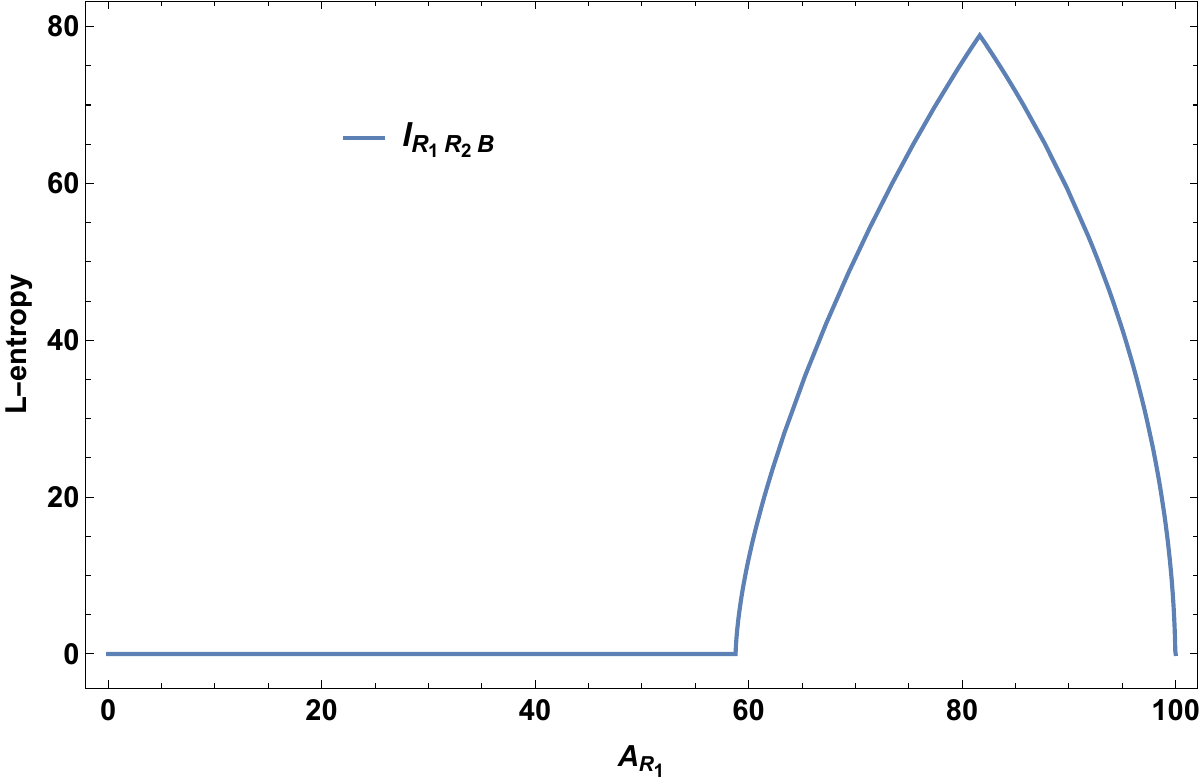}
	\caption{Page curve for L-entropy.}  
	\label{l_mb_wh}
\end{figure}
 We have plotted the L-entropy in Fig.~\ref{l_mb_wh} for the increasing area of the HRT surfaces of $R_1$ and $R_2$ where both these boundaries are considered to be of the same size for simplicity. Note that in Fig.~\ref{l_mb_wh}, L-entropy starts from zero and finally becomes zero again, indicating unitary evolution. This is expected as initially we only have a black hole without any radiation region, and finally, the black hole has evaporated completely leaving two radiation regions where genuine tripartite entanglement is zero. We will call the characteristics shown in Fig.~\ref{l_mb_wh} the \textit{Page curve} for L-entropy following the developments in \cite{Page:1993df,Page:1993wv,Page:2013dx}. Interestingly, the L-entropy remains zero until the Page time for the entanglement entropy of the total radiation region where $\mathcal{A}_{B}=\mathcal{A}_{R_1}+\mathcal{A}_{R_2}$. This suggests that the degrees of freedom beyond the black hole's horizon are inaccessible, rendering the entire spacetime as a biseparable system where entanglement exists solely between the two radiation regions. After the Page time, the information from the black hole interior becomes accessible and the L-entropy shows a significant increase. However, considering the three-party scenario, the L-entropy is bounded by the degrees of freedom of the subsystems as indicated in \cref{labbound3p}. As a result, the maximum value of this measure is obtained when $\mathcal{S}_{R_1}=\mathcal{S}_{R_2}=\mathcal{S}_{B}$. At this time $t=t_{max}$, the value of L-entropy is same to the entanglement entropy of either the black hole or the radiation subsystems if we consider $\mathcal{A}_0$, $\mathcal{A}_{R_1}$ and $\mathcal{A}_{R_2}$ to be large. In the same limit, the Markov gap also results to be infinitesimally small. 

\subsubsection*{ Three party random state and three boundary wormhole}

In this section, we undertake a comparative analysis of the multipartite entanglement configuration in a 3-boundary wormhole (with equal horizon lengths) and that found in a three-party random state. Our findings will illustrate that  the L-entropy displays significantly different scaling behavior in these two scenarios, providing evidence that the 3-boundary wormhole (with equal horizon lengths) does not correspond to the three-party random state. As elaborated in \cref{randomsec}, the entanglement entropy for individual parties $A$ and $B$ can be approximated using~\cite{Page:1993df}
\begin{align}
	 S_A\,=\,S_B\,\approx\, \log d - {1\over 2 d} \label{eq: gamma in 3 partite random state}
\end{align}
where $\gamma$ denotes the length of the horizon corresponding to $S_A$ and $S_B$. The reflected entropy and hence the L-entropy are given by
\begin{align}
  S_R(A:B)&=  2 \log[d]-\frac{1}{2} + \frac{3-2 d \log[d]}{d}+O(\frac{1}{d^2})\\
  \end{align}
  which led to the following L-entropy
  \begin{align}
 \ell_{AB}&=  \frac{1}{2} +\frac{2\log [d]-5}{2 d}+O(\frac{1}{d^2})
\end{align}
Hence, at the leading correction comes at $O(d^0)$.

Now, we evaluate the reflected entropy from the entanglement wedge cross section in the 3-party state which is holographically dual to the 3-boundary wormhole. In this three boundary wormhole, one can evaluate the entanglement wedge cross section $\mathcal{A}_{\gamma'}$ by 
\begin{align}
    \mathcal{A}_{\gamma^\prime}\,=\,2 \sinh ^{-1}\left[\textrm{csch}\left(\frac{\mathcal{A}_C}{2}\right) \sqrt{2 \cosh \left(\frac{\mathcal{A}_C}{2}\right) \cosh \left(\frac{\mathcal{A}_{A}}{2}\right) \cosh \left(\frac{\mathcal{A}_{B}}{2}\right)+\frac{\cosh \left(\mathcal{A}_{A}\right)}{2}+\frac{\cosh \left(\mathcal{A}_{B}\right)}{2}+1}\right]
\end{align}

When the horizon lengths are considered to be equal (which is equivalent to taking $d_A=d_B=d_C$ in the random state) i.e
 $\mathcal{A}_A=\mathcal{A}_B=\mathcal{A}_B=\gamma$ as the length of the horizon in each asymptotic boundary then the above expression can be approximated in the large $\gamma$ limit to be 
 \begin{align}
       \mathcal{A}_{\gamma^\prime}\,\approx \,2 \sinh ^{-1}[e^{\frac{\gamma}{4}}]\approx \frac{\gamma}{2}
 \end{align}
The reflected entropy is therefore given by
\begin{align}
    S_R(A:B)&=  2 \mathcal{A}_{\gamma^\prime} \approx \gamma
\end{align}
Hence we obtain the L-entropy to be 
\begin{align}
	\ell_{AB}\,\approx \gamma=\log[d_{eff}]
\end{align}
If we consider $d_{eff}=e^{\gamma}$, the L-entropy in three boundary wormhole is $O(\log[d])$ which was not true in three party random state where the leading contribution comes at $O(d^0)$. This clearly  indicates that the three party state is not dual to the three boundary wormhole with equal horizon lengths.

 \subsubsection{Four-boundary wormhole}

 In this section, we consider a four-boundary ($B_i$ for $i=1,\cdot\cdot\cdot,4$) wormhole by sewing two pairs of ``pants" along $\gamma_{14}$ following the construction given in \cite{Balasubramanian:2014hda}. One can apply twists $\theta$ along the patching surface, resulting in different spacetime geometries \cite{Marolf:2015vma}. These possibilities indicate a rich phase structure of L-entropy and multipartite entanglement. Here we consider a genus zero four-boundary wormhole geometry with $\theta=0$ where the L-entropy can be explained utilizing the parameter space of five independent parameters $\mathcal{A}_{i}$ for $i=1,\cdot\cdot\cdot,4$ and $\mathcal{A}_{14}$. Here, $\mathcal{A}_{i}$ and $\mathcal{A}_{14}$ corresponds to the area of the surfaces $\gamma_{i}$ and $\gamma_{14}$ respectively.
 
 \begin{figure}[h]
	\centering
	\includegraphics[width=.4\linewidth]{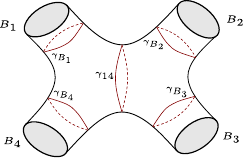}
	\caption{Four-boundary wormhole.}  
	\label{mb_wh_4}
\end{figure}
 Let us first consider the limit where any $\mathcal{A}_i$ is smaller than the area of any surface situated in the interior bulk. In this specific limit, the mutual information $I(B_1:B_2)$, $I(B_2:B_3)$, $I(B_3:B_4)$ and $I(B_4:B_1)$ are all identically zero. Furthermore, considering $\mathcal{A}_{i}=\mathcal{A}$ the entanglement entropy for $B_1\cup B_3$ is $\frac{1}{4G}(A_1+A_3)=\frac{1}{4G}(A_2+A_4)$. Consequently, the mutual information $I(B_1:B_3)$ and $I(B_2:B_4)$ are also found to be zero. Interestingly, for this specific limit, we observe all the reflected entropies to be zero as the wedge becomes disconnected. As a result, the L-entropy measuring the four-party entanglement becomes $\ell_{B_1B_2B_3B_4}=2\mathcal{A}$ which is the maximum permissible value of L-entropy for given $\mathcal{A}$. Note that, in this specific parameter space, \cite{Balasubramanian:2014hda} explored the tripartite mutual information $I_{3}$ as a measure of $n$-partite entanglement which also obtained the highest possible value. However, the maximum value of L-entropy can only be observed when $\mathcal{A}_i+\mathcal{A}_j<\mathcal{A}_{ij}$. Beyond this limit, the L-entropy decreases significantly. The exact computation for a generic four-boundary geometry is expected to show a richer phase structure of the multipartite entanglement.


\subsubsection*{Four party random state and four boundary wormhole}

Analogous to our previous analysis of the tripartite scenario, we shall now examine the distinctions in the multipartite entanglement structure between a randomly chosen 4-party quantum state and a 4-boundary wormhole. Focusing on the parties $A$, $B$, $C$, and $D$ that form a random state, the entanglement entropy concerning individual parties $A$ and $B$ may once again be approximated by~\cite{Page:1993df} 
\begin{align}
	 S_A\,=\,S_B\,\approx\, \log d - {1\over 2 d^2} \label{eq: gamma in 4 partite random state}
\end{align}
 Moreover, the entanglement entropy associated with the $AB$ subsystem is equivalently expressed as
\begin{align}
	S_{AB}\,\approx\, 2\log d - {1\over 2} 
\end{align}
Hence, the mutual information between $A$ and $B$ is given by
\begin{align}
	I(A:B)\approx\,  {1\over 2} +\cO (d^{-2})\,>\,0
\end{align}
Observe that the presence of non-zero mutual information indicates a lack of factorization for the state $\rho_{AB}$. This subsequently leads to the conclusion that the 4-party random state cannot be classified as a $2$-uniform state, since mutual information between any two parties would vanish in such a state. Furthermore, the L-entropy of the 4-party random state is below $2 \log [d]$.
\begin{align}\label{4ran}
	\ell_{AB}\,\approx\, 1.44 \log [d] \,<\, 2\log [d]\ .
\end{align}

Now, we evaluate the reflected entropy from the entanglement wedge cross section in the state holographically dual to the 4-boundary wormhole. When the entanglement wedge of $ABCD$ is in connected phase (similar to Fig.~\ref{mb_wh_4}), we may cut the 4-boundary wormhole along the inner horizon of length $L$ to obtain the 3-boundary wormhole for $ABE$ (similar to Fig.~\ref{mb_wh}). The asymptotic boundary $E$ may be considered as the purification of the reduced density matrix $\rho_{AB}$. In this three boundary wormhole, one may evaluate the entanglement wedge cross section $\mathcal{A}_{\gamma'}$ by 
\begin{align}
    \mathcal{A}_{\gamma^\prime}\,=\,2 \sinh ^{-1}\left[\textrm{csch}\left(\frac{\mathcal{A}_E}{2}\right) \sqrt{2 \cosh \left(\frac{\mathcal{A}_E}{2}\right) \cosh \left(\frac{\mathcal{A}_{A}}{2}\right) \cosh \left(\frac{\mathcal{A}_{B}}{2}\right)+\frac{\cosh \left(\mathcal{A}_{A}\right)}{2}+\frac{\cosh \left(\mathcal{A}_{B}\right)}{2}+1}\right]
\end{align}
Let us denote $\mathcal{A}_E=L$ and $\mathcal{A}_A=\mathcal{A}_B=\gamma$ as the length of the horizon in each asymptotic boundary. Using the estimation of $\gamma$ and $L$ for the 4-partite random state~\eqref{eq: gamma in 4 partite random state} and \eqref{eq: L in 4 partite random state}, we find 
\begin{align}\label{4bdywh}
    \mathcal{A}_{\gamma^\prime}  \,\approx\,& 2\sinh^{-1} \big[e^{{1\over 4}(2\gamma- L)}\big] 
\end{align}

On the other hand in the holographic dual of the four boundary wormhole if $S_{AB}$($=S_E$ in the purified three boundary wormhole) corresponds to the sum of the outermost horizon lengths $\gamma$ of  $A$ and $B$ (by ignoring $\cO(1)$ term),  then the entanglement wedge of $AB$ is disconnected and is simply given by the composition of the individual entanglement wedges of $A$ and $B$. This in turn implies that the reduced density matrix $\rho_{AB}$ is factorized into $\rho_A\otimes \rho_B\,\approx\, \mathbb{I}\otimes \mathbb{I}$. Note that this outcome contradicts the scenario involving a random state, where the L-entropy did not reach its peak. An alternative possibility for the RT surface associated with $S_{AB}$ is characterized by the inner horizon's length $L$. This length can be approximately expressed as
\begin{align}
	L\,\approx\,  2\log [d_{eff}] - {1\over 2}  \label{eq: L in 4 partite random state}
\end{align}
where we have denoted the large effective dimension $d_{eff}=e^\gamma$, and hence $L$ is almost identical to $2\gamma$. Consequently, it remains uncertain whether the entanglement wedge associated with the subsystem $AB$ in a state dual to a 4-boundary wormhole is in the connected or disconnected EW phase under the aforementioned conditions. Furthermore, utilizing the above expression for $L$ in \cref{4bdywh} we get
\begin{align}
   S_R(A:B)=2 \mathcal{A}_{\gamma^\prime}  \,\,\approx\, \cO(d_{eff}^0)
\end{align}
The corresponding L-entropy is given by
\begin{align}
   \ell_{AB}\approx2 \gamma=2\log[d_{eff}] 
\end{align}
which implies the L-entropy saturates to its maximal value like that of a 2-uniform state.
Hence, the above result is not consistent with the L-entropy of the 4-partite random state~\eqref{eq: l entropy 4 partite random state} calculated by the resolvent technique.
Therefore, we conclude that the 4-partite random state is not holographic dual to the 4-boundary wormhole. However, this does not mean that there is no state holographically dual to the 4-boundary wormhole. The existence of 4-party 2-uniform states has been established for various Hilbert space dimensions, although it remains unproven for specific cases like d=6 \cite{Pang:2019wod,goyeneche2014genuinely}. Since such states appear to be dual to 4-boundary wormholes based on our earlier arguments, it would be intriguing to identify a state that corresponds to the 4-boundary wormhole. We leave these fascinating questions for future exploration.

\section{ Multipartite entanglement at finite temperature}
\label{sec_temp_multi}

In this section, we introduce the concept of temperature into states exhibiting genuine multipartite entanglement. We begin with a brief review of the canonical thermal pure quantum state (TPQ) and its key property that leads to thermal-like behavior. Following this, we propose a multipartite extension of the TPQ state (MTPQ) and demonstrate that, for the two-party case, it reproduces the left-right correlations observed in the Thermo Field Dynamics/Double (TFD) state when state dependence is appropriately implemented. Subsequently, we apply our proposed MTPQ extension to the multi-copy Sachdev-Ye-Kitaev (SYK) model and show that the individual subsystems exhibit thermal behavior. This is evidenced by analyzing the entanglement entropy, relative entropy, and the eigenvalues of the single-party Hamiltonian. Furthermore, we generalize the notion of $k$-uniform states to finite temperatures and demonstrate that a 5-party state closely resembles a thermal 2-uniform state, as indicated by its L-entropy behavior.

\subsection{Multipartite Thermal Pure Quantum State (MTPQ)}

For a given Hamiltonian $H$, the canonical thermal pure quantum (TPQ) state~\cite{Sugiura:2013pla} is defined by
\begin{align}\label{CTPQ}
    |\Psi(\beta) \rangle \,\equiv \, e^{-{\beta\over 2}H}|\psi\rangle 
\end{align}
where $\beta$ is the inverse temperature, and $|\psi\rangle$ is a random state. The thermal expectation value of an operator $\mathcal{O}$ can be obtained as the random average of its expectation value with respect to the canonical TPQ state:
\begin{align}
    {\overline{ \langle \Psi_\beta | \cO |\Psi_\beta \rangle }\over \overline{ \langle \Psi_\beta |\Psi_\beta \rangle } }\,=\, {1\over Z(\beta)}\tr \big(\cO \; e^{-\beta H}\big)
\end{align}
Even without random averaging, the canonical TPQ state can approximate the thermal behavior of the system. In \cite{Sugiura:2013pla}, the random state $|\psi\rangle$ was chosen within the Hilbert space $\cH$ of the system. Consequently, the TPQ state in \cite{Sugiura:2013pla} does not serve as a purification of the thermal state. In this article, we relax this restriction, allowing the random state to exist outside the Hilbert space $\mathcal{H}$. For instance, a random state in the doubled Hilbert space $\mathcal{H} \otimes \mathcal{H}$ can be used to define a canonical extended TPQ state:
\begin{align}
    |\Psi(\beta)\rangle\,=\, e^{-{\beta\over 2} \big(H\otimes \mathbb{I}+\mathbb{I}\otimes H\big)}|\psi\rangle
\end{align}
where $|\psi\rangle \in \cH\otimes \cH$. For operators $\cO_L\equiv \cO \otimes \mathbb{I}$ or $\cO_R\equiv  \mathbb{I}\otimes \cO$, acting on one of the doubled Hilbert space, the random average of their expectation values with respect to the expended TPQ state also reproduce the thermal expectation value: 
\begin{align}
    {\overline{ \langle \Psi_{\beta} | \cO_R |\Psi_\beta \rangle }\over \overline{ \langle \Psi_\beta |\Psi_\beta \rangle } }\,=\, {1\over [Z(\beta)]^2}\tr \big(\cO_R \; e^{-\beta (H_L+H_R)}\big)\,=\, {1\over Z(\beta)}\tr \big(\cO \; e^{-\beta H}\big)\label{eq: average thermal}
\end{align}
Therefore, the expended TPQ state can be viewed as an approximate purification of thermal state. In holographic CFTs, one may conjecture that this extended TPQ state could correspond to a black hole microstate. However, the random average of the expectation value of $\cO_L\cO_R$ with respect to the expended TPQ state is factorized:
\begin{align}
    {\overline{ \langle \Psi(\beta) | \cO_L\cO_R |\Psi_\beta \rangle }\over \overline{ \langle \Psi_\beta |\Psi_\beta \rangle } }\,=\, {1\over [Z(\beta)]^2}\tr \big(\cO_L\cO_R \; e^{-\beta (H_L+H_R)}\big)\,=\, \bigg[{1\over Z(\beta)}\tr \big(\cO \; e^{-\beta H}\big)\bigg]^2\ ,\label{eq: average factorization}
\end{align}
This result is inconsistent with the non-factorized nature of the corresponding correlation function in the dual black hole. This discrepancy arises from the confusion between the state-dependence and the state independence of a mirror operator. For a given operator $\cO_R$, the construction of the corresponding mirror operator $\cO_L$ is state-dependent in the black hole~\cite{Papadodimas:2013wnh,Papadodimas:2013jku}. The factorization~in \cref{eq: average factorization} assumes a state-independent mirror operator $\cO_L$, leading to a factorized correlation function. However, when a state-dependent $\cO_L$ is properly taken into account, it becomes part of the random averaging process, ensuring that the random average of the correlation function $\langle \cO_L \cO_R\rangle$ is no longer factorized, thereby resolving the apparent contradiction.

We propose the construction of a multi-entangled state at finite temperature from a random state. Let us begin with a reference Hamiltonian $H$ for a single-party system, with energy eigenstates $|E_j\rangle $ corresponding to energy eigenvalue $E_j$ ($j=1,2,\cdots, d$):
\begin{align}
    H|E_j\rangle \,=\, E_j |E_j\rangle
\end{align}
For a $n$-party random state $|\psi\rangle$, re represent it as:
\begin{align}
    |\psi\rangle \,=\, \sum_{i_1,\cdots,i_n}c_{i_1i_2\cdots i_n} |i_1\rangle \otimes|i_2\rangle\otimes \cdots\otimes |i_n\rangle\ ,
\end{align}
Newt, we consider the Schmidt decomposition between $k$-th party and the rest:
\begin{align}
    |\psi\rangle \,=\, \sum_j c_j^{(k)} |\sigma^{(k)}_j \rangle_{\{k\}} \otimes |\omega^{(k)}_j\rangle_{\{1,2,\cdots, n\}/\{k\} }  \ ,\label{eq: schmidt decomposition of random state}
\end{align}
where $\{ |\sigma^{(k)}_j \rangle_{\{k\}}|j=1,\cdots, d\}$ and $\{|\omega^{(k)}_j\rangle_{\{1,\cdots, n\}/\{k\} } |j=1,\cdots, d\} $ are orthonormal sets. Here, although we position the $k$th party state $|\sigma^{(k)}_j \rangle_{\{k\}}$ on the leftmost side in the Schmidt decomposition for presentation purposes only, the original order remains unchanged in the actual calculation. 

Recall that the reduced density matrix of $k$-th party, derived from the pure random state $|\psi\rangle$, is approximately the identity matrix. Consequently, the coefficients $c_j^{(k)}$ are close to $d^{-{1\over 2}}$, where $d$ is the dimension of the $k$-th party. Thus, we approximate:
\begin{align}
    |\psi\rangle \,\approx\, {1\over \sqrt{d}} \sum_j |\sigma^{(k)}_j \rangle_{\{k\}} \otimes |\omega^{(k)}_j\rangle_{\{1,2,\cdots, n\}/\{k\} }  \ .
\end{align}
We now define a unitary operator $U^{(k)}$ which maps the basis $\{ |\sigma^{(k)}_j\rangle \}$ in the Schmidt decomposition to the energy eigen-basis $|E_j\rangle$ of the reference Hamiltonian:
\begin{align}
    U^{(k)}\,\equiv \,  \sum_j |E_j\rangle \langle \sigma^{(k)}_j|\ .
\end{align}
Using the reference Hamiltonian, we construct the Hamiltonian for the $k$-th party as:
\begin{align}
    H^{(k)}\,\equiv\, \underbrace{\mathbb{I}\otimes \cdots\otimes \mathbb{I}}_{k-1}\otimes \big(\underbrace{U^{(k) \dag }H U^{(k)}}_{k\text{\tiny th}}\big)\otimes \underbrace{\mathbb{I}\otimes \cdots \otimes \mathbb{I}}_{n-k}
\end{align}
By construction, the state $ |\sigma^{(k)}_j \rangle_{\{k\}}$ is the energy eigenstate of $H^{(k)}$ with the energy eigenvalue $E_j$:
\begin{align}
    H^{(k)} |\sigma^{(k)}_j\rangle \,=\, E_j |\sigma^{(k)}_j\rangle 
\end{align}
With the Hamiltonian 
$H^{(k)}$ ($k=1,\cdots, n$), we define a state $|\Psi_\alpha\rangle $ by
\begin{align}
    |\Psi_\alpha\rangle\,\equiv\,\prod_{i=1}^n e^{-{\alpha \over 2}H^{(k)}}  \;|\psi\rangle\label{def: psi alpha}
\end{align}
where $\alpha$ is a non-negative real parameter. We refer to this state as the multi-partite thermal pure quantum~(MTPQ) state to distinguish it from the canonical TPQ state~\cref{CTPQ} introduced in \cite{Sugiura:2013pla}, emphasizing the distinction that the MTPQ extends the Hilbert space beyond the original framework.

When selecting the basis $\{\sigma_j^{(k)}\}$ for the energy eigenstate of each party in the Schmidt decomposition, the ordering of the energy eigenstate is ambiguous. To address this, one can determine the ordering of $\{\sigma_j^{(k)}\}$ in the following procedure. Starting with the Schmidt decomposition of the 1st party, assign the energy eigenvalue $E_j$ to $|\sigma^{(1)}_j\rangle$. Next, for the Schmidt decomposition of the $k$-th party, consider the expectation value of the projection operator $|\sigma_i^{(1)}\rangle \langle \sigma_i^{(1)}| $for each $i$ with respect to $|\omega_j^{(k)}\rangle$~\eqref{eq: schmidt decomposition of random state}:
\begin{align}
    \cM_{ij}\,\equiv \, |\langle \sigma_i^{(1)} |\omega_j^{(k)}\rangle|^2\ . 
\end{align}
If the matrix $\cM$ has its maximum value in the $i$th row and $j$th column, assign the energy eigenvalue $E_i$ to $|\omega_j^{(k)}\rangle$. Subsequently, delete $i$-th row and $j$-th column of $\cM$, and repeat the allocation process by finding the maximum element in the reduced matrix. Using this iterative procedure, one can assign the energy eigenvalues $\{E_j\}$ to $\{|\omega_j^{(k)}\rangle\}$. By reordering $\{|\omega_j^{(k)}\rangle\}$, the energy eigenstate $|\omega_j^{(k)}\rangle\}$ can be obtained with the corresponding eigenvalue $E_j$ ($j=1,\cdots, d$).

Alternatively, without using Schmidt decomposition, the state $|\Psi_\alpha\rangle$ can be constructed directly by acting with the reference Hamiltonian:
\begin{align}
    H^{(k)}\,=\, \underbrace{\mathbb{I}\otimes \cdots\otimes \mathbb{I}}_{k-1}\otimes \underbrace{H}_{k\text{\tiny th}}\otimes \underbrace{\mathbb{I}\otimes \cdots \otimes \mathbb{I}}_{n-k}\hspace{5mm} (k=1,2,\cdots, n)
\end{align}
We find that the three methods--(i) constructing the Hamiltonian via Schmidt decomposition with reallocation of energy eigenvalues, (ii) constructing it via Schmidt decomposition without reallocation, and (iii) using the reference Hamiltonian--yield similar results (see Appendix~\ref{app: three methods}).

Unlike the random average calculation in Eq.~\eqref{eq: average thermal}, the parameter $\alpha$ is not generally identical to the inverse temperature. To determine the effective temperature for each party in the state \eqref{def: psi alpha}, we compare the reduced density matrix of each party with a thermal density matrix. Specifically, for the $k$-th party, the reduced density matrix is defined as:
\begin{align}
    \rho^{(k)}\,\equiv \tr_{\{1,2,\cdots, n\}/\{k\}} \rho
\end{align}
where the density matrix $\rho$ is obtained from the pure state $|\Psi_\alpha\rangle$.\:
\begin{align}
    \rho\,=\,|\Psi_\alpha\rangle \langle \Psi_\alpha|
\end{align}
The reduced density matrix $\rho^{(k)}$ can be diagonalized as:
\begin{align}
    \rho^{(k)}\,=\, \sum_j | \zeta_j^{(k)} \rangle \lambda^{(k)}_j \langle \zeta_j^{(k)} |
\end{align}
where  $\{|\zeta_j^{(k)}\rangle\}$ are the eigenstate and $\{\lambda^{(k)}_j \}$ are the eigenvalue. Using these eigenstate $|\zeta_j^{(k)}\rangle $, we define a thermal density matrix $\rho^{(k)}_{\text{\tiny thermal}}(\beta)$ where the energy eigenstate is $|\zeta_j^{(k)}\rangle $ and the corresponding energy eigenvalue is $E_j$: 
\begin{align}
    \rho^{(k)}_{\text{\tiny thermal}}(\beta) \,\equiv \, | \zeta_j^{(k)} \rangle {e^{-\beta E_j } \over Z(\beta)} \langle \zeta_j^{(k)} |
\end{align}
The effective temperature $\beta_k$ for $\rho^{(k)}$ is determined by minimizing the relative entropy between $\rho^{(k)}$ and $\rho^{(k)}_{\text{\tiny thermal}}(\beta)$:
\begin{align}
    S(\rho^{(k)}|| \rho^{(k)}_{\text{\tiny thermal}}(\beta))\,=\, \tr \big( \rho^{(k)}\log \rho^{(k)}\big) - \tr \big( \rho^{(k)}\log \rho^{(k)}_{\text{\tiny thermal}}(\beta)\big)
\end{align}
Since two density matrices are simultaneously diagonalized, the relative entropy can be simplified as:
\begin{align}
    S(\rho^{(k)}|| \rho^{(k)}_{\text{\tiny thermal}}(\beta))\,=\, \sum_j \lambda^{(k)}_j\log \lambda^{(k)}_j +\beta \sum_j \lambda_j^{(k)} E_j +\log Z(\beta)
\end{align}
After determining the effective temperature $\beta_k$,the reduced density matrix $\rho^{(k)}$ can be expressed as a thermal density matrix with effective temperature $\beta_k$:
\begin{align}
    \rho^{(k)}\,=\, \sum_j | \zeta_j^{(k)} \rangle {e^{-\beta_k \widetilde{E}_j } \over \widetilde{Z}(\beta_k) } \langle \zeta_j^{(k)} |
\end{align}
where $\{\widetilde{E}_j\}$ is the energy spectrum associated with $\rho^{(k)}$. Numerical calculations for concrete examples indicate that $\{\widetilde{E}_j\}$ closely approximates the reference spectrum  $\{E_j\}$. Thus, the state  $|\Psi_\alpha\rangle$ is a good approximation of a multipartite entangled state, where the reduced density matrix for each party is close to a thermal density matrix with an effective inverse temperature $\beta_k $ ($k=1,2,\cdots, n$).

To connect the expectation value of $|\Psi_\alpha \rangle $ with holographic result, we define the unitary operator $W^{(k)}$, which maps from the eigenstate $|\zeta_j^{(k)}\rangle$ to the reference energy eigenstate $E_j$, as follows:
\begin{align}
    W^{(k)}\,\equiv \, \sum_j | E_j \rangle  \langle \zeta_j^{(k)}|
\end{align}
Using an operator $\cO$ in the reference system, we construct the state-dependent operator $\cO^{(k)}$ by
\begin{align}
    \cO^{(k)}\,\equiv \,   W^{(k)\dag}\cO W^{(k)}
\end{align}
The expectation value of the state-dependent operator acing only on a single party reproduces its thermal expectation value:
\begin{align}
    \langle \Psi_\alpha |\cO^{(k)}|\Psi_\alpha \rangle \,=\, {1\over \widetilde{Z}(\beta_k)}\sum_{j} \langle E_j | \cO | E_j\rangle 
\end{align}
While the analytic result for the correlation function among different parties for the $n$-partite state $|\Psi_\alpha\rangle$ ($n>2$), it is intriguing open question to explore whether such correlation functions are consistent with holographic dual gravity calculations.

As a simple example, let us consider a bipartite state. From the Schmidt decomposition of the random state $|\psi\rangle$, we have
\begin{align}
    |\psi \rangle \,=\, \sum_j c_j | \sigma_j^{(1)}\rangle | \sigma_j^{(2)} \rangle \,\approx \, {1\over \sqrt{d}} \sum_j  | \sigma_j^{(1)}\rangle | \sigma_j^{(2)} \rangle \ .
\end{align}
The unitary operator $U^{(k)}$, defined by the reference energy eigenstate $|E_j\rangle$, is given by
\begin{align}
    U^{(k)}\,\equiv\, \sum_j |E_j \rangle \langle \sigma_j^{(k)} |    \ ,
\end{align}
Using $U^{(k)}$, we define the local Hamiltonian $H^{(k)}$ for the $k$-th party ($k=1,2$) as
\begin{align}
    H^{(k)}\,=\, U^{(k)\dag} H U^{(k)}
\end{align}
It follows that the bi-partite thermal state $|\Psi_\alpha\rangle$ corresponds to the thermofield dynamics~(TFD) state:
\begin{align}
    |\Psi_\alpha \rangle\,=\, e^{-{\alpha\over 2}\big(H^{(1)}+H^{(2)}\big)} |\psi\rangle \,\approx\,  {1\over \sqrt{d}} \sum_j  e^{-\alpha E_j} | \sigma_j^{(1)}\rangle | \sigma_j^{(2)} \rangle 
\end{align}
where the effective temperature $\beta$ is related to $\alpha$ by $\beta=2\alpha$. Thus, we can express
\begin{align}
    |\Psi_{\beta\over 2} \rangle \,\approx\, {\sqrt{Z(\beta)}\over \sqrt{d}} |TFD(\beta)\rangle
\end{align}
Since the reduced density matrix is already diagonalized in the Schmidt basis, the state-dependent operators can be constructed directly using $U^{(k)}$:
\begin{align}
    \cO^{L}\,=\,U^{(1)\dag} \cO U^{(1)} \\
    \cO^{R}\,=\,U^{(2)\dag} \cO U^{(2)} \\
\end{align}
The expectation values of $\mathcal{O}^L$  and  $\mathcal{O}^R$ reproduce their thermal values:
\begin{align}
    \langle \Psi_{\beta\over 2} |\cO^{R}   | \Psi_{\beta\over 2}\rangle\,=\, {1\over Z} \sum_{j} e^{-\beta E_j} \langle E_j | \cO |E_j \rangle 
\end{align}
Moreover,the correlation function of two operators acting on opposite parties also matches the result from the TFD state:
\begin{align}
    \langle \Psi_{\beta\over 2} |\cO^{L}_{1} \cO^{R}_2  | \Psi_{\beta\over 2}\rangle\,=\, {1\over Z} \sum_{j,k} e^{-{\beta\over 2}(E_j+E_k) }\langle E_j| \cO_1 |E_k\rangle  \langle E_j| \cO_2 |E_k\rangle  
\end{align}
For the thermal $k$-partite entangled state ($k>2$), we will present the numerical results in the next section, using the SYK model as an example.

\subsection{Finite temperature multipartite entanglement in the SYK model}
\label{sec: multiboundary wormhole and syk model}
In this section, we investigate the multipartite L-entropy and the thermal $k$-uniformity characteristic of a multi-copy SYK model, employing a methodology predicated on a multipartite variant of the TPQ state (MTPQ), as defined in \cref{def: psi alpha}. Subsequently, we calculate the L-entropy corresponding to the MTPQ state within the context of many-copy SYK models, specifically for 3-party, 4-party, and 5-party configurations. We illustrate that, although the behavior of entanglement entropy and relative entropy for 3 and 4-party scenarios indicates that each party approximates a thermal state, the L-entropy does not follow the expectations associated with a thermal $k\geq 2$ uniform state. Conversely, our numerical results clearly demonstrate that the behavior L-entropy for the 5-party case very closely resembles that of a thermal 2-uniform state.

\subsubsection{ MTPQ State in SYK and thermality}

\begin{figure}[H]
	\centering
	\begin{subfigure}[h]{0.3\textwidth}
		\centering
		\includegraphics[width=55mm]{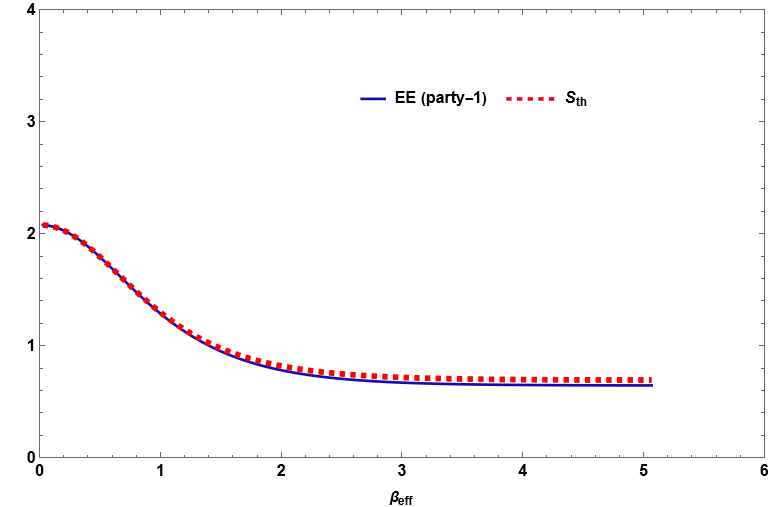}
		\caption{Party-1}
		\label{EE_5_1}
	\end{subfigure}
	\begin{subfigure}[h]{0.3\textwidth}
		\centering
		\includegraphics[width=55mm]{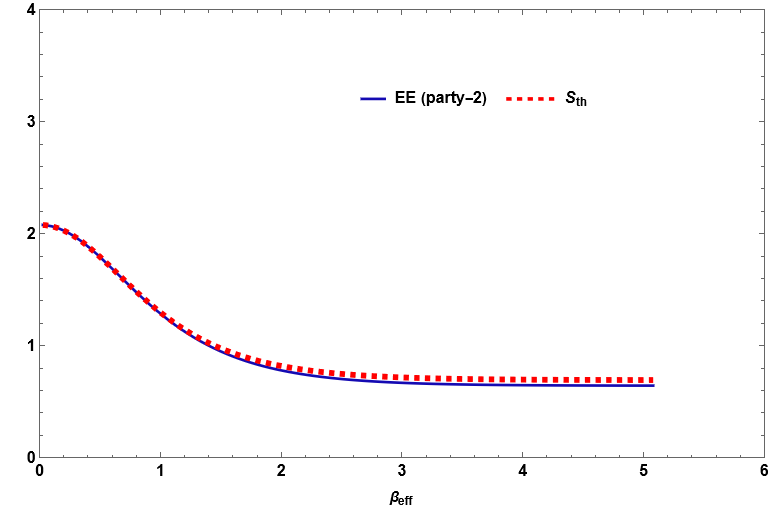}
		\caption{Party-2}
		\label{EE_5_2}
	\end{subfigure}
	\begin{subfigure}[h]{0.3\textwidth}
		\centering
		\includegraphics[width=55mm]{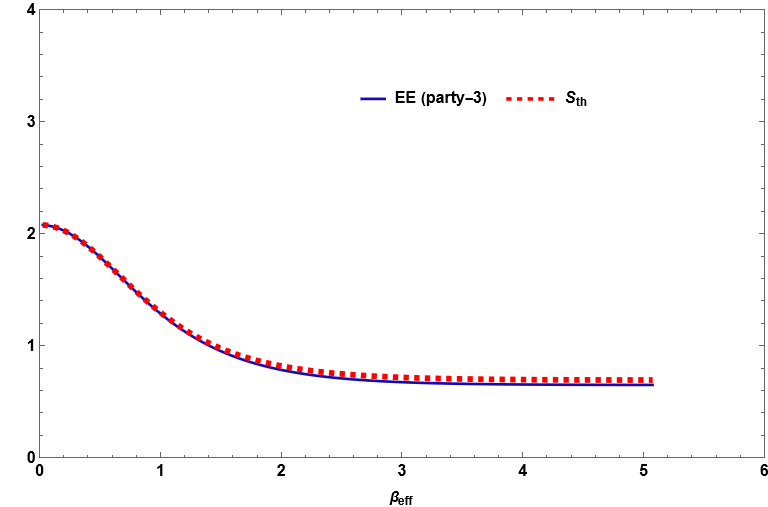}
		\caption{Party-3}
		\label{EE_5_3}
  	\end{subfigure}
	\begin{subfigure}[h]{0.3\textwidth}
		\centering
		\includegraphics[width=55mm]{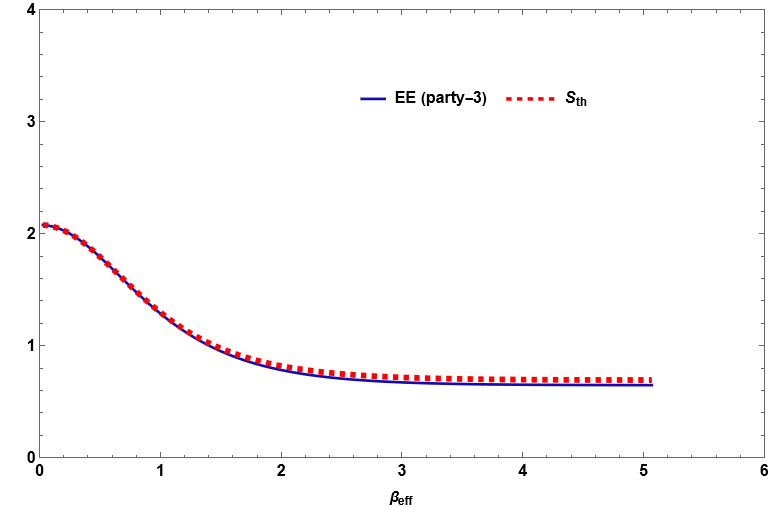}
		\caption{Party-4}
		\label{EE_5_4}
        \end{subfigure}
	\begin{subfigure}[h]{0.3\textwidth}
		\centering
		\includegraphics[width=55mm]{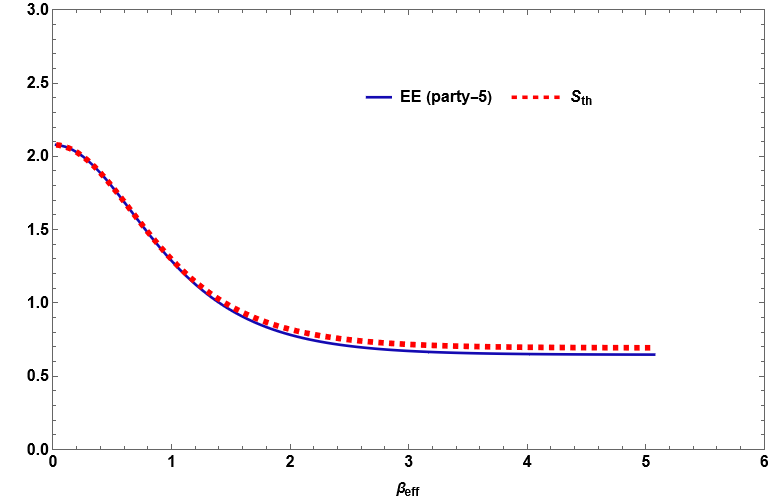}
		\caption{Party-5}
		\label{EE_5_5}
	\end{subfigure}
	\caption{ 5 parties of 3-qubits from SYK. Comparision between single party entanglement entropy with the corresponding thermal entropy for a single copy SYK}
	\label{EEthplots5}
\end{figure}

\begin{figure}[H]
	\centering
	\begin{subfigure}[h]{0.42\textwidth}
		\centering
		\includegraphics[width=0.9\textwidth]{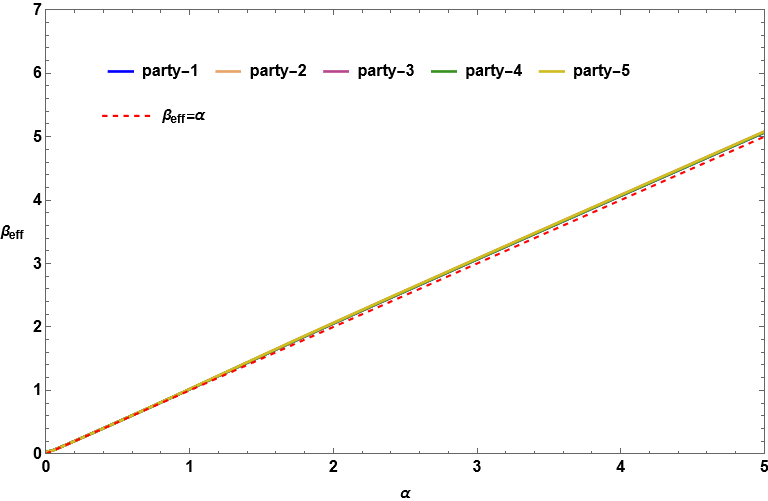}
		\caption{Average Effective Temperature vs $\alpha$ parameter}
		\label{Effective_Temp_5}
	\end{subfigure}
    \begin{subfigure}[h]{0.42\textwidth}
		\centering
		\includegraphics[width=1\textwidth]{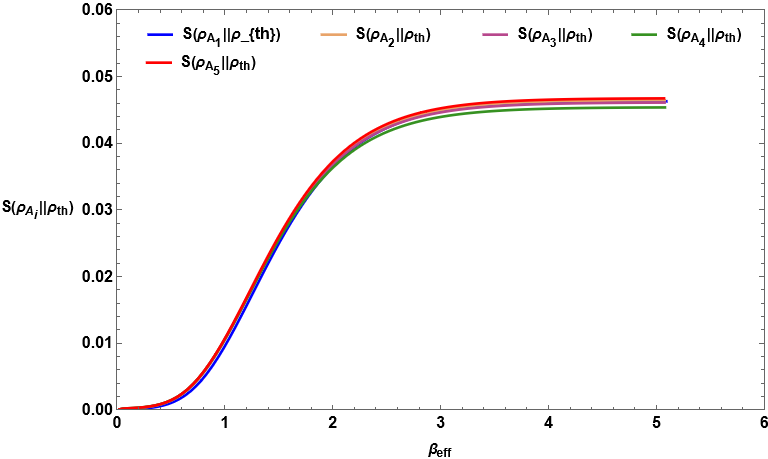}
		\caption{Relative entropy}
		\label{Relative_entropy_5}
	\end{subfigure}
    \caption{}
	\end{figure}

We have provided an overview of the standard notation alongside the qubit realization via the Fock space representation of a single-copy SYK model in \cref{app:SYK}. Initially, we substantiate the assertion that in the multipartite formulation of the TPQ state, each component exhibits a quasi-thermal profile with distinct effective temperatures. As illustrated in Fig.~\ref{EEthplots5}, the entanglement entropies of individual parties within a 5-party quantum configuration approximate their thermal entropy levels corresponding to specific effective temperatures. The dependence of each party's effective temperature on the $\alpha$ parameter of the TPQ state is analyzed in Fig.~\ref{Effective_Temp_5}. While the entanglement entropy and effective temperature display expected behaviors analogous to a thermal state, for additional corroboration that each separate party is thermal, we evaluate the relative entropy between each party and its thermal counterpart with the appropriate effective temperatures, as shown in Fig.~\ref{Relative_entropy_5}. The characteristics of the graphs are analogous in cases involving three and four parties, thus they are compiled in \cref{sec:MTPQ34}.



\begin{figure}[H]
    \centering
    \begin{subfigure}[h]{0.3\textwidth}
        \centering
        \includegraphics[width=52mm]{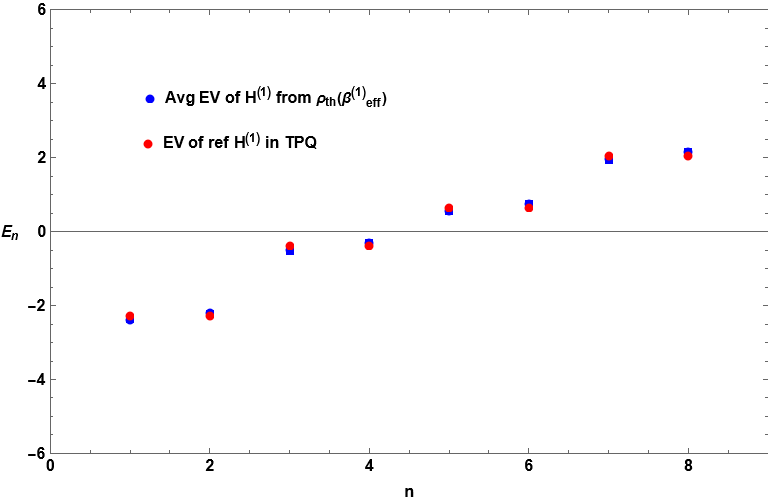}
        \caption{Party-1}
        \label{En_5_1}
    \end{subfigure}
    \begin{subfigure}[h]{0.3\textwidth}
        \centering
        \includegraphics[width=52mm]{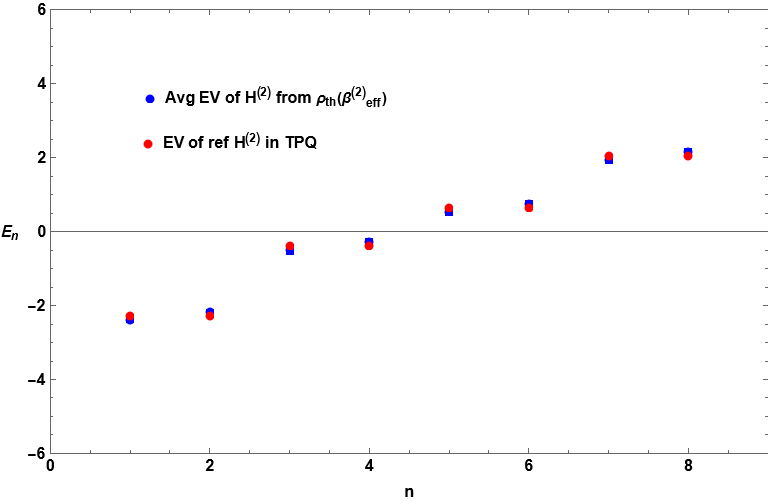}
        \caption{Party-2}
        \label{En_5_2}
    \end{subfigure}
    \begin{subfigure}[h]{0.3\textwidth}
        \centering
        \includegraphics[width=52mm]{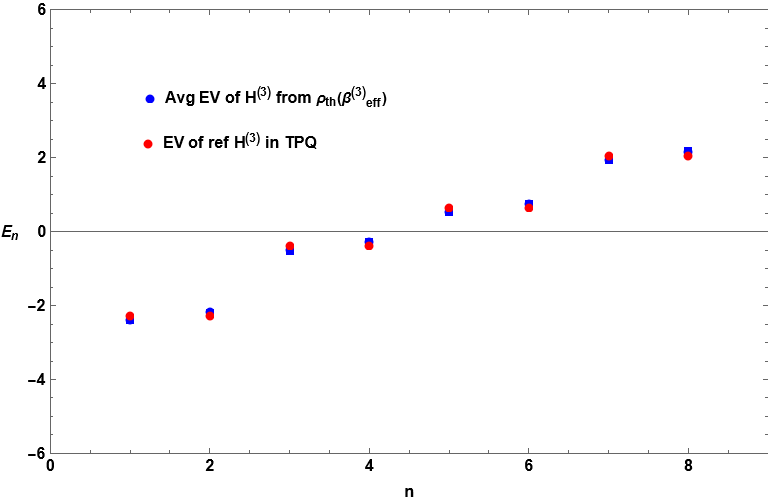}
        \caption{Party-3}
        \label{En_5_3}
    \end{subfigure}
    \begin{subfigure}[h]{0.3\textwidth}
        \centering
        \includegraphics[width=52mm]{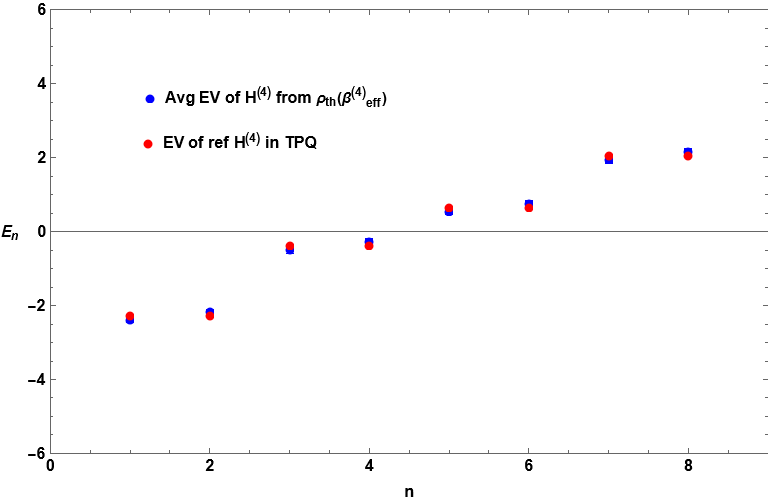}
        \caption{Party-4}
        \label{En_5_4}
    \end{subfigure}
    \begin{subfigure}[h]{0.3\textwidth}
        \centering
        \includegraphics[width=52mm]{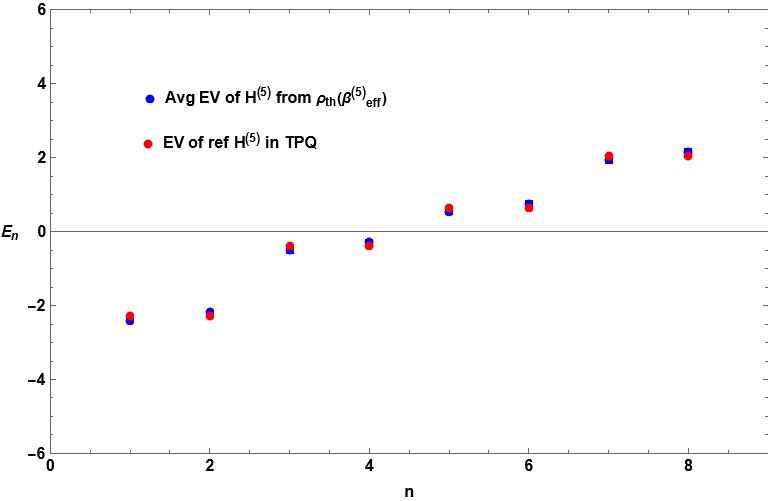}
        \caption{Party-5}
        \label{En_5_5}
    \end{subfigure}
    \caption{5 parties of 3-qubits from SYK. Comparison between eigenvalues of the reference Hamiltonian-$H^{(i)}$ in the TPQ state with the corresponding thermal Hamiltonian $H^{(i)}$ with corresponding effective temperature.}
    \label{Evplots5pT}
\end{figure}

\subsubsection{Multipartite L-entropy in MTPQ state}
\label{sec:5party}
\begin{figure}[H]
	\centering
	\begin{subfigure}[h]{0.42\textwidth}
		\centering
		\includegraphics[width=1\textwidth]{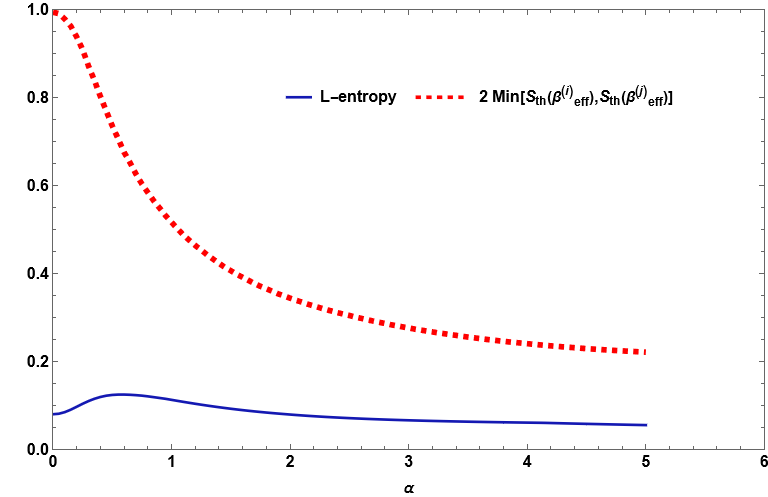}
		\caption{3 Party}
		\label{L_3}
	\end{subfigure}
  \begin{subfigure}[h]{0.42\textwidth}
		\centering
		\includegraphics[width=1\textwidth]{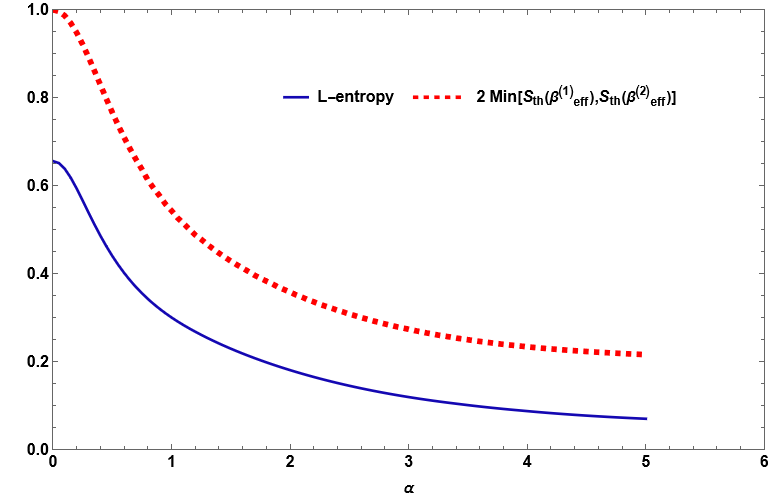}
		\caption{4 party}
		\label{L_4}
	\end{subfigure}
     \begin{subfigure}[h]{0.47\textwidth}
		\centering
		\includegraphics[width=1\textwidth]{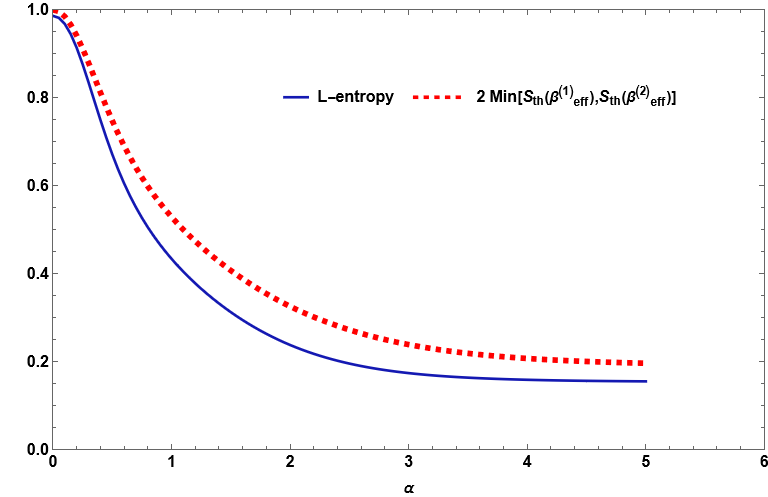}
		\caption{5 party}
		\label{L_5}
	\end{subfigure}
    \caption{A comparative analysis of L-entropy in 3, 4, and 5-party MTPQ states versus a thermal 2-uniform state is conducted within the framework of a multi-copy SYK model.}
	\end{figure}

In this subsection we analyze the behaviour of the L-entropy in the MTPQ state within the multi-copy SYK model as a function of the $\alpha$ parameter in three-, four-, and five-party quantum systemss. Notably, in the scenario involving three parties, the L-entropy is significantly less than that of the thermal 2-uniform state, evident in Fig.~\ref{L_3}. In contrast, for the four-party MTPQ state, the L-entropy  closer to the thermal 2-uniform state's value compared to the three-party configuration, although a considerable discrepancy persists, as illustrated in Fig.~\ref{L_4}. Meanwhile, in the five-party case, the L-entropy aligns very closely with the thermal 2-uniform state's behavior, as depicted in Fig.~\ref{L_5}.

\subsubsection{Phase transitions in MTPQ state}
\label{sec: phase transition in multi-entangled state}
\begin{figure}[H]
	\centering
	\begin{subfigure}[h]{0.3\textwidth}
		\centering
		\includegraphics[width=53mm]{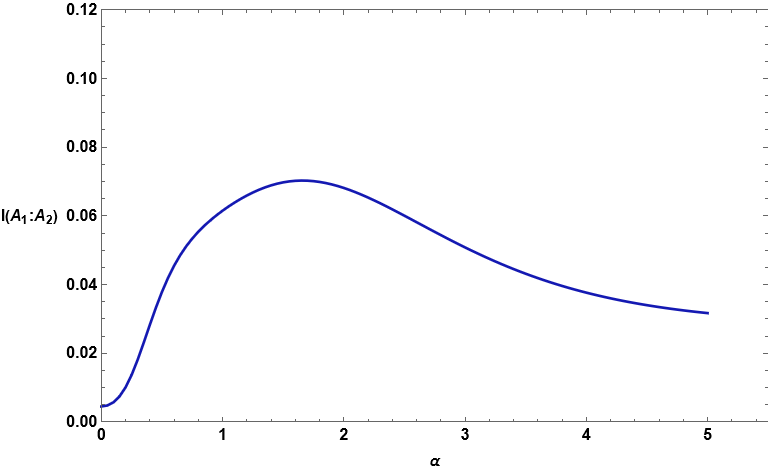}
		\caption{Mutual Information}
		\label{MI}
	\end{subfigure}
	\hspace{0.05cm}
	\begin{subfigure}[h]{0.3\textwidth}
		\centering
		\includegraphics[width=53mm]{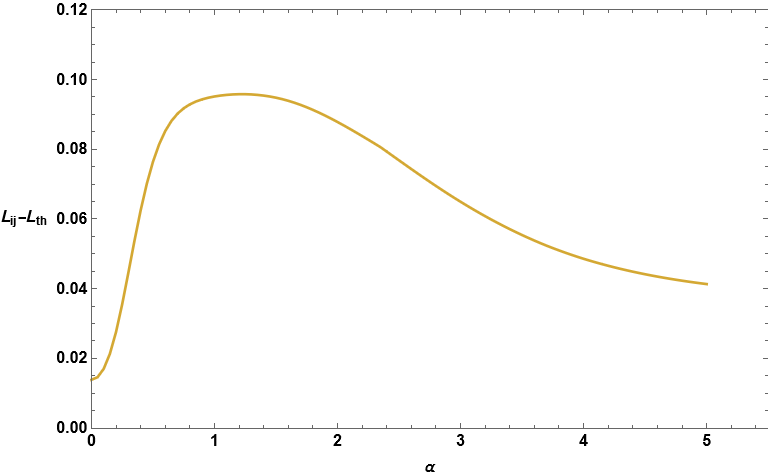}
		\caption{L-entropy difference}
		\label{Ldiff}
	\end{subfigure}
 \hspace{0.05cm}
	\begin{subfigure}[h]{0.3\textwidth}
		\centering
		\includegraphics[width=53mm]{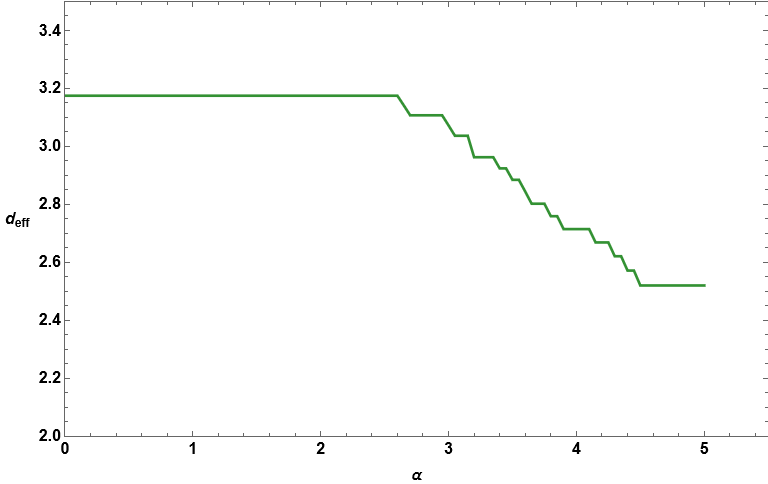}
		\caption{Effective dimension}
		\label{dimeff}
  	\end{subfigure}
    \caption{ Figures depicting the behaviour of mutual information, the difference between L-entropy in the MTPQ state and the thermal 2-uniform state and the effective dimension of the reduced state of each party as a function of the $\alpha$ parameter.}
\end{figure}
We have seen that to a good approximation the random state can be considered as a $n$-partite $2$-uniform state $(n\geqq 5)$. The reduced density matrix of any two parties in the $n$-partite random state can be approximately factorized into the direct product of the reduced density matrices of the individual parties. This behavior is consistent with the mutual information between any two parties in the $n$-partite random state, which is of order $\cO(e^{-(n-4)S})$. Similarly, for the MTPQ state $|\Psi_\alpha\rangle$ with an infinitesimal $\alpha$, the reduced density matrix of any two parties remains approximately factorized into the product of thermal density matrices of the individual subsystems.

As the $\alpha$ in the MTPQ grows, there is a corresponding rise in the  mutual information between any two parties in the MTPQ state (See Fig.~\ref{MI}). This implies that the reduced density matrix of the two parties fails to factorize. Furthermore, the difference between the L-entropy in the MTPQ state and the thermal 2-uniform state continues to widen as $\alpha$ grows (See Fig.~\ref{Ldiff}). Consequently, the MTPQ state undergoes a phase transition from the thermal $2$-uniform phase to the thermal $1$-uniform phase after a critical value of the parameter $\alpha$ is reached. 

The MTPQ state is constructed by applying the Boltzmann factor $e^{-\frac{1}{2}\alpha H}$ on each party of a random state. From the above discussion it also clear that MTPQ state has the property that the reduced density matrix of each single party is approximately thermal. Notice that the  holographic state describing the asymptotic boundaries of a multi-boundary wormhole also share this property, where each single-party density matrix is thermal. Assuming that the holographic state dual to the multi-boundary wormhole can be described by such an MTPQ state, we observe the following: as the parameter $\alpha$ increases, the horizon lengths in the wormhole decrease. For example, the Boltzmann factors in the MTPQ state reduce the lengths  $\gamma_1$ and $\gamma_2$, corresponding to the entropy of parties $A_1$ and $A_2$, respectively, as well as the length $\gamma_{12}$ of the internal horizon, corresponding to the entropy of the subsystem  $A_1A_2$. The phase transition from the thermal 2-uniform phase to the thermal 1-uniform phase implies that the inner horizon length $\gamma_{12}$ shrinks faster than the outermost horizon lengths $\gamma_1$ and $\gamma_2$ with increasing $\alpha$.

As $\alpha$ continues to increase beyond a certain threshold, a sufficiently large $\alpha$ causes a reduction in mutual information (see Fig.~\ref{MI}). Furthermore, the difference between the L-entropies corresponding to the MTPQ state and the thermal 2-uniform state reduces as well (see Fig.~\ref{Ldiff}). However, this reduction does not lead to another phase transition back to the thermal 2-uniform state. This is because, for sufficiently large  $\alpha$, the Boltzmann factors significantly suppress the reduced density matrix, effectively lowering its rank. This reduction in the effective dimensionality of the reduced density matrix is the primary cause of the decreased mutual information and reduced L-entropy difference. This behavior can also be observed through the diminished rank of the reduced density matrix for a single party (see Fig.~\ref{dimeff}).

The behavior of the MTPQ state $|\Psi_\alpha\rangle$ with respect to changes in the $\alpha$ parameter crucially depends on our definition of the MTPQ state. Alternative constructions of the MTPQ state from the random state could exist, where the internal horizon shrinks more slowly than the outermost horizon, thereby preventing the occurrence of a phase transition. We leave further exploration of this interesting issue for future research investigations.

\subsubsection{Thermal k-uniform State}
\label{sec: thermal k uniform state}

Similar to the TFD state, which can be obtained from the maximally entangled state by introducing temperature, we generalize the $k$-uniform state to define the thermal $k$-uniform state. For given inverse temperatures $\beta_j$ ($j=1,2,\cdots, k$) corresponding to each party, the thermal $k$-uniform state $| \Psi\big(\{\beta_k\}\big) \rangle$ is defined as a pure state where the reduced density matrix of any $k$ parties is factorized into the thermal density matrix of the individual parties:
\begin{align}
    \rho_{A_{j_1}\cdots A_{j_k}}\,=\, \rho_{\text{\tiny th}}(\beta_{j_1})\otimes\cdots \otimes \rho_{ \text{\tiny th}}(\beta_{j_k})\ .
\end{align}
Furthermore, tracing out this reduced density matrix $\rho_{A_{j_1}\cdots A_{j_k}}$ results in a reduced density matrix that remains factorized into the thermal density matrices of the remaining subsystems. Therefore, the thermal $k$-uniform state $(k\geq 2)$ is also thermal 2-uniform state. Since the reduced density matrix of any two parties in the thermal $k$-uniform state is factorized, the L-entropy of $k$-uniform state is given by
\begin{align}
    \ell_{A_iA_j}\,=\, 2\min\big(S_{\text{\tiny th}}(\beta_i),S_{\text{\tiny th}}(\beta_j)\big)\ .
\end{align}
where $S_{\text{\tiny th}}(\beta)$ denotes the thermal entropy. 

In the context of the multi-boundary wormhole, which is holographically dual to the thermal $k$-uniform state, the factorization of the reduced density matrix implies that the entanglement wedge of the  $k$ boundaries is the union of the disconnected entanglement wedges of the individual boundaries.

For local operators $\cO_i$ and $\cO_j$ which belongs to a party $A_i$ and $A_j$, respectively, the two point function with respect to the thermal $k$-uniform state is factorized.
\begin{align}
    \langle \Psi(\{\beta_k\}) | \cO_i \cO_j |  \Psi(\{\beta_k\})\rangle\,=\, \Tr\big( \rho_{A_i A_j} \cO_i \cO_j\big)\,=\,\tr\big(\rho_{A_i, \text{\tiny th}}(\beta_i)\cO_i\big)\tr\big(\rho_{A_j, \text{\tiny th}}(\beta_j)\cO_j\big)
\end{align}
On the other hand, the two-point function of boundary operators can be holographically obtained through the geodesic distance between the two operators:
\begin{align}
    \langle \cO_i (x_1) \cO_j(x_2) \rangle \sim \exp \big[-\Delta\, L(x_1;x_2)\big]
\end{align}
However, this result seemingly appears inconsistent with the geometry of the multi-boundary wormhole, as a geodesic could exist between two operators even if the entanglement wedge is disconnected. To address this, we propose that the two-point function in such cases should be holographically evaluated by extremizing the length of paths supported by the entanglement wedge of the two parties $A_i\cup A_j$.
\begin{align}
    \langle \cO_i (x_1) \cO_j(x_2) \rangle\sim \underset{\gamma}{\text{Ext}} \big\{  e^{-\Delta L[\gamma]} \big|\; \mbox{$\gamma$  is supported by the EW of $A_i$ and $A_j$} \;\big\}
\end{align}
According to this prescription, when the entanglement wedge of two party $A_1$ and $A_2$ is disconnected, no geodesic connects the two operators, ensuring consistency with the factorization of the correlation function in the thermal $k$-uniform state.

\section{Summary and discussion}
\label{Summary and Discussion}

In this work, we introduce the Latent Entropy (L-entropy) as a novel measure of genuine multipartite entanglement upto five-party pure states, based on the upper bound of reflected entropy. First, we focus on tripartite states and demonstrate that the tripartite L-entropy  is local unitary invariant function which vanishes for separable and biseparable states, while attaining its maximum value for the GHZ state.  Furthermore, it ranks GHZ states higher than W states in the 3-party scenario. These properties establish L-entropy as a valid measure of genuine multipartite entanglement. Building on its expected behavior in tripartite states, we extend L-entropy to four- and higher-party systems, defining a multipartite generalization. Using this construction, we explore the characterization of the Hilbert space for four-party states, which includes nine distinct classes. In this context, we compare multipartite L-entropy with two other measures: the tripartite mutual information and the Markov gap. While the tripartite mutual information and the Markov gap suggest zero multipartite entanglement for certain representative states,  L-entropy consistently identifies multipartite entanglement across all classes. Notably, the cluster state is shown to possess the highest multipartite entanglement, a result aligning with expectations from quantum information theory. Finally, we attempted to establish LOCC monotonicity of the proposed measure. This analysis reveals a technical obstruction arising from the nonlinear dependence of the canonical purification on the reduced density matrix, specifically through the matrix square-root operation. Consequently, an analytic proof of LOCC monotonicity appears nontrivial within our present framework. We therefore complement our analytical considerations with numerical tests. These simulations show that while violations can occur for three-qubit states, no counterexamples were found for four- or five-qubit states within the sampled state space.

We then utilize the multipartite L-entropy to examine the behavior of multipartite entanglement in various spin chain models. For the Ising model with nearest-neighbor interactions, we observe that both the multipartite L-entropy and the Markov gap exhibit oscillatory behavior. Interestingly, the time evolution of tripartite entanglement heavily depends on the type of interactions and the initial states. Similar dynamics are observed for spin chains governed by nearest-neighbor random Hamiltonians, with $k$-party L-entropy showing consistent characteristics across different $k$. Next, we consider the SYK model, where the results are particularly intriguing. The tripartite L-entropy initially grows and saturates at late times, though it does not reach the maximum tripartite value. However, for large $n$, the saturation value approaches the maximum. For higher-party entanglement, the saturation value of $n$-party L-entropy converges more rapidly to its maximum with increasing $n$. Subsequently, we analyze multipartite L-entropy for Haar random states, revealing distinctive behavior in the large Hilbert space dimension limit. The tripartite L-entropy attains a leading constant value after a large-$d$ expansion, whereas the 4-party L-entropy approaches approximately $1.44\log[d]$, below the maximum value of $2\log[d]$. In contrast, for 5 parties, the multipartite L-entropy reaches the maximum value of $2\log[d]$ in the large-$d$ limit.

Next, we explore the holographic scenario for the multipartite L-entropy. The bipartite L-entropy is defined as the difference between the minimum entanglement entropy of the individual subsystems and the reflected entropy of the bipartite system. A multipartite L-entropy is then constructed using the bipartite L-entropies of all possible bipartitions. In holography, the entanglement entropy and reflected entropy correspond to the areas of the Ryu-Takayanagi surface and the entanglement wedge cross-section, respectively. Using these dualities, we define the holographic bipartite L-entropy and, consequently, the multipartite L-entropy. However, the existence of a single bulk quantity dual to the multipartite L-entropy remains an open question. 

In this article, we adopt the above construction and investigate multiboundary wormholes with three and four boundaries. In the three-boundary wormhole scenario, we consider one boundary as a black hole and the other two as radiation regions. We compute the tripartite L-entropy for a black hole evaporation process in this framework and propose a Page curve for our measure. Initially, the tripartite L-entropy is zero until the sum of the entanglement entropies of the radiation regions exceeds that of the black hole, corresponding to the Page time for entanglement entropy. After the Page time, the black hole's interior information becomes accessible to the radiation regions through the formation of islands. Consequently, the tripartite L-entropy increases, signifying the emergence of tripartite entanglement in the system. At time $t=t_{max}$, when all boundaries are of equal size, the tripartite L-entropy reaches its maximum, equal to the entanglement entropy of any individual subsystem. This indicates that all available degrees of freedom contribute fully to tripartite entanglement. Beyond $t_{max}$, the tripartite L-entropy decreases and eventually becomes zero when the black hole evaporates completely, rendering the system effectively bipartite. In the four-boundary wormhole scenario, we analyze the 4-party L-entropy across different parameter regimes. Interestingly, the multipartite L-entropy attains its maximum value when all boundaries are small and equal in size. Notably, the tripartite mutual information also indicates maximal 4-party entanglement in this parameter regime, providing strong consistency for the validity of our measure.

Finally, we define the multipartite thermal pure quantum (MTPQ) state as a multipartite extension of the thermal pure quantum state and examine the dynamics of the L-entropy. We demonstrate that the MTPQ state can be constructed using three methods: (i) a specific Schmidt decomposition of the k-th party and the rest, with or without redistributing the Hamiltonian’s energy eigenvalues, or (ii) using a reference Hamiltonian without Schmidt decomposition. Remarkably, all three approaches yield the same MTPQ state. We apply this construction to the multi-copy SYK model and compute the entanglement entropy, relative entropy, and L-entropy for 3-party, 4-party, and 5-party configurations. These measures exhibit two distinct behaviors: while the entanglement entropy and relative entropy suggest that all configurations approach thermal states, the L-entropy indicates the MTPQ state is not a thermal $2$-uniform states for the 3- and 4-party cases. However, for the 5-party configuration, the L-entropy suggests a very close resemblance of MTPQ state to that  of a  thermal 2-uniform state. Furthermore, we propose that MTPQ states can be interpreted as holographic states dual to multiboundary wormholes. In this framework, the phase transition of the MTPQ states from thermal 1-uniform states to thermal 2-uniform states can be described by the relative growth of the inner and the outer horizons of a pair of boundaries in the multiboundary wormhole model. Finally, we propose the construction of thermal $k$-uniform states and posit that the multiple-point function of the boundary operators can only be nonzero when the wedge of the corresponding boundaries is connected.

In this article, we explored various intriguing properties of multipartite entanglement using the L-entropy. These findings can serve as a guiding framework for investigating a wide range of physical phenomena. A natural extension of this work could involve constructing a similar quantity to the L-entropy based on the multipartite reflected entropy and its bounds, to determine whether a new genuine multipartite entanglement measure can be defined. Additionally, the L-entropy could be computed for various CFTs to analyze the structure of multipartite entanglement and the potential existence of universal properties. Here,  we demonstrated that the bipartite L-entropy achieves its maximum value for a 2-uniform state. However, a bottom-up approach could be taken to define new quantities that achieve maximum values for higher uniform states, potentially uncovering generalized characteristics of such measures. In this direction, a novel construction of MTPQ states could be developed, allowing for the study of the growth dynamics of internal and external horizons in multiboundary wormhole models.
Moreover, the current formulation of L-entropy possesses a holographic dual represented by a combination of the areas of two distinct surfaces. Inspired by the developments of the holographic Markov gap and recent advances in \cite{Mori:2024gwe}, it would be compelling to identify a single bulk quantity as the holographic dual of the L-entropy. Furthermore, extensive studies of L-entropy in multiboundary wormholes across various parameter regimes could be conducted, with comparisons to recent developments in the field \cite{Balasubramanian:2024ysu,Ju:2024hba}.

\section*{Acknowledgement}

JKB and VM would like to thank Chris Akers and Takato Mori for their valuable discussions and remarks at the Quantum Extreme Universe: Matter, Information, and Gravity conference hosted by the Okinawa Institute of Science and Technology. The authors would like to thank  Hyukjoon Kwon, Eunok Bae and  Minjin Choi for their interesting feed back and comments on our work during Quantum Threads conference held at Kyung Hee university. The authors would like to thank Abhijit Gadde for constructive comments and important feedback.

The research work of JKB is supported by the grant 110-2636-M-110-008 by the National Science and Technology Council (NSTC) of Taiwan. JKB was supported by he Brain Pool program funded by the Ministry of Science and ICT through the National Research Foundation of Korea (RS-2024-00445164) V.M. and J.Y. was supported by the National Research Foundation of Korea (NRF) grant funded by the Korean government (MSIT) (No.\ 2022R1A2C1003182) and by the Brain Pool program funded by the Ministry of Science and ICT through the National Research Foundation of Korea (RS-2023-00261799). V.M. and J.Y. are supported by an appointment to the JRG Program at the APCTP through the Science and Technology Promotion Fund and Lottery Fund of the Korean Government. This is also supported by the Korean Local Governments - Gyeongsangbuk-do Province and Pohang City.

\appendix

\section{Unitary evolution from Bell to GHZ }
\label{app:bisep_to_ghz}

\begin{figure}[H]
	\centering
	\includegraphics[width=.45\linewidth]{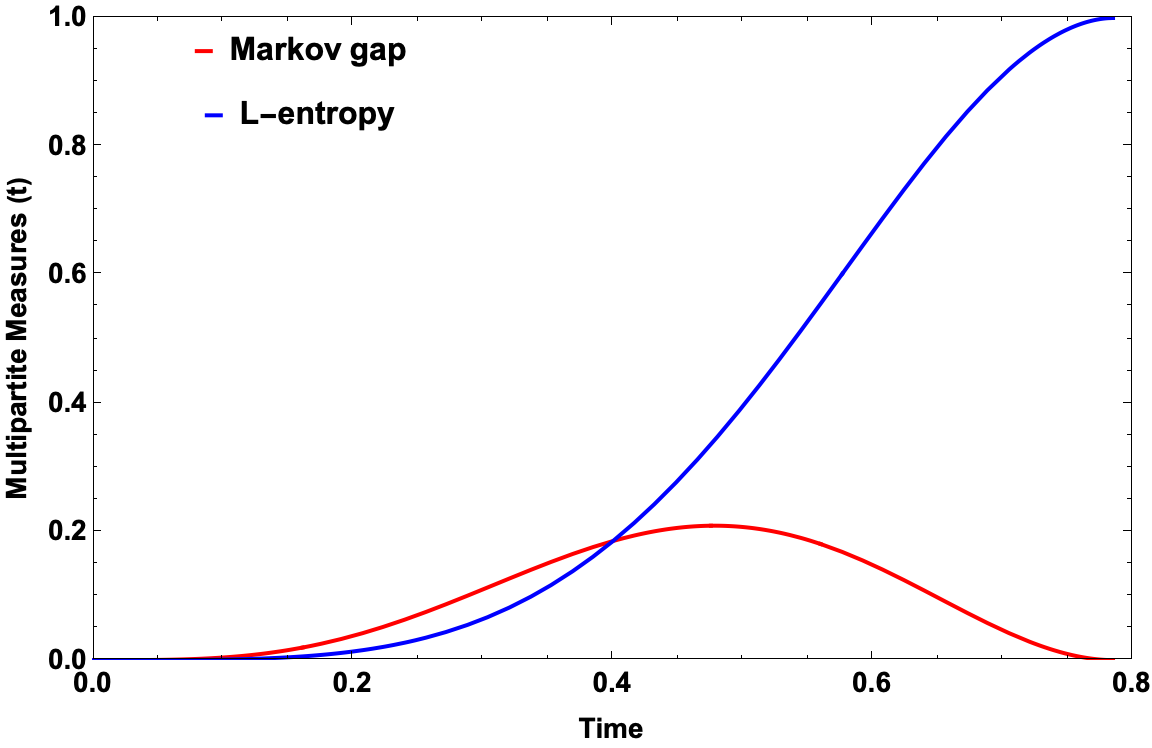}
	\caption{This figure depicts the evolution of L-entropy ($\ell_{ABC}$) in blue and the Markov gap ($h_{ABC}$) in red during the unitary evolution from a Bell state to a GHZ state.}  
	\label{BellToGHZ}
\end{figure}

In this appendix, we examine the behavior of the L-entropy in a simple three-qubit system, where a Bell state evolves into a GHZ state. Specifically, we consider a system of three qubits initially in a biseparable state, with two of the qubits maximally entangled while remaining in a product state with the third qubit. Our goal is to construct a unitary operator that evolves the system from the Bell state to the GHZ state.  It is well-known that the application of a CNOT gate directly transforms a Bell state into a GHZ state\footnote{This occurs because the CNOT gate is a two-qubit gate that flips the target qubit if and only if the control qubit is $\ket{1}$ and leaves it unchanged if the control qubit is $\ket{0}$.}. Here, we consider a generalization of this unitary operator, allowing for a continuous transformation from the Bell state to the GHZ state. As discussed in \cite{Susskind:2014yaa}, this type of operation is particularly intriguing because it serves as a toy model for measuring the second qubit through its interaction with the third qubit, which acts as an environment.

\begin{align}
	\ket{\psi(\theta)}=U(\theta)\ket{\psi(0)}
\end{align}
In the above equation $\ket{\psi(0)}$ is a biseparable state
\begin{align}\label{bisepu}
	\ket{\psi(0)}=\frac{1}{\sqrt{2}}(\ket{00}+\ket{11})\otimes \ket{0}
\end{align}
The unitary operator $U(\theta)$ involve successive operation of three unitary operators as expressed bellow
\begin{align}
	U(\theta)=U_x(\theta)U_H(t)U_y(\theta)
\end{align}
where $U_x$ and $U_y$ can be thought of as rotation matrices in the block sphere of first two Qubits
\begin{align}
	U_x&=I^A \otimes I^B \otimes R_x^C(\theta) \, \, \quad R_x(\theta)=\left(
	\begin{array}{cc}
		\cos \left(\frac{\theta }{2}\right) & -i \sin \left(\frac{\theta }{2}\right) \\
		-i \sin \left(\frac{\theta }{2}\right) & \cos \left(\frac{\theta }{2}\right) \\
	\end{array}
	\right)\\
	U_y&=I^A \otimes I^B \otimes R^C_y(\theta)  \, \,\quad R_y(\theta)=\left(
	\begin{array}{cc}
		\cos \left(\frac{\theta }{2}\right) & -\sin \left(\frac{\theta }{2}\right) \\
		\sin \left(\frac{\theta }{2}\right) & \cos \left(\frac{\theta }{2}\right) \\
	\end{array}
	\right)
\end{align}
Note that a generic rotation matrix in any direction in bloch sphere can be written in an exponential form
\begin{align}
	R_{\hat{n}}(\theta)=e^{-i\frac{\theta}{2} \hat{n}.\hat{\sigma} }
\end{align}
Quite interestingly, $U_H(t)$ corresponds to an experimentally realizable unitary operator with the Hamiltonian given by
\begin{align}
	U_H(t)=e^{-I H t}, H=I^A \otimes\sigma_{z}^B\otimes\sigma_{z}^C
\end{align}
It is easy to check that if we set $\theta =2 t$ then the full unitary is a physical realization of the C-Not gate for $t=\frac{\pi}{4}$. In other words the unitary operator of our interest is a product of three unitaries
\begin{align}
	U(t)=e^{-i H_1 t}e^{-i H_2 t}e^{-i H_3 t}
\end{align}
where the $H_1,H_2,H_3$ are given by
\begin{align}
	H_1=I^A \otimes I^B \otimes\sigma^C_x\\
 H_2=I^A \otimes\sigma_{z}^B\otimes\sigma_{z}^C\\
 H_3=I^A \otimes I^B \otimes \sigma^C_y
\end{align}
The  unitary operators are constructed such that at $t = 0$, the system is in a biseparable Bell state in \cref{bisepu}, and by $t = \pi/4$, it evolves into a GHZ state. We have numerically studied the behavior of both the L-entropy, denoted as $\ell_{ABC}$, and the Markov gap, $h_{ABC}$, under this time evolution. The resulting plots are shown in Fig.~\ref{BellToGHZ}. As can be observed, $\ell_{ABC}$ increases monotonically from 0 to 1 as $t$ progresses from 0 to $\pi/4$. In contrast, the Markov gap $h_{ABC}$ initially increases, reaches a maximum, and then returns to zero.

\section{2-Uniform states found by optimization}\label{2unioptim}
Below we liste out the 2-uniform states we found by optimization described in \cref{optimization}
 $\ket{\psi_1}$ represents a 2-uniform state for 5-qubits.  Similarly, $ \ket{\psi_2}, \ket{\psi_3}$ are representatives of 6-qubit 2-uniform states, and $ \ket{\psi_4}, \ket{\psi_5}$ pertain to 7-qubit 2-uniform states. Meanwhile, $\ket{\psi_6}$ corresponds to a 2-uniform state for 4 qutrits.

\begin{align}
    \ket{\psi_1} &= {1\over \sqrt{8}}\big(|00000\rangle +|00110\rangle + |01111\rangle + |10101\rangle - |01001\rangle - |10011\rangle - |11010\rangle - |11100\rangle \big) \nonumber\\
    \ket{\psi_2} &= {1\over \sqrt{8}}\big(|000100\rangle+|011000\rangle+|011111\rangle+|101110\rangle+|110010\rangle-|000011\rangle-|101001\rangle-|110101\rangle\big) \nonumber\\
    \ket{\psi_3} &= {1\over \sqrt{8}}\big(|001000\rangle+|010100\rangle+|100010\rangle+|100101\rangle-|001111\rangle-|010011\rangle-|111001\rangle-|111110\rangle\big)\nonumber \\
    \ket{\psi_4} &= {1\over \sqrt{8}}\big(|0000011\rangle+|0010100\rangle+|0101110\rangle+|0111001\rangle\nonumber\\&\quad\quad\quad\quad\quad\quad\quad\quad\quad\quad\quad\quad\quad\quad\quad\quad\quad+|1001101\rangle+|1011010\rangle+|1100000\rangle+|1110111\rangle\big) \nonumber\\
    \ket{\psi_5} &= {1\over \sqrt{16}} \big(|0000000\rangle+|0001011\rangle+|0011001\rangle+|0110010\rangle+|0110100\rangle\big) \nonumber \\
    &\quad +{1\over \sqrt{16}} \big(|1000110\rangle+|1010011\rangle+|1100001\rangle+|1101010\rangle+|1111111\rangle\big) \nonumber \\
    &\quad -{1\over \sqrt{16}} \big(|0011110\rangle+|0100111\rangle+|0101101\rangle+|1001100\rangle+|1010101\rangle+|1111000\rangle\big) \nonumber\\
    \ket{\psi_6} &= {1\over \sqrt{9}}\big(|0121\rangle +|0202\rangle + |1022\rangle + |1100\rangle + |2001\rangle + |2112\rangle - |0010\rangle - |1211\rangle  - |2220\rangle \big)
\end{align}

\section{Numerical evidence for agreement of three methods}
\label{app: three methods}
\begin{figure}[H]
	\centering
	\begin{subfigure}[h]{0.22\textwidth}
		\centering
		\includegraphics[width=4.5cm,height=4cm]{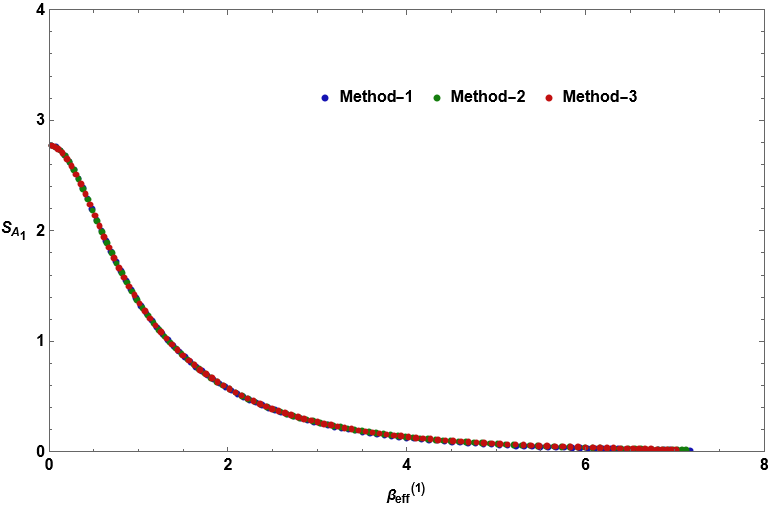}
		\caption{Party-1}
		\label{MEE41}
	\end{subfigure}
	\hfill
	\begin{subfigure}[h]{0.22\textwidth}
		\centering
		\includegraphics[width=4.5cm,height=4cm]{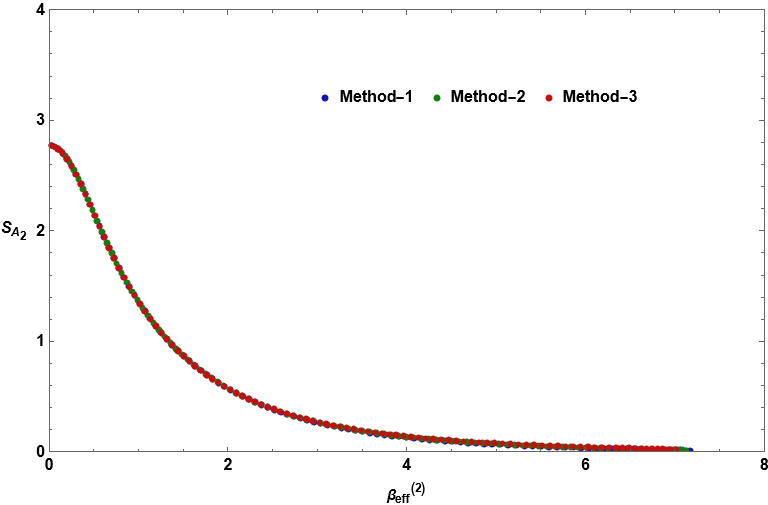}
		\caption{Party-2}
		\label{MEE42}
	\end{subfigure}
 \hfill
	\begin{subfigure}[h]{0.22\textwidth}
		\centering
		\includegraphics[width=4.5cm,height=4cm]{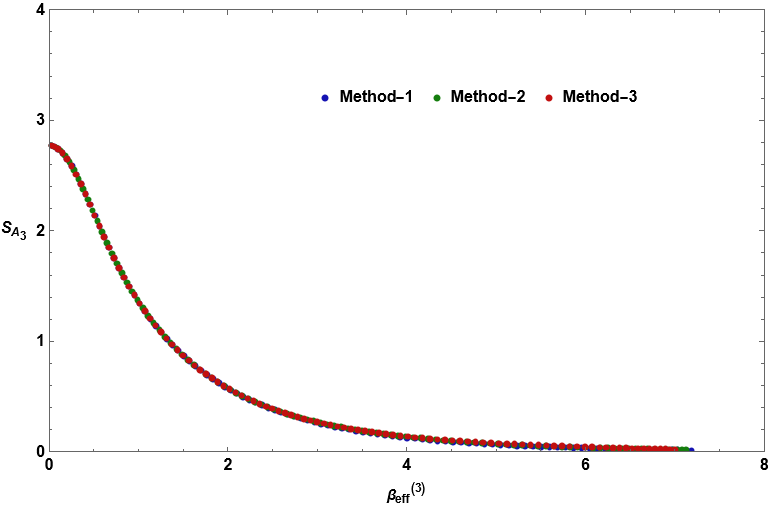}
		\caption{Party-3}
		\label{MEE43}
  	\end{subfigure}
  \hfill
	\begin{subfigure}[h]{0.22\textwidth}
		\centering
		\includegraphics[width=4.5cm,height=4cm]{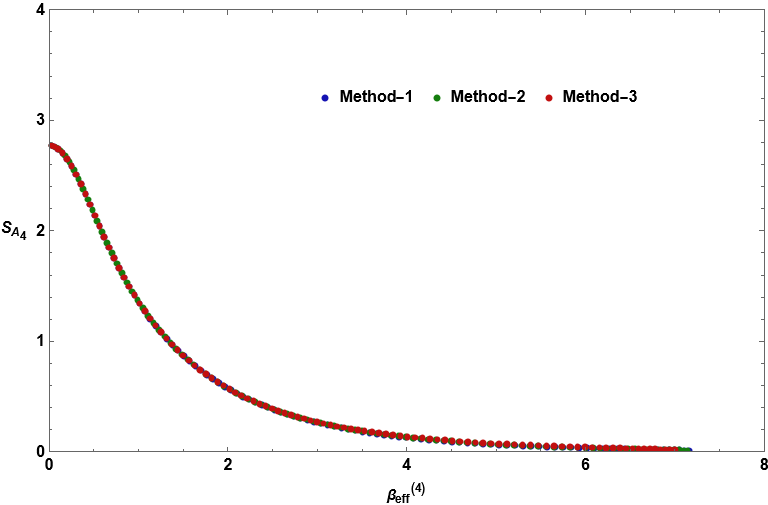}
		\caption{Party-4}
		\label{MEE44}
	\end{subfigure}
	\caption{ EE of four parties quantum system where each party is given by 4-qubit systems derived from the multicopy SYK model. Comparison between entanglement entropies of all parties using different methods.}
	\label{EEthplotsM4}
\end{figure}

\begin{figure}[H]
	\centering
	\begin{subfigure}[h]{0.22\textwidth}
		\centering
		\includegraphics[width=5cm,height=4.3cm]{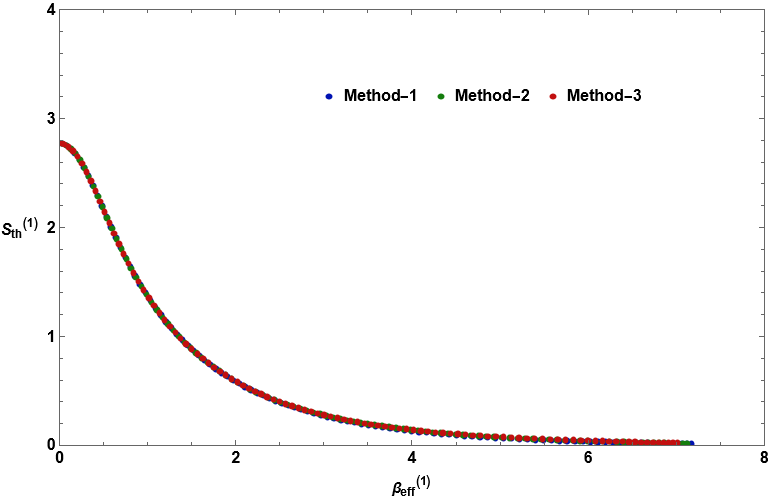}
		\caption{Party-1}
		\label{sM41}
	\end{subfigure}
	\hfill
	\begin{subfigure}[h]{0.22\textwidth}
		\centering
		\includegraphics[width=5cm,height=4.5cm]{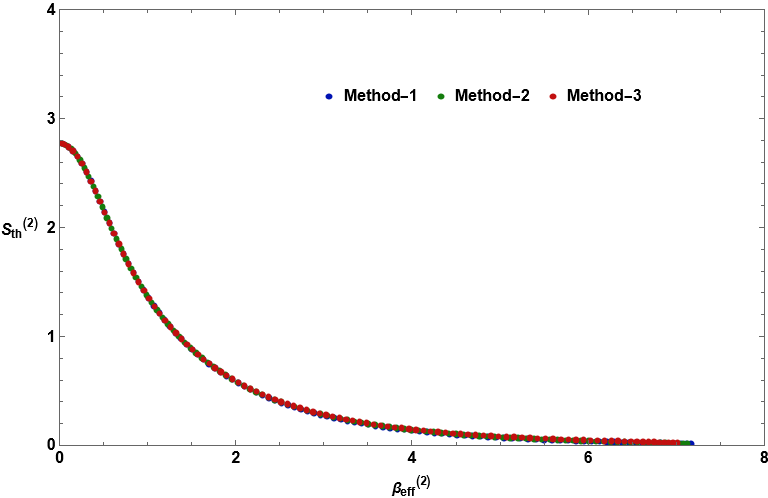}
		\caption{Party-2}
		\label{sM42}
	\end{subfigure}
 \hfill
	\begin{subfigure}[h]{0.22\textwidth}
		\centering
		\includegraphics[width=5cm,height=4.5cm]{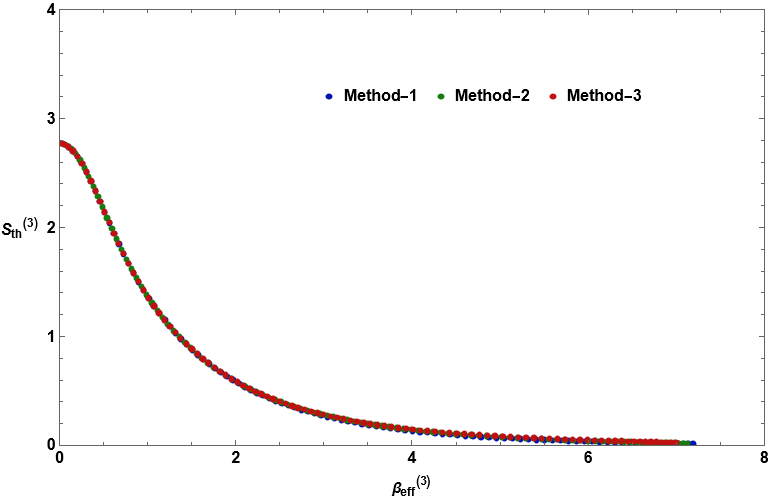}
		\caption{Party-3}
		\label{sM43}
  	\end{subfigure}
  \hfill
	\begin{subfigure}[h]{0.22\textwidth}
		\centering
		\includegraphics[width=5cm,height=4.5cm]{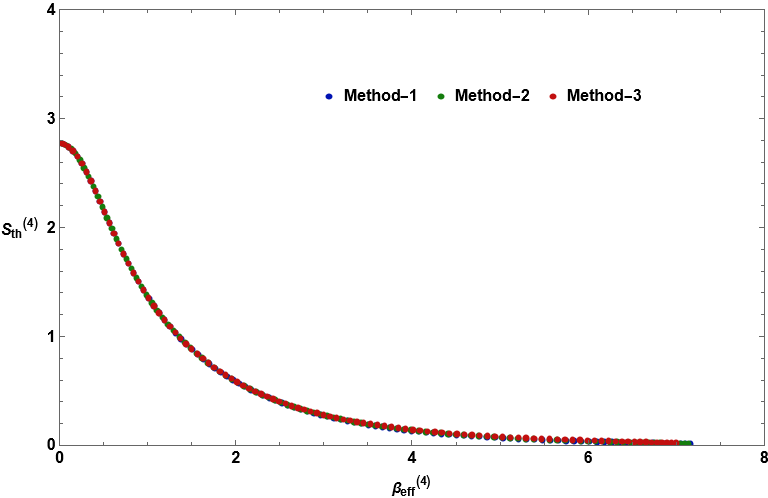}
		\caption{Party-4}
		\label{sM44}
	\end{subfigure}
	\caption{  Comparison between thermal entropies of all parties using different methods. }
	\label{EvplotsM4}
\end{figure}

\begin{figure}[H]
	\centering
	\begin{subfigure}[h]{0.22\textwidth}
		\centering
		\includegraphics[width=4.5cm,height=4cm]{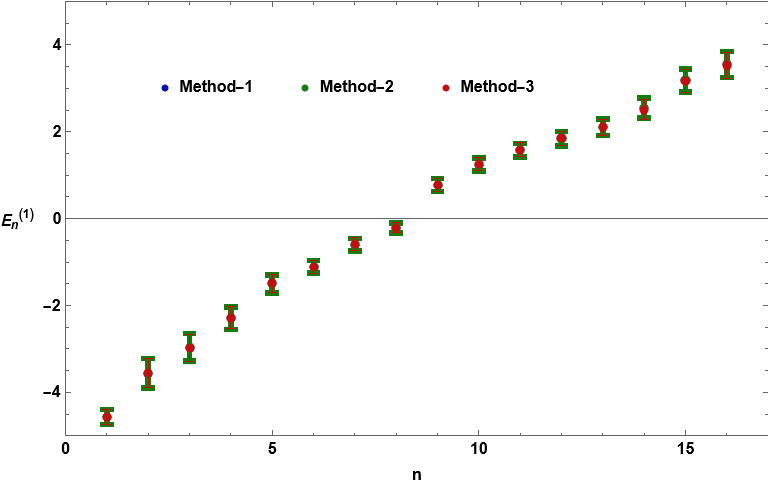}
		\caption{Party-1}
		\label{EE41}
	\end{subfigure}
	\hfill
	\begin{subfigure}[h]{0.22\textwidth}
		\centering
		\includegraphics[width=4.5cm,height=4cm]{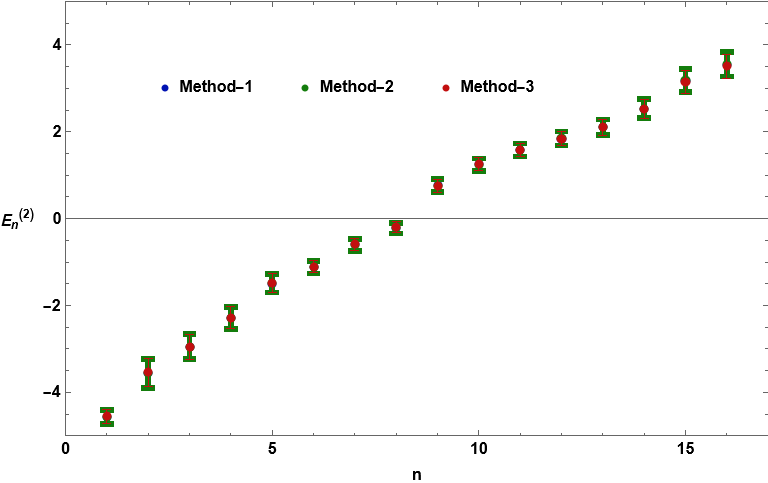}
		\caption{Party-2}
		\label{EE42}
	\end{subfigure}
 \hfill
	\begin{subfigure}[h]{0.22\textwidth}
		\centering
		\includegraphics[width=4.5cm,height=4cm]{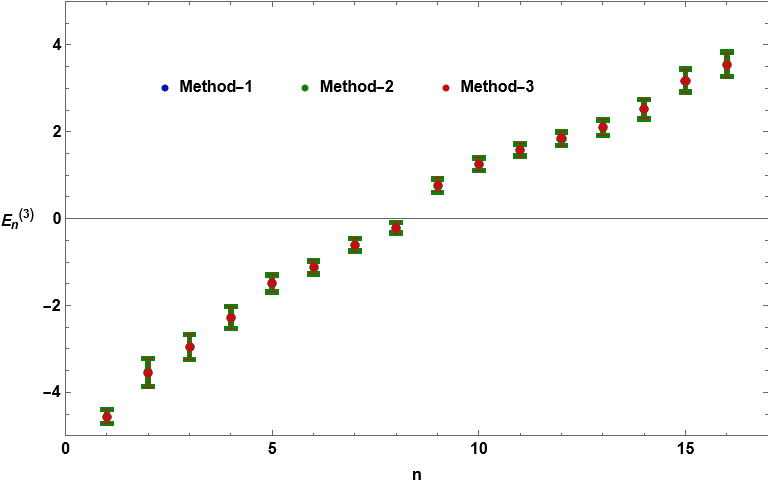}
		\caption{Party-3}
		\label{EE43}
  	\end{subfigure}
  \hfill
	\begin{subfigure}[h]{0.22\textwidth}
		\centering
		\includegraphics[width=4.5cm,height=4cm]{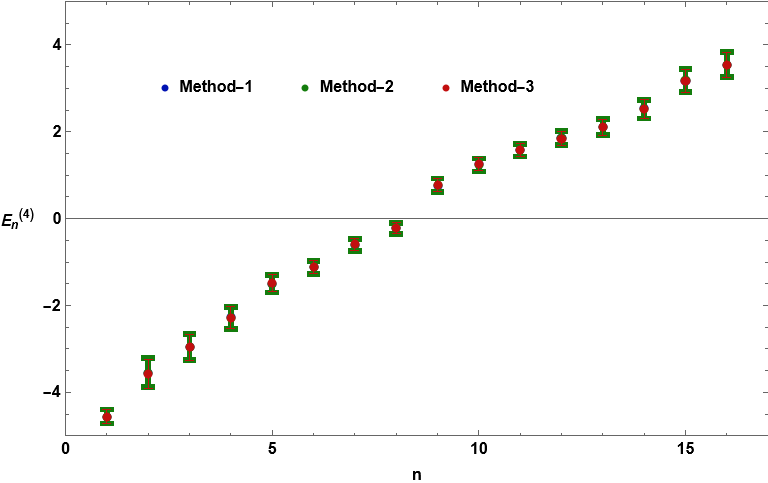}
		\caption{Party-4}
		\label{EE44}
	\end{subfigure}
	\caption{ A comparison of the eigenvalues of the Hamiltonian for each party in a thermal state, employing three different methods. }
	\label{Evplots4_app}
\end{figure}

\begin{figure}[H]
	\centering
	\begin{subfigure}[h]{0.22\textwidth}
		\centering
		\includegraphics[width=4.5cm,height=4cm]{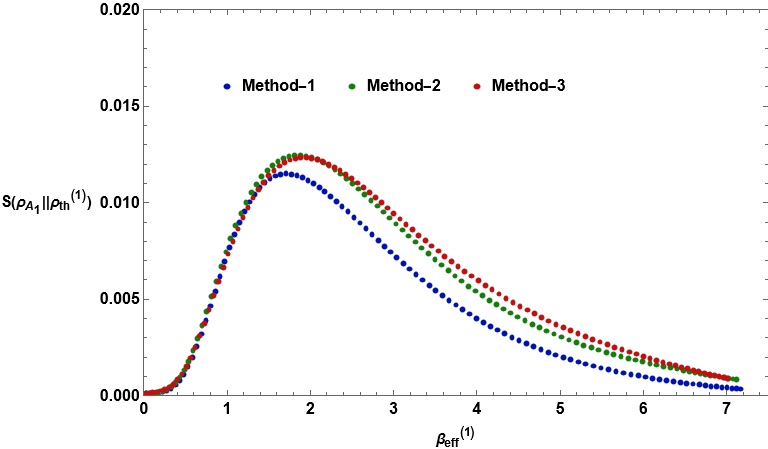}
		\caption{Party-1}
		\label{Relm41}
	\end{subfigure}
	\hfill
	\begin{subfigure}[h]{0.22\textwidth}
		\centering
		\includegraphics[width=4.5cm,height=4cm]{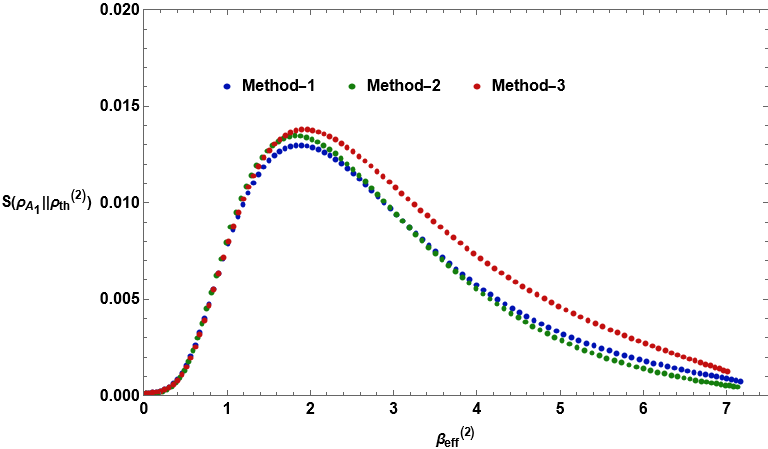}
		\caption{Party-2}
		\label{Relm42}
	\end{subfigure}
 \hfill
	\begin{subfigure}[h]{0.22\textwidth}
		\centering
		\includegraphics[width=4.5cm,height=4cm]{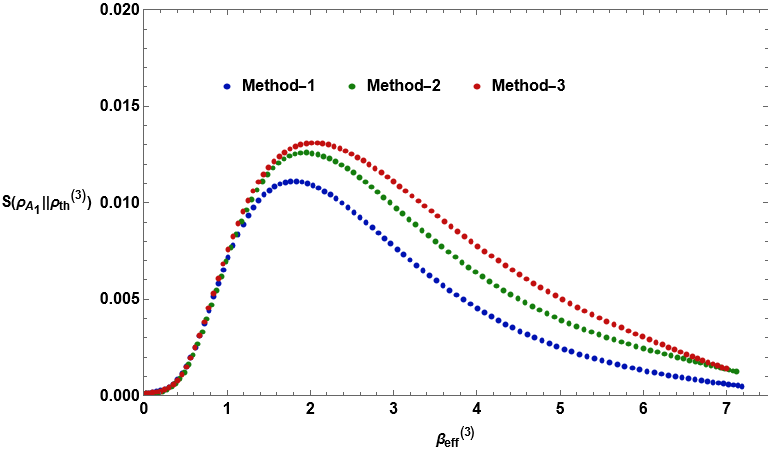}
		\caption{Party-3}
		\label{Relm43}
  	\end{subfigure}
  \hfill
	\begin{subfigure}[h]{0.22\textwidth}
		\centering
		\includegraphics[width=4.5cm,height=4cm]{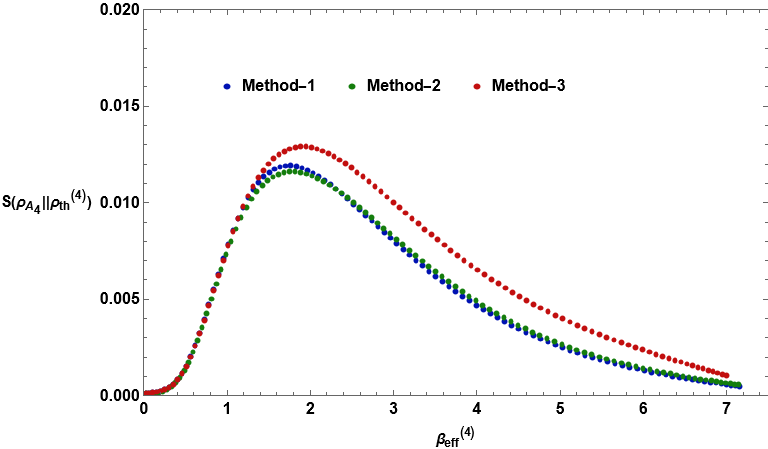}
		\caption{Party-4}
		\label{Relm44}
	\end{subfigure}
	\caption{ A comparison of the relative entropy between the reduced state for each party and its corresponding thermal state, employing three different methods. }
	\label{Relmplots}
\end{figure}

\section{SYK model}
\label{app:SYK}

\subsection*{Single copy SYK model}
\label{sec: single SYK model}

The SYK model involves $N$ Majorana fermionic fields denoted by $\chi^i(t)\hspace{2mm} (i=1,2,\cdots, N)$ in a $(0+1)$-dimensions which obey the anti-commutation rule
\begin{align}
	\{\chi^i, \chi^j\} =\delta^{ij}
\end{align}
The Hamiltonian for the SYK model is expressed as
\begin{align}
	H=\sum_{i<j<k<l} J_{ijkl} \chi^i \chi^j \chi^k \chi^l
\end{align}
where $J_{ijkl}$ represents a random coupling constant sampled from a Gaussian distribution, with its variance specified by
\begin{align}
	\langle J_{ijkl} J_{ijkl}\rangle ={6\over N^3}
\end{align}
%

\subsection*{Fock space representation of SYK Model}
\label{sec: fock space}

To determine the (multipartite) L-entropy, we focus on the Hilbert space associated with the SYK model. In this context, employing the standard representation of the Majorana fermion $\chi_i$ ($i=1,2,\cdots, N$) (or equivalently, the gamma matrices) is convenient
\begin{equation}
	\chi_j={1\over \sqrt{2}} \gamma_i\hspace{5mm} (j=1, 1,2,\cdots, N)
\end{equation}
The corresponding fermionic oscillators, denoted as $b_j$ and $\bar{b}_j$ for index $j=1,2,\ldots,n\equiv N/2$, can be characterized in terms of the gamma matrices as follows: %
\begin{align}
	b_j\equiv&{1\over \sqrt{2}}(\chi_{2j-1} - i \chi_{2j})={1\over 2}(\gamma_{2j-1} - i \gamma_{2j})\\
	\bar{b}_j\equiv&{1\over \sqrt{2}}(\chi_{2j-1} + i \chi_{2j})={1\over 2}(\gamma_{2j-1} + i \gamma_{2j})
\end{align}
which in turn satisfy the following anti-commutation relations
\begin{equation}
	\{ b_j, \bar{b}_k\}= \delta_{jk}\hspace{5mm},\hspace{5mm} \{b_j,b_k\}=\{\bar{b}_j,\bar{b}_k\}=0
\end{equation}
In this study, we focus exclusively on the scenario where $N$ is even $(N=2n)$. Analogous to how spin chains are assessed, it is convenient to utilize Fock space for the computation of the L-entropy and other measures, leading to the following definition of the state
\begin{equation}
	\bar{b}_{j_1}\bar{b}_{j_2}\cdots \bar{b}_{j_a}|0\rangle\hspace{5mm}( j_1>j_2\cdots >j_a)
\end{equation}
We will systematically organize the states within the Fock space by employing a specific sequence. This involves assigning labels ranging from $j = 0, 1, \ldots, 2^n-1$ to each of the $2^n$ states. Subsequently, these labels are interpreted as binary numbers (for instance, $[\nu_n \nu_{n-1} \cdots \nu_2 \nu_1]$, where each $\nu_k$ is either $0$ or $1$). Consequently, the state is defined as $\bar{b}_n^{\nu_n} \cdots \bar{b}_2^{\nu_1} \bar{b}_1^{\nu_1} |0\rangle$. Refer to Table~\ref{tab: state} for further illustration.
\begin{table}[h!]
\centering
{\centering
{\renewcommand{\arraystretch}{1.5}
\begin{tabular}{>{\centering\arraybackslash}m{3cm}|>{\centering\arraybackslash}m{5cm}|>{\centering\arraybackslash}m{5cm}}
\hline
State Number &  Binary Label  & State\\       
\hline
\hline
0 & $|0 \cdots  000\rangle$ & $|0\rangle$\\
\hline
1 & $|0 \cdots  001\rangle$ & $\bar{b}_1|0\rangle$\\
\hline
2 & $|0 \cdots  010\rangle$ & $\bar{b}_2|0\rangle$\\
\hline
3 & $|0 \cdots  011\rangle$ & $\bar{b}_2\bar{b}_1|0\rangle$\\
\hline
4 & $|0 \cdots  100\rangle$ & $\bar{b}_3|0\rangle$\\
\hline
$\vdots$ & $\vdots$ & $\vdots$\\
\hline
$j$ & $|\nu_n \cdots  \nu_3\nu_2\nu_1\rangle $ & $\bar{b}_n^{\nu_n} \cdots \bar{b}_2^{\nu_1}\bar{b}_1^{\nu_1} |0\rangle$\\
\hline
$\vdots$ & $\vdots$ & $\vdots$\\
\hline
$2^n-1$ & $|1 \cdots  111\rangle$ & $\bar{b}_n \cdots \bar{b}_2 \bar{b}_1|0\rangle$\\
\hline
\end{tabular}}
}
\caption{State ordering in the Fock space}
\label{tab: state}
\end{table}
In this basis, one can easily take partial trace of density matrix for entanglement entropy. In addition, the gamma matrices in this basis are given in terms of the Pauli matrices as follows

\begin{align}
	\gamma_{2j-1}=&\underbracket{\sigma_3 \otimes\cdots \otimes \sigma_3}_{n-j} \otimes \underbracket{\sigma_1}_{j\text{\tiny th} } \otimes  \underbracket{\mathbb{I} \otimes\cdots \otimes \mathbb{I}}_{j-1} \\
	\gamma_{2j}=&\underbracket{\sigma_3 \otimes\cdots \otimes \sigma_3}_{n-j} \otimes \underbracket{\sigma_2}_{j\text{\tiny th} } \otimes  \underbracket{\mathbb{I} \otimes \cdots \otimes \mathbb{I}}_{j-1}
\end{align}
where $\sigma_i$ are the Pauli matrices and identity matrix are 
\begin{equation}
	\sigma_1=\begin{pmatrix}
	0 & 1\\
	1 & 0\\
	\end{pmatrix}\quad,\quad \sigma_2=\begin{pmatrix}
	0 & -i\\
	 i & 0\\
	\end{pmatrix}\quad,\quad \sigma_3= \begin{pmatrix}
	1 & 0\\
	0 & -1\\
	\end{pmatrix}\quad,\quad \mathbb{I} =\begin{pmatrix}
	1 & 0 \\
	0 & 1\\
	\end{pmatrix}
\end{equation}

\section{Multipartite TPQ states: 3 and 4 party} \label{sec:MTPQ34}
\begin{figure}[H]
	\centering
	\begin{subfigure}[h]{0.3\textwidth}
		\centering
		\includegraphics[width=45mm]{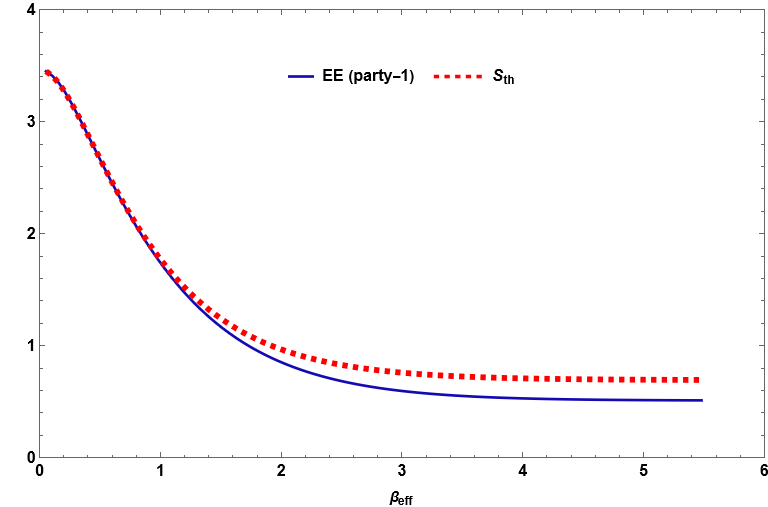}
		\caption{Party-1}
		\label{EE_3_1}
	\end{subfigure}
	\hspace{0.05cm}
	\begin{subfigure}[h]{0.3\textwidth}
		\centering
		\includegraphics[width=45mm]{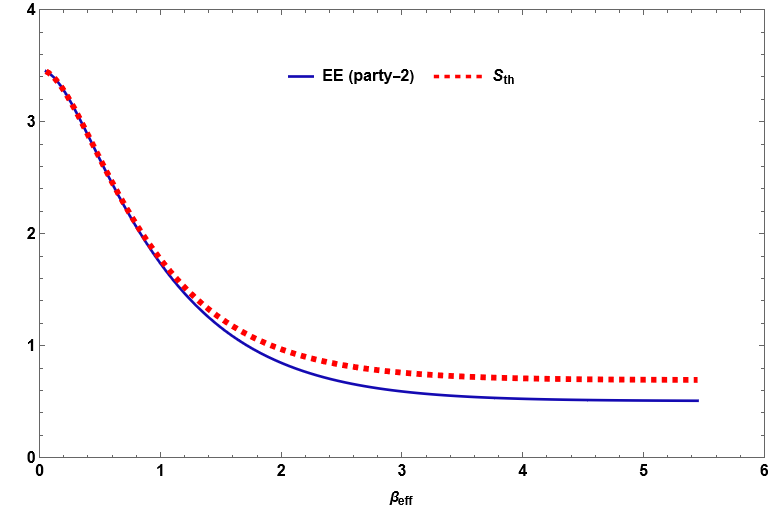}
		\caption{Party-2}
		\label{EE_3_2}
	\end{subfigure}
 \hspace{0.05cm}
	\begin{subfigure}[h]{0.3\textwidth}
		\centering
		\includegraphics[width=45mm]{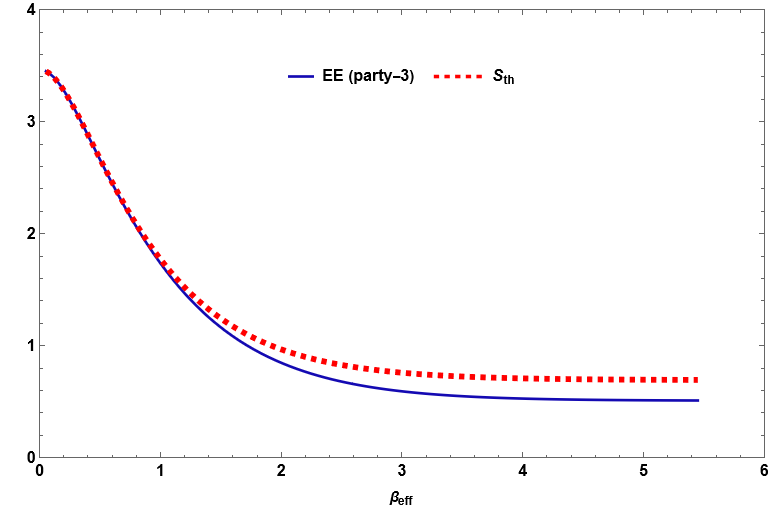}
		\caption{Party-3}
		\label{EE_3_3}
  	\end{subfigure}

	\caption{ 3 parties of 5-qubits from SYK. Comparision between single party entanglement entropy with the corresponding thermal entropy for SYK}
	\label{EEthplots3}
\end{figure}

\begin{figure}[H]
	\centering
	\begin{subfigure}[h]{0.3\textwidth}
		\centering
		\includegraphics[width=53mm]{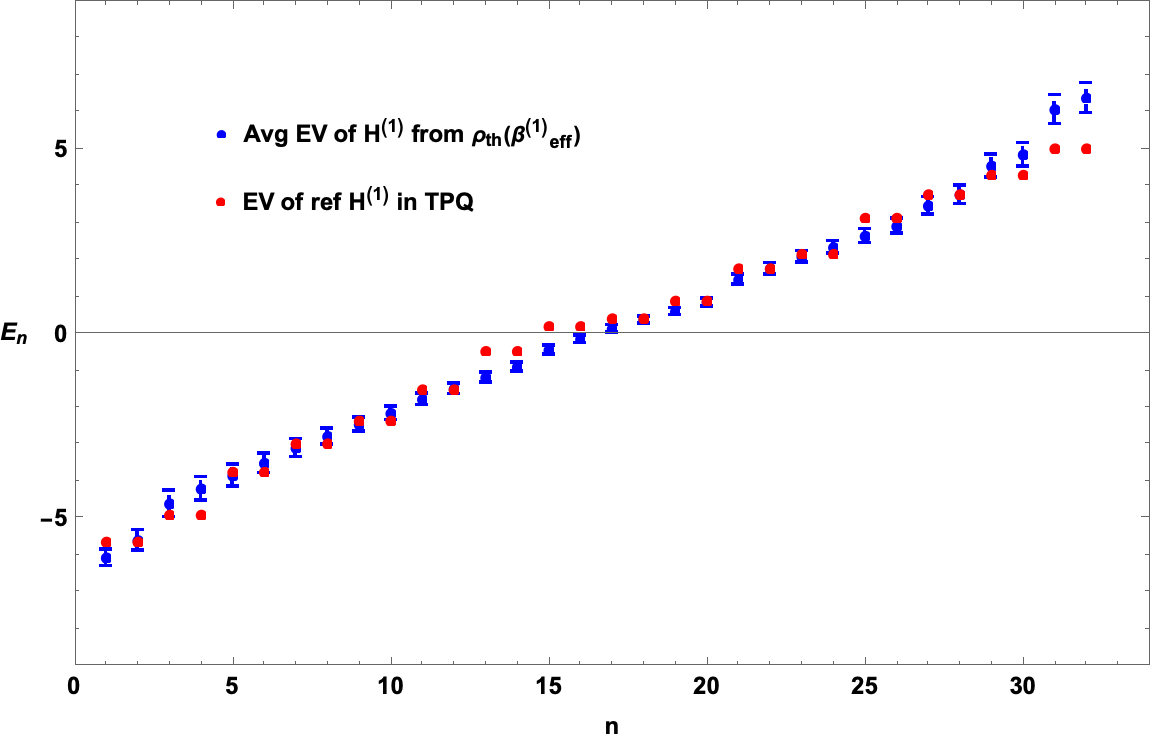}
		\caption{Party-1}
		\label{En_3_1}
	\end{subfigure}
	\hspace{0.05cm}
	\begin{subfigure}[h]{0.3\textwidth}
		\centering
		\includegraphics[width=53mm]{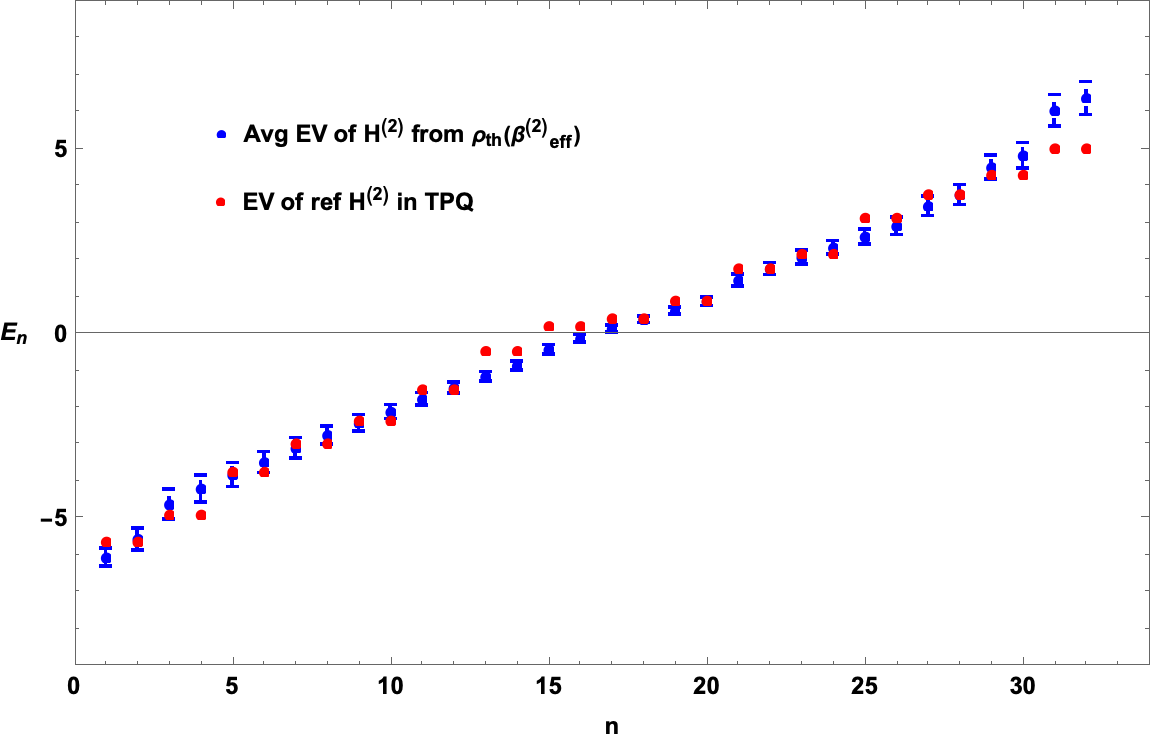}
		\caption{Party-2}
		\label{En_3_2}
	\end{subfigure}
 \hspace{0.05cm}
	\begin{subfigure}[h]{0.3\textwidth}
		\centering
		\includegraphics[width=53mm]{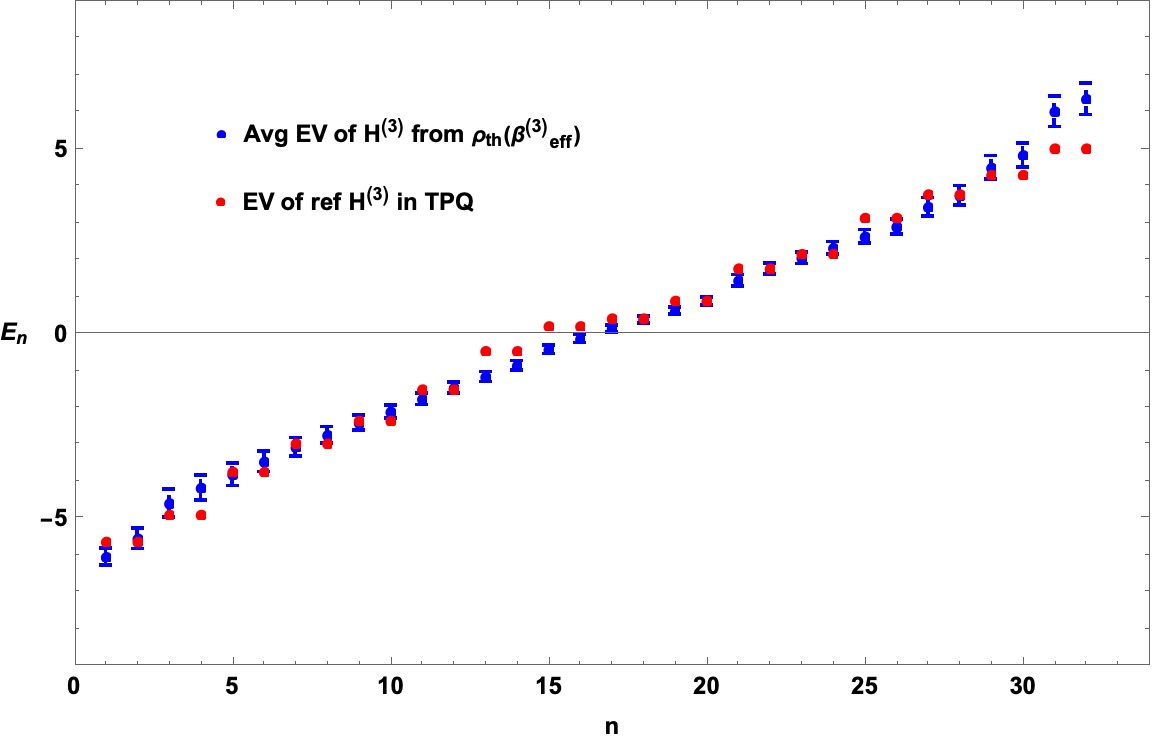}
		\caption{Party-3}
		\label{En_3_3}
  	\end{subfigure}
	\caption{ 3 parties of 5-qubits from SYK. Comparision between eigen values of the reference Hamiltonian-$H^(i)$ in the TPQ state with the corresponding thermal Hamiltonian $H^(i)$ with corresponding eff temperature. }
	\label{Evplots3}
\end{figure}

\begin{figure}[H]
	\centering
	\begin{subfigure}[h]{0.45\textwidth}
		\centering
		\includegraphics[width=1\textwidth]{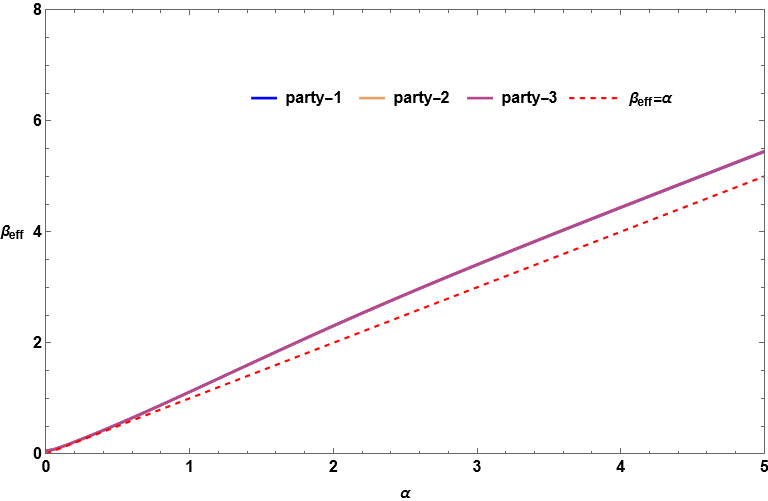}
		\caption{Effective Temperature vs $\alpha$ parameter}
		\label{Effective_Temp_3}
	\end{subfigure}
	\hfill
    \begin{subfigure}[h]{0.45\textwidth}
		\centering
		\includegraphics[width=1\textwidth]{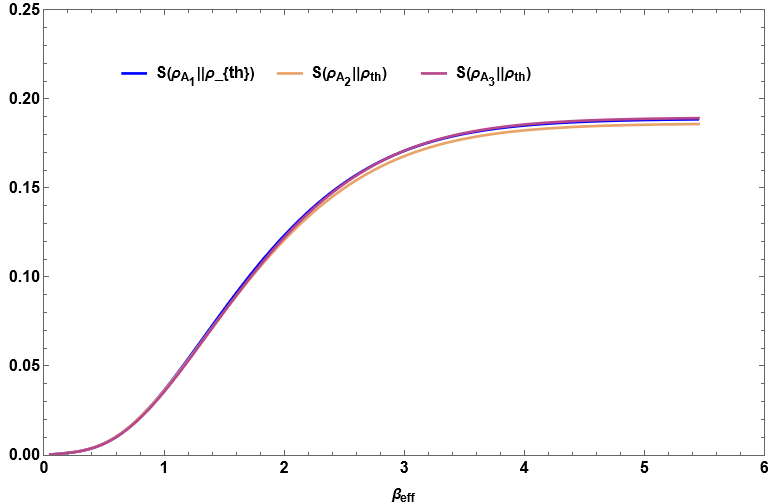}
		\caption{Relative entropy between single party reduced state and thermal state .}
		\label{Relative_entropy_3}
	\end{subfigure}
    \caption{}
	\end{figure}

\begin{figure}[H]
	\centering
	\begin{subfigure}[h]{0.22\textwidth}
		\centering
		\includegraphics[width=45mm]{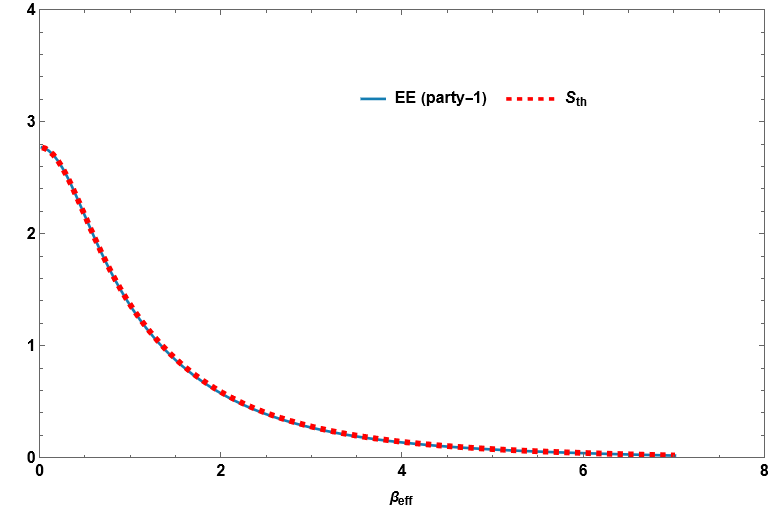}
		\caption{Party-1}
		\label{EE_4_1}
	\end{subfigure}
	\hspace{0.05cm}
	\begin{subfigure}[h]{0.22\textwidth}
		\centering
		\includegraphics[width=45mm]{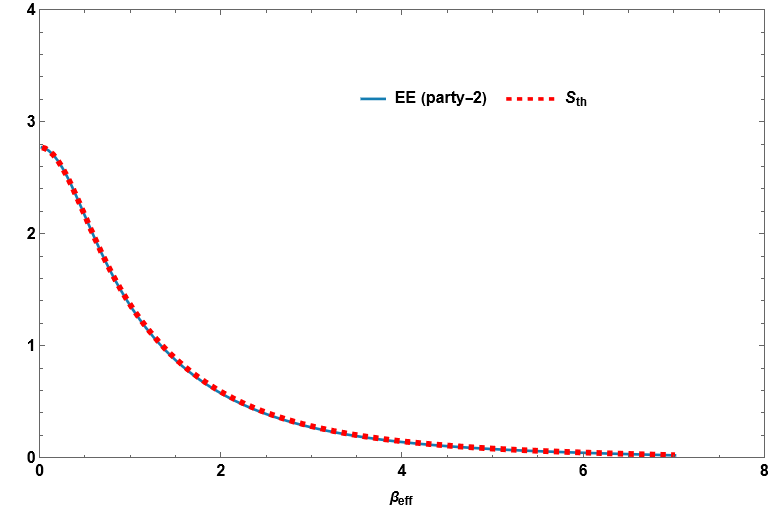}
		\caption{Party-2}
		\label{EE_4_2}
	\end{subfigure}
 \hspace{0.05cm}
	\begin{subfigure}[h]{0.22\textwidth}
		\centering
		\includegraphics[width=45mm]{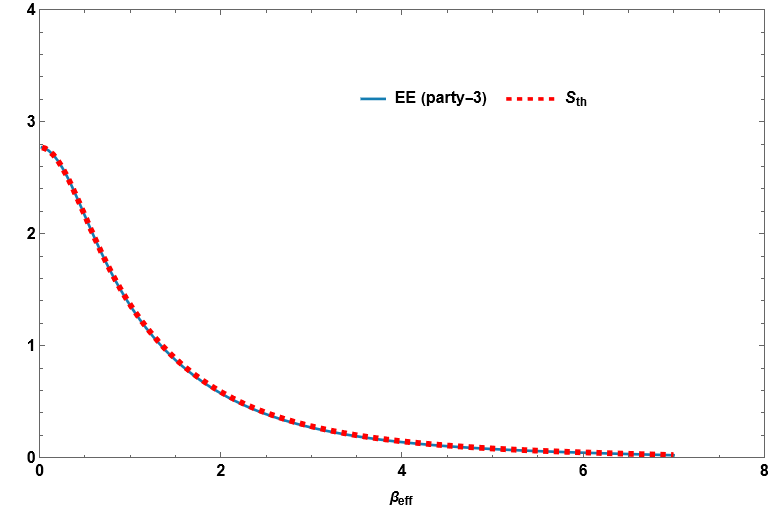}
		\caption{Party-3}
		\label{EE_4_3}
  	\end{subfigure}
  \hspace{0.05cm}
	\begin{subfigure}[h]{0.22\textwidth}
		\centering
		\includegraphics[width=45mm]{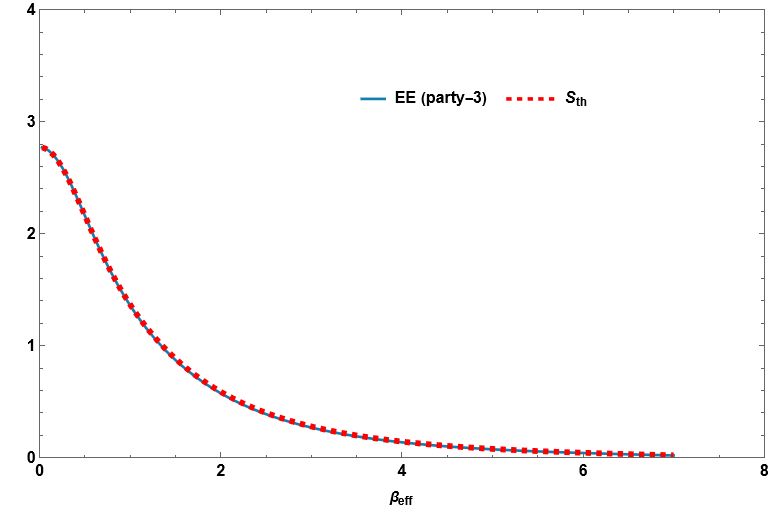}
		\caption{Party-4}
		\label{EE_4_4}
	\end{subfigure}
	\caption{ 4 parties of 4-qubits from SYK. Comparision between single party entanglement entropy with the corresponding thermal entropy for SYK}
	\label{EEthplots4}
\end{figure}

\begin{figure}[H]
	\centering
	\begin{subfigure}[h]{0.22\textwidth}
		\centering
		\includegraphics[width=4.5cm,height=4cm]{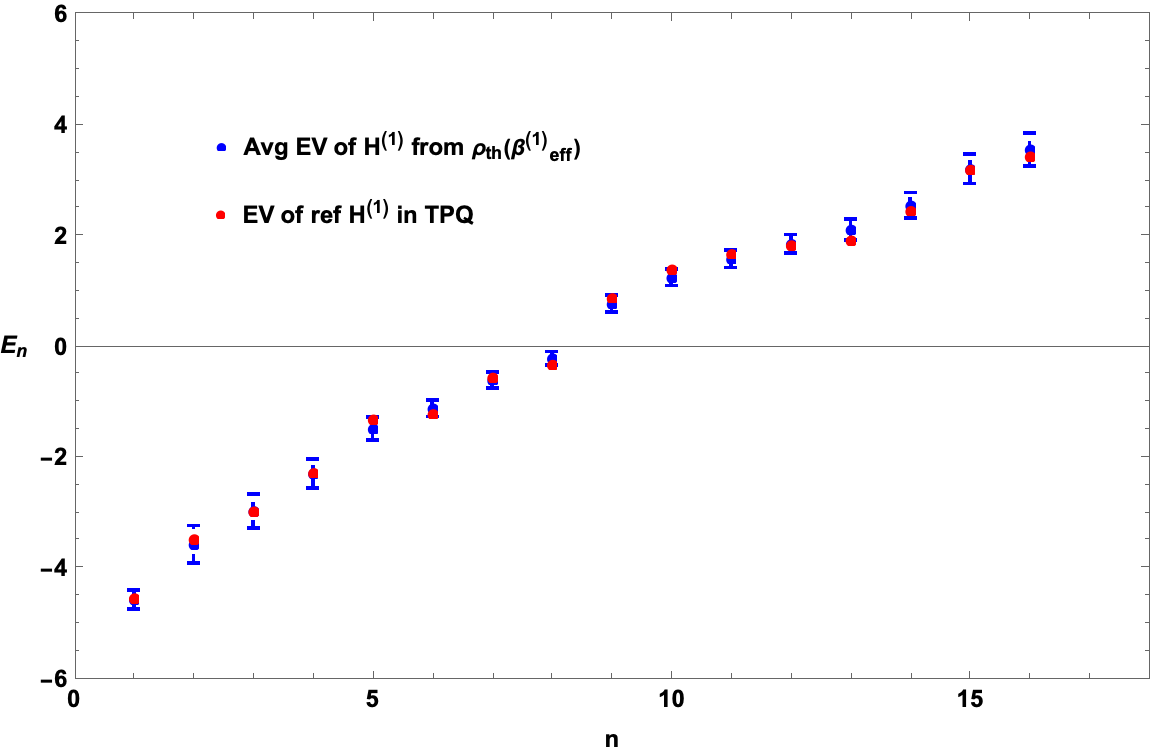}
		\caption{Party-1}
		\label{En_4_1}
	\end{subfigure}
	\hfill
	\begin{subfigure}[h]{0.22\textwidth}
		\centering
		\includegraphics[width=4.5cm,height=4cm]{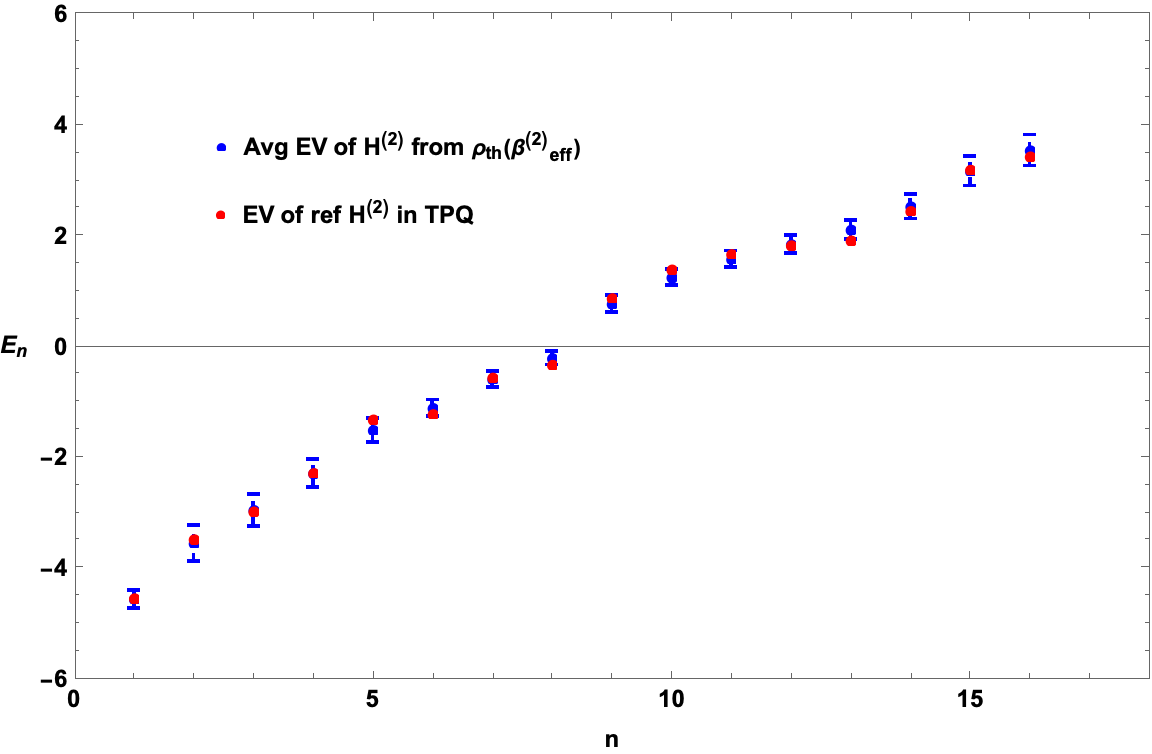}
		\caption{Party-2}
		\label{En_4_2}
	\end{subfigure}
 \hfill
	\begin{subfigure}[h]{0.22\textwidth}
		\centering
		\includegraphics[width=4.5cm,height=4cm]{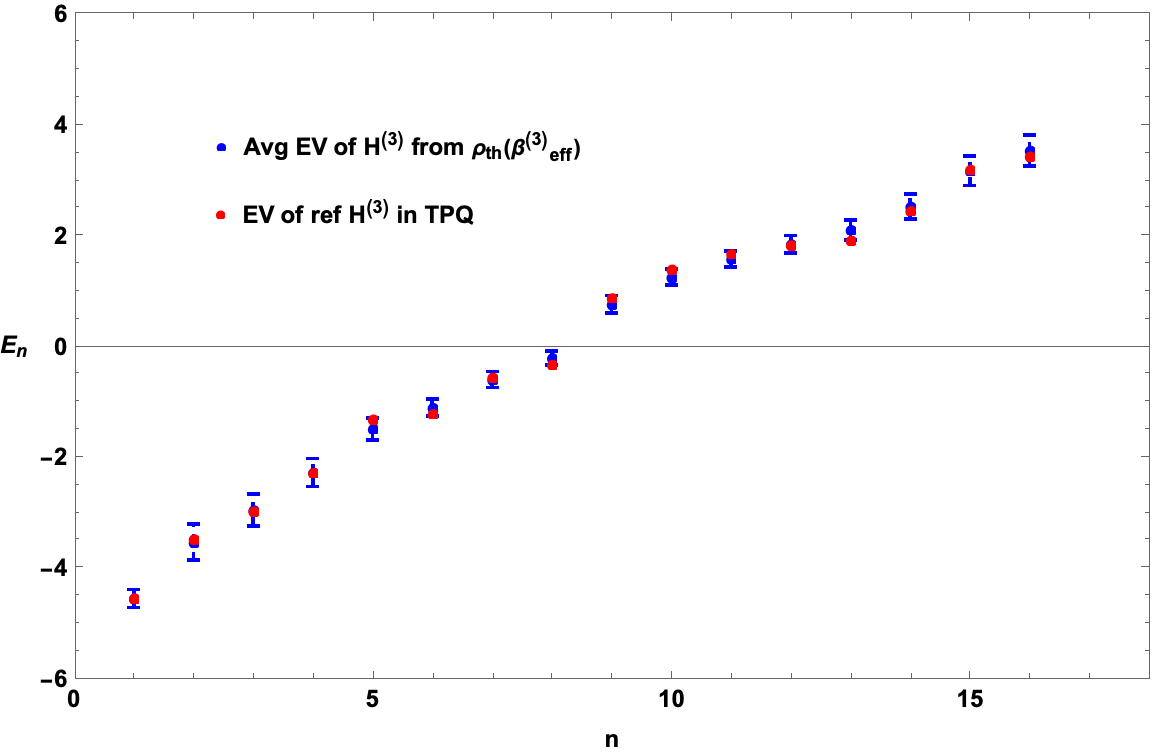}
		\caption{Party-3}
		\label{En_4_3}
  	\end{subfigure}
  \hfill
	\begin{subfigure}[h]{0.22\textwidth}
		\centering
		\includegraphics[width=4.5cm,height=4cm]{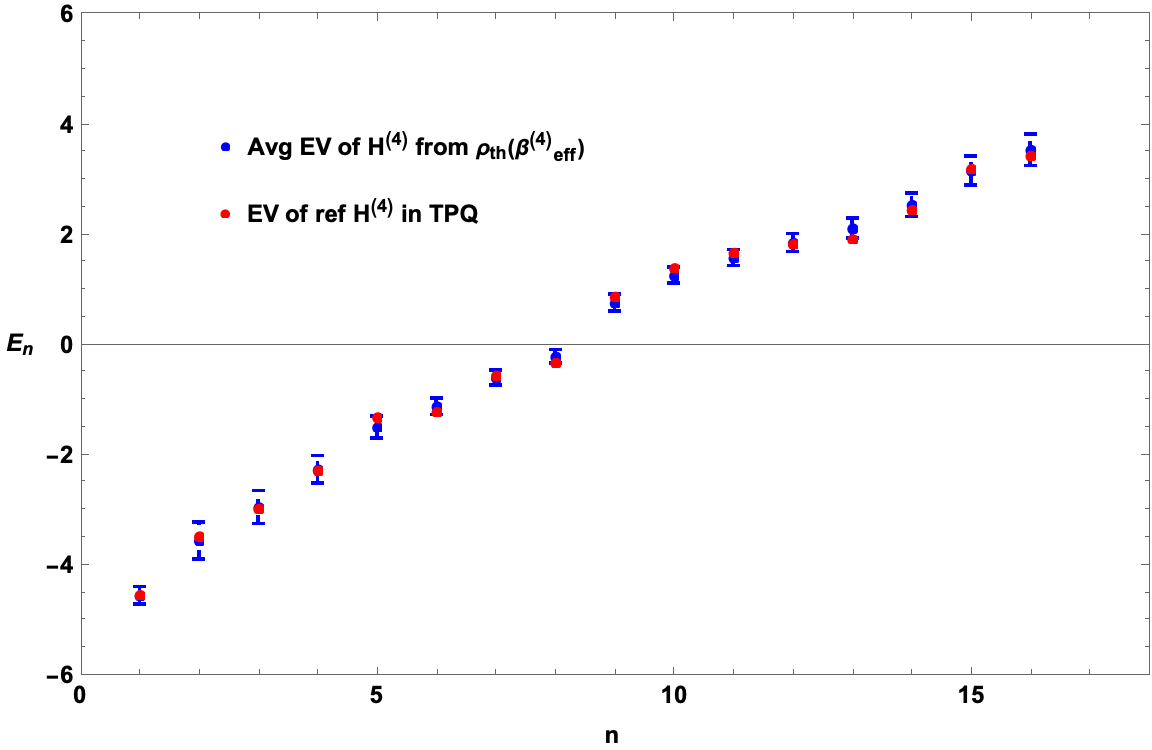}
		\caption{Party-4}
		\label{En_4_4}
	\end{subfigure}
	\caption{ 4 parties of 4-qubits from SYK. Comparision between eigen values of the reference Hamiltonian-$H^(i)$ in the TPQ state with the corresponding thermal Hamiltonian $H^(i)$ with corresponding eff temperature. }
	\label{Evplots4}
\end{figure}

\begin{figure}[H]
	\centering
	\begin{subfigure}[h]{0.45\textwidth}
		\centering
		\includegraphics[width=1\textwidth]{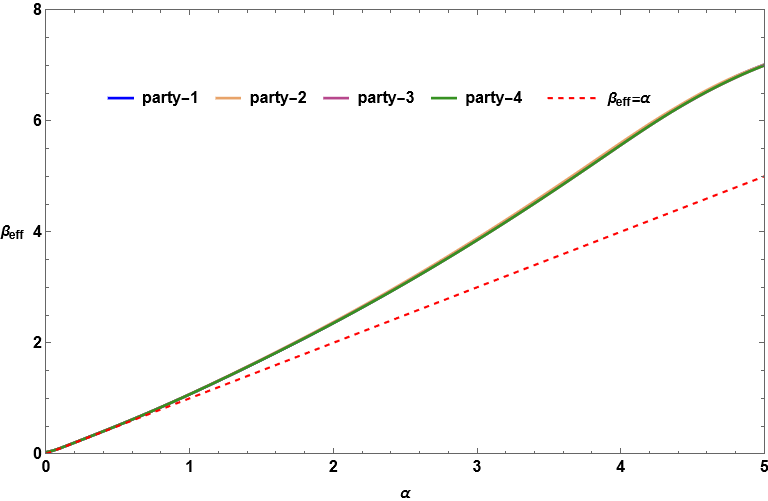}
		\caption{Effective Temperature vs $\alpha$ parameter}
		\label{Effective_Temp_4}
	\end{subfigure}
    \hfill
    \begin{subfigure}[h]{0.45\textwidth}
		\centering
		\includegraphics[width=1\textwidth]{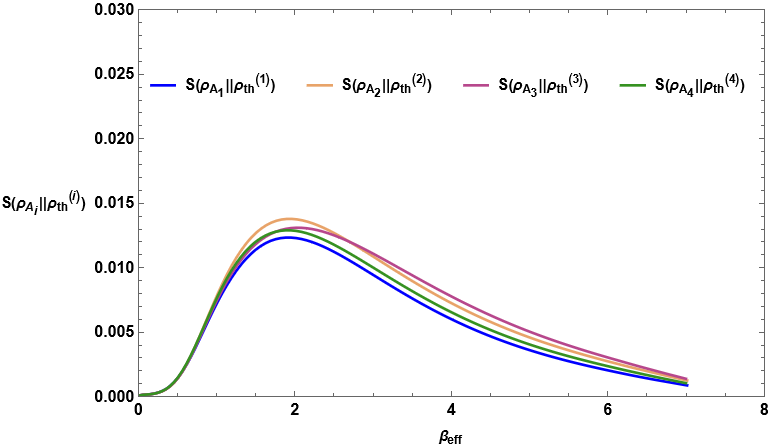}
		\caption{Relative entropy between single party reduced state and thermal state .}
		\label{Relative_entropy_4}
	\end{subfigure}
    \caption{}
	\end{figure}

Below, we present the behavior of various quantities for the $3$- and $4$-party MTPQ states in the multi-copy SYK model. For each individual party, we plot the entanglement entropy and the corresponding thermal entropy in Fig.~\ref{EEthplots3} and Fig.~\ref{EEthplots4}, for the three- and four-party states, respectively. Next, we analyze the eigenvalues of the single-party Hamiltonians used in the construction of the MTPQ states, as well as the eigenvalues of the Hamiltonian for each party obtained by assuming that it is in thermal state  with an effective temperature. These results are shown in Fig.~\ref{Evplots3} and Fig.~\ref{Evplots4}. The effective temperature is plotted as a function of the $\alpha$ parameter in Fig.~\ref{Effective_Temp_3} and Fig.~\ref{Effective_Temp_4}. Finally, the relative entropy between the reduced state of each party and the corresponding thermal state is depicted in Fig.~\ref{Relative_entropy_3} and Fig.~\ref{Relative_entropy_4} for the three- and four-party systems, respectively.

\subsection*{3-partite}
\label{sec:3party}

\subsection*{4-partite}
\label{sec:4party}

\bibliographystyle{JHEP}

\bibliography{arxiv_3}

\end{document}